\documentclass[10pt]{iopart}

\usepackage{iopams}
\expandafter\let\csname equation*\endcsname\relax
\expandafter\let\csname endequation*\endcsname\relax
\usepackage{amsmath}
\usepackage{graphicx}
\usepackage{xcolor}
\usepackage{pifont}
\usepackage{bm}
\usepackage{tocloft}

\let\oldr\r

\newcommand{\cmVs}{cm$^2$/Vs}
\renewcommand{\equiv}{=}                

\begin{document}

\review{First-principles calculations of charge carrier mobility and conductivity
in bulk semiconductors and two-dimensional materials}

\author{Samuel Ponc\'e$^1$, Wenbin Li$^1$, Sven Reichardt$^1$, Feliciano Giustino$^1$}

\address{$^1$ Department of Materials, University of Oxford, Parks Road, Oxford, OX1 3PH, UK}
\ead{samuel.ponce@materials.ox.ac.uk}
\ead{wenbin.li@materials.ox.ac.uk}
\ead{sven.reichardt@materials.ox.ac.uk}
\ead{feliciano.giustino@materials.ox.ac.uk}

\vspace{10pt}
\begin{indented}
\item[]June 2019
\end{indented}

\begin{abstract}
One of the fundamental properties of semiconductors is their ability to support
highly tunable electric currents in the presence of electric fields or carrier concentration gradients.
These properties are described by transport coefficients such as electron and hole mobilities.
Over the last decades, our understanding of carrier mobilities has largely been shaped by
experimental investigations and empirical models.
Recently, advances in electronic structure methods for real materials have made it possible
to study these properties with predictive accuracy and without resorting to empirical parameters.
These new developments are unlocking exciting new opportunities,
from exploring carrier transport in quantum matter to \textit{in silico} designing new semiconductors with tailored transport properties.
In this article, we review the most recent developments in the area of \textit{ab initio} calculations of carrier mobilities of semiconductors.
Our aim is threefold:
to make this rapidly-growing research area accessible to a broad community of condensed-matter theorists and materials scientists;
to identify key challenges that need to be addressed in order to increase the predictive power of these methods;
and to identify new opportunities for increasing the impact of these computational methods on the science and technology of advanced materials.
The review is organized in three parts.
In the first part, we offer a brief historical overview of approaches to the calculation of carrier mobilities,
and we establish the conceptual framework underlying modern \textit{ab initio} approaches.
We summarize the Boltzmann theory of carrier transport and we discuss its
scope of applicability, merits, and limitations in the broader context of many-body Green's function approaches.
We discuss recent implementations of the Boltzmann formalism within the context of density functional theory
and many-body perturbation theory calculations, placing an emphasis on the key computational challenges and suggested solutions.
In the second part of the article, we review applications of these methods to materials of current interest,
from three-dimensional semiconductors to layered and two-dimensional materials.
In particular, we discuss in detail recent investigations of classic materials such as silicon, diamond, gallium arsenide, gallium nitride, gallium oxide, and lead halide perovskites
as well as low-dimensional semiconductors such as graphene, silicene, phosphorene, molybdenum disulfide, and indium selenide.
We also review recent efforts toward high-throughput calculations of carrier transport.
In the last part, we identify important classes of materials for which an \textit{ab initio} study of carrier mobilities is warranted.
We discuss the extension of the methodology to study topological quantum matter and materials for spintronics
and we comment on the possibility of incorporating Berry-phase effects and many-body correlations beyond the standard Boltzmann formalism.
\end{abstract}

%
\vspace{2pc}
\submitto{\RPP}
%
\maketitle
%
\ioptwocol

\begin{footnotesize}
\tableofcontents
\end{footnotesize}


\section{Introduction}

The carrier mobility $\mu$ quantifies how fast an electron or hole can travel in a metal or
in a  semiconductor when subjected to an external electric field $E$.
The average velocity is called the \emph{drift velocity} $v_{\mathrm{d}}$ and can be
determined, for example, via Hall measurements.
The change of drift velocity with electric field defines the charge carrier mobility $\mu \equiv v_{\mathrm{d}}/E$.
In the Drude model of carrier transport,
an electron with effective mass $m^*$ experiences a force $eE$ when subjected to a uniform electric field
and completely looses its momentum $m^* v_{\mathrm{d}}$ in a time $\tau/2$, due to scattering from material defects, impurities, or lattice vibrations~\cite{Drude1900,Kittel2004}.
By equating the force and the rate of momentum loss at equilibrium, one obtains
$eE = m^* v_{\mathrm{d}}/\tau$, which leads to the celebrated Drude formula for the electron mobility: $\mu = e \tau / m^*$.
From this simple relation, it is already clear that a predictive theory for the
calculation of carrier mobilities requires accurate calculations of the scattering rate $1/\tau$ as well as the carrier effective mass $m^*$.

The mobility plays a central role in semiconductor devices:
it determines the switching frequency in transistors, the photoconductive gain in photodetectors, and transport properties in solar cells and light-emitting devices.
Therefore it is not surprising that considerable efforts have been devoted to make
accurate predictions of carrier mobilities ever since the beginning of solid state physics.

The first quantum-mechanical description of electron transport in crystals was provided by Bloch, who discussed how the fluctuations of the crystal potential
arising from lattice vibrations act as a source of scattering for electrons travelling through the solid~\cite{Bloch1929}.
Since then, many analytical approaches have been  developed to describe
the main scattering mechanisms arising from the electron-phonon interaction (EPI)~\cite{Vogl1976}.
Key mechanisms include:
(i) acoustic-deformation potential scattering~\cite{Bardeen1950,Herring1956}, which links the change of the electronic band structure with the macroscopic strain;
(ii) optical deformation potential scattering, which describes the interaction of long-wavelength
optical phonons with electrons in nonpolar crystals~\cite{Harrison1956};
(iii) piezoelectric scattering, where a lattice distortion is induced by a piezoelectric
field in a material lacking inversion symmetry~\cite{Meijer1953};
and (iv) polar-optical phonon scattering or Fr\"ohlich coupling,
whereby long-wavelength longitudinal-optical phonons in polar crystals induce macroscopic electric fields~\cite{Frohlich1954}.

More refined theories of carrier transport started to appear in the 1960s
and include work on non-equilibrium Green's functions~\cite{Kadanoff1962,Keldysh1965},
the Kubo formalism~\cite{Kubo1966}, the Landauer-B\"uttiker formalism~\cite{Landauer1981,
Buettiker1986,Landauer1989}, and the Boltzmann transport equation (BTE).
Traditionally, the BTE has been employed in the context of iterative,
finite-difference techniques~\cite{Budd1967,Rode1970,Rode1970a,Rode1975},
variational approaches~\cite{Howarth1953,Ehrenreich1960,Ehrenreich1961},
or Monte Carlo sampling~\cite{Jacoboni1983,Fischetti1988,Fischetti1993,Smirnov2003}.
Most of these previous approaches rely on analytical models to describe the
 scattering due to specific manifestations of the EPI, hence their applicability is limited to certain classes of materials.

Besides electron-phonon scattering, other important scattering processes can be grouped in two categories:
(i) scattering by lattice defects, such as for example impurities in semiconductors
and (ii) carrier-carrier scattering.
Some of the historically significant models to investigate these effects include
the theory of ionized-impurity scattering by Conwell, Brooks, Norton, and
others~\cite{Conwell1950,Brooks1951,Norton1973,Li1977,Chattopadhyay1981,Johnson1951},
and the theory of electron-electron scattering by Matulionis, Po{\v z}ela, and Reklaitis~\cite{Matulionis1975}.
Recent work aimed at recasting these earlier models for defect-induced scattering
and carrier-carrier scattering in the framework of $\mathbf{k}\cdot\mathbf{p}$
perturbation theory~\cite{Miller2011} and \textit{ab initio} calculations~\cite{Faghaninia2015}.

Among the scattering mechanisms described above, only the scattering theory of charged carriers by phonons
has been developed far enough that predictive calculations are now possible.
At the heart of the modern theory of electron-phonon scattering processes is the calculation of EPIs from first principles.
These calculations have been enabled by the development of density functional perturbation theory (DFPT) starting in the
1980s~\cite{Baroni1987,Savrasov1992,Gonze1997,Baroni2001}.
First-principles-based methods to study EPIs have become popular in recent years,
possibly as a result of the increased availability of high-performance computing,
new theoretical developments, and advanced software implementations~\cite{Giustino2017}.

This review focuses on modern \textit{ab initio} calculations of carrier transport
in metals and semiconductors, with an emphasis on the role of electron-phonon interactions and the temperature dependence of transport coefficients.

The manuscript is organized as follows.
In Sec.~\ref{Sec2} we review the \textit{ab initio} theory of carrier transport.
We start from a general, many-body quantum mechanical framework based on the Kadanoff-Baym formalism in Sec.~\ref{Sec2.1}
and we make the link with the popular BTE approach in Sec.~\ref{Sec2.2}.
Common approximations employed for solving the BTE are discussed in Sec.~\ref{Sec2.3},
including the response to electric and magnetic fields in Sec.~\ref{Sec2.4}.
Section~\ref{Sec2.5} establishes the relation between the BTE approach and the Kubo formula
and in Sec.~\ref{Sec2.6} we discuss the hierarchy of approximations used in
calculations of carrier transport and the tradeoff between complexity and accuracy.
Section~\ref{Sec3} provides an overview of the implementations of the BTE formalism in modern electronic structure codes and summarizes available software.
In Sec.~\ref{Sec4} we discuss recent \textit{ab initio} calculations of carrier mobilities, with a focus on bulk semiconductors and two-dimensional (2D) materials.
In the case of bulk semiconductors, we cover work on silicon, diamond, gallium arsenide, gallium nitride, gallium oxide, and
hybrid organic-inorganic halide perovskites (Sec.~\ref{Sec4.1}).
In the case of 2D materials, we review recent works on graphene, silicene, phosphorene,
molybdenum disulfide, and indium selenide (Sec.~\ref{Sec4.2}) as well as recent efforts in the direction of high-throughput calculations (Sec.~\ref{Sec4.3}).
Finally in Sec.~\ref{Sec4.4}, we gather experimental and theoretical results and discuss the predictive accuracy of first-principles mobility calculations.
In Sec.~\ref{Sec5} we offer our perspective on interesting new directions and opportunities in this field.
In particular, we discuss spin transport (Sec.~\ref{Sec5.1}), topological materials (Sec.~\ref{Sec5.2}),
the influence of the Berry phase on velocities and scattering rates (Sec.~\ref{Sec5.3}), and transport in correlated electron systems (Sec.~\ref{Sec5.4}).
We present our conclusions and outlook in Sec.~\ref{Sec6}.
The appendices report mathematical details of the derivations provided in Sec.~\ref{Sec2}.


\section{\textit{Ab initio} theory of electron transport}\label{Sec2}

\subsection{Quantum theory of mobility}\label{Sec2.1}

In this section we review the current theoretical description of charge transport from a modern, Green's function-based point of view,
expressed in a field-theoretic language.
The present derivation rests on seminal works by Martin, Schwinger, Kadanoff, Baym, Keldysh, Mahan, Datta, Haug, Kita, Stefanucci, van Leeuwen, and others~\cite{Martin1959,Kadanoff1962,Keldysh1965,Danielewicz1984,Rammer1986,Mahan1987,Khan1987,Datta1995,Mahan2000,Tatara2008,Haug2008,Kamenev2009,Kita2010,Stefanucci2013}.

\begin{figure}[t]
  \centering
  \includegraphics[width=\columnwidth]{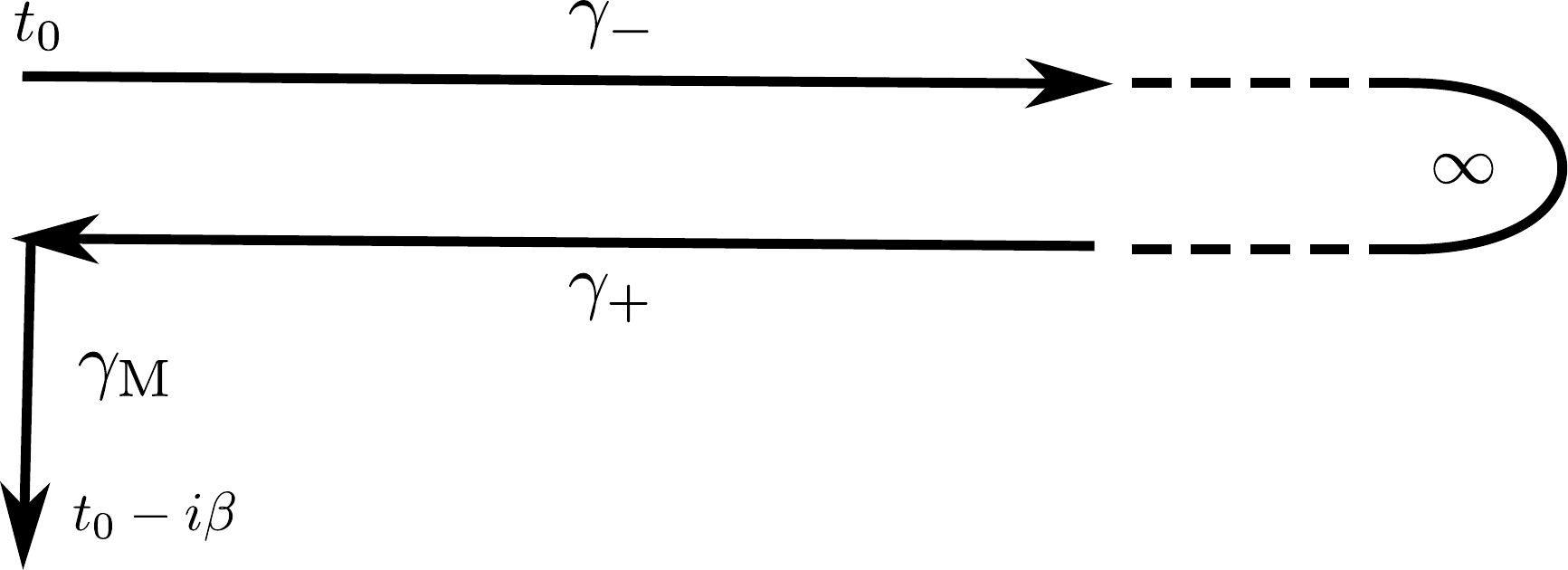}
  \caption{\label{fig:contour}
  The Keldysh-Schwinger contour consists of three pieces:
  $\gamma_-$, which runs from $z={t_{0-}}$ to $z=\infty_-$,
  $\gamma_+$, from $z=\infty_+$ to $z=t_{0+}$,
  and $\gamma_{\mathrm{M}}$ from $z=t_0$ to $z=t_0-i\beta$.
  }
\end{figure}

The central quantity in the description of charge transport is the current density $\mathbf{J}(\mathbf{r},t)$, which is represented by the Schr\"odinger-picture operator
\begin{equation}\label{eq:currentdef}
    \hat{\mathbf{J}}(\mathbf{r}) = \frac{ie\hbar}{2m} \Big\{ \hat{\psi}^{\dagger}(\mathbf{r})\big[\nabla\hat{\psi}(\mathbf{r})\big]
    -\big[\nabla\hat{\psi}^{\dagger}(\mathbf{r})\big] \hat{\psi}(\mathbf{r}) \Big\},
\end{equation}
where  $\hat{\psi}(\mathbf{r})$ denotes the electron field operator, $-e$ the electron charge, and $m$ the electron mass.
We assume that the system is in thermodynamic equilibrium at time $t_0$ with a heat bath at temperature $T$.
The expectation value of $\hat{\mathbf{J}}(\mathbf{r})$ at a later time $t$ is then given by
\begin{equation}\label{eq:expvaluecurrent}
  \mathbf{J}(\mathbf{r},t) = \big\langle\hat{\mathbf{J}}_{\mathrm{H}}(\mathbf{r},t)\big\rangle
  \equiv \frac{1}{Z}\mathrm{tr}\left[\mathrm{e}^{-\beta\hat{H}(t_0)}\hat{\mathbf{J}}_{\mathrm{H}}(\mathbf{r},t)\right],
\end{equation}
where $\beta^{-1} \equiv k_{\mathrm{B}}T$, $\hat{H}(t)$ denotes the total Hamiltonian of the system,
and $Z \equiv \mathrm{tr}\{\exp[-\beta \hat{H}(t_0)]\}$ is the canonical partition function.
In Eq.~\eqref{eq:expvaluecurrent}, the current density operator in the Heisenberg picture reads
\begin{equation}
  \hat{\mathbf{J}}_{\mathrm{H}}(\mathbf{r},t) \!= \!
  \overline{\mathcal{T}}\left[\mathrm{e}^{\frac{i}{\hbar}\! \int^{t}_{t_0}\! \mathrm{d} t'  \hat{H}(t')}\right]\!
  \hat{\mathbf{J}}(\mathbf{r})\mathcal{T} \! \left[\mathrm{e}^{\frac{-i}{\hbar}\!\int^{t}_{t_0}\! \mathrm{d} t'  \hat{H}(t')}\right]\!,
\end{equation}
with $\mathcal{T}$($\overline{\mathcal{T}}$) being the (anti-) time-ordering symbol.
To compute $\mathbf{J}(\mathbf{r},t)$, we make use of the Green's function formalism.
To start with, we introduce the \emph{lesser} Green's function as
\begin{equation}\label{eq:greenslesser}
  G^<(\mathbf{r}_1,\mathbf{r}_2;t_1,t_2) \equiv \frac{i}{\hbar}\left\langle\hat{\psi}^{\dagger}_{\mathrm{H}}(\mathbf{r}_2,t_2)\hat{\psi}_{\mathrm{H}}(\mathbf{r}_1,t_1)\right\rangle,
\end{equation}
in terms of which the expectation value of the current density can be written as
\begin{equation}\label{eq:currentdensity}
  \mathbf{J}(\mathbf{r},t) = \frac{-e \hbar^2}{2m} \lim_{\mathbf{r}'\to\mathbf{r}} (\nabla'-\nabla)G^<(\mathbf{r},\mathbf{r}';t,t).
\end{equation}
The Hamiltonian $\hat{H}(t)$ appears both in the thermodynamic weights and as part of the Heisenberg-picture field operators $\hat{\psi}_{\rm H}$.
This makes the use of perturbation theory difficult, since the Hamiltonians at two different times in general do not commute.
One way to overcome this difficulty is by making use of the Keldysh-Schwinger contour formalism.
In this formalism, the three occurrences of the Hamiltonian in the expression
for the time-dependent expectation value of an operator are merged into one single exponential
under a \emph{contour-ordering symbol} $\mathcal{T}_{\mathrm{C}}$.
This operation leads to the definition of the contour-ordered Green's function:
\begin{multline}\label{eq:greensfuncontour}
  G(\mathbf{r}_1,\mathbf{r}_2;z_1,z_2) = \frac{-i}{\hbar} \frac{1}{Z}
  \mathrm{tr}\bigg\{ \mathcal{T}_{\mathrm{C}} \Big[ \mathrm{e}^{\frac{-i}{\hbar}\int_{\gamma} \mathrm{d} z \, \hat{H}(z)} \\
  \times [\hat{\psi}(\mathbf{r}_1)]_{z_1}[\hat{\psi}^{\dagger}(\mathbf{r}_2)]_{z_2} \Big] \bigg\}.
\end{multline}
The contour-ordering symbol orders the operators inside the square brackets
according to their place on the contour $\gamma$, depicted in Fig.~\ref{fig:contour}.
The subscripts $z_1$ and $z_2$  denote the placement of the field operators in the string of operators under the contour-ordering symbol.
For instance, the lesser Green's function can be recovered by choosing $z_2=t_{2+}$ and $z_1=t_{1-}$,
so that $\hat{\psi}^{\dagger}(\mathbf{r}_2)$ is placed in front of $\hat{\psi}(\mathbf{r}_1)$, irrespective of the value of the times $t_1$ and $t_2$,
since all operators on $\gamma_+$ are considered to be later on the contour than operators on $\gamma_-$.
We use this notation exclusively for time-independent operators as for time-dependent operators, their placement under the contour-ordering symbol is already determined by their time argument.
The strength of the Keldysh-Schwinger formalism is that it naturally leads to a convenient perturbative expansion
of the contour-ordered Green's function $G(\mathbf{r}_1,\mathbf{r}_2;z_1,z_2)$.

We start from a Hamiltonian $\hat{H}_0$ for which we can compute the eigenstates (for example
the Kohn-Sham Hamiltonian), and divide the total Hamiltonian into three pieces:
\begin{equation}
  \hat{H}(z) = \hat{H}_0 + \hat{H}_{\mathrm{int}} + \hat{H}_{\mathrm{ext}}(z),
\end{equation}
where $\hat{H}_{\mathrm{int}}$ captures all \emph{internal} interactions not contained in $\hat{H}_0$
and $\hat{H}_{\mathrm{ext}}(z)$ contains the interaction with external electromagnetic fields,
such as the applied bias voltage in experiments.
Note that, under the contour-ordering symbol, all parts of the Hamiltonian can be treated as if they mutually commuted with each other.
We can thus perform a perturbative expansion by Taylor-expanding the exponential in powers of $\hat{H}_{\mathrm{int}}$ and $\hat{H}_{\mathrm{ext}}(z)$.
To start with, we write the non-interacting one-particle Hamiltonian in the form
\begin{equation}
  \hat{H}_0 = \int \mathrm{d}^3 r \, \hat{\psi}^{\dagger}(\mathbf{r}) h_0(\mathbf{r},-i\hbar\nabla) \hat{\psi}(\mathbf{r}),
\end{equation}
where
\begin{equation}\label{eq:h0}
  h_0(\mathbf{r},-i\hbar\nabla) = \frac{-\hbar^2 \nabla^2}{2m} + V(\mathbf{r}).
\end{equation}
Here $V(\mathbf{r})$ is a one-particle potential, such as for example the Kohn-Sham potential in density functional theory (DFT)~\cite{Hohenberg1964,Kohn1965}.
We also take the external Hamiltonian to be of the form
\begin{equation}\label{eq:Hext}
  \hat{H}_{\mathrm{ext}}(z) = \int \mathrm{d}^3 r \, \hat{\psi}^{\dagger}(\mathbf{r}) (-e) \phi_{\mathrm{ext}}(\mathbf{r},z)  \hat{\psi}(\mathbf{r}),
\end{equation}
where $\phi_{\mathrm{ext}}(\mathbf{r},t)$ denotes an externally applied scalar potential. For simplicity
we here ignore the possibility of a vector potential.
The part $\hat{H}_{\mathrm{int}}$ is understood to contain all inter-particle interactions,
such as the inter-electron Coulomb interaction and the interaction of electrons with lattice degrees of freedom
in case of a solid.
The internal interaction Hamiltonian can also be expressed in terms of $\hat{\psi}(\mathbf{r})$ and $\hat{\psi}^{\dagger}(\mathbf{r})$.

The contour-ordered Green's function from Eq.~\eqref{eq:greensfuncontour} can be expanded in a Taylor series in powers of $\hat{H}_{\mathrm{int}}$ and $\hat{H}_{\mathrm{ext}}(z)$:
\begin{multline}\label{eq:greenfunTayl}
  G(\mathbf{r}_1,\mathbf{r}_2;z_1,z_2) = G_0(\mathbf{r}_1,\mathbf{r}_2;z_1,z_2) +\!\! \sum_{n,m=1}^{\infty}\!\!\frac{(-i/\hbar)^{n+m}}{n!m!} \\
  \times \int_\gamma \! \mathrm{d} z'_1 \ldots \int_\gamma \! \mathrm{d} z'_n \int_\gamma \! \mathrm{d} z''_1 \ldots \! \int_\gamma \! \mathrm{d} z''_m  \frac{1}{Z} \mathrm{tr}\Big[  \mathcal{T}_{\mathrm{C}} \mathrm{e}^{\frac{-i}{\hbar}\int_{\gamma} \mathrm{d} z \, [\hat{H}_{0}]_{z}} \\
  \times  \big[\hat{H}_{\mathrm{int}}\big]_{z'_1} \ldots \big[\hat{H}_{\mathrm{int}}\big]_{z'_n} \hat{H}_{\mathrm{ext}}(z''_1) \ldots \hat{H}_{\mathrm{ext}}(z''_m)  \\
  \times \big[\hat{\psi}(\mathbf{r}_1)\big]_{z_1} \big[\hat{\psi}^{\dagger}(\mathbf{r}_2)\big]_{z_2} \Big].
\end{multline}
Here we have defined the non-interacting Green's function as
\begin{multline}\label{eq:g0}
    G_0(\mathbf{r}_1,\mathbf{r}_2;z_1,z_2) = \frac{-i}{\hbar}\frac{1}{Z_0}
    \mathrm{tr}\Big[ \mathcal{T}_{\mathrm{C}} \mathrm{e}^{\frac{-i}{\hbar}\int_{\gamma} \mathrm{d} z \, [\hat{H}_{0}]_z} \\
    \times \big[\hat{\psi}(\mathbf{r}_1)\big]_{z_1} \big[\hat{\psi}^{\dagger}(\mathbf{r}_2)\big]_{z_2} \Big],
\end{multline}
with $Z_0 \equiv \mathrm{tr}\{\exp[-\beta \hat{H}_0]\}$.
Expressing $\hat{H}_{\mathrm{int}}$ and $\hat{H}_{\mathrm{ext}}(z)$ in Eq.~\eqref{eq:greenfunTayl} in terms of the field operators leads to terms of the form
\begin{multline}
  \mathrm{tr}\Big[ \mathcal{T}_{\mathrm{C}} \mathrm{e}^{\frac{-i}{\hbar}\int_{\gamma} \mathrm{d} z \, [\hat{H}_0]_{z}}
  [\hat{\psi}(\mathbf{r}_1)]_{z_1} [\hat{\psi}(\mathbf{r}_2)]_{z_2} \ldots \\
  \ldots [\hat{\psi}^{\dagger}(\mathbf{r}'_1)]_{z'_1} [\hat{\psi}^{\dagger}(\mathbf{r}'_2)]_{z'_2}\ldots \Big].
\end{multline}
These terms can be evaluated using a generalized version of Wick's theorem~\cite{Kita2010}
and written in terms of products of $G_0$.
The exact perturbation series for $G$ can be analyzed with the help of Feynman diagrams, which results in Dyson's equation on the contour:
\begin{multline}\label{eq:dysoncontour}
    G(1,2) = G_0(1,2) \\
    + \int_{\gamma} \mathrm{d} 3 \int_{\gamma} \mathrm{d} 4 \, G_0(1,3) \Sigma[G](3,4) G(4,2).
\end{multline}
In the equation above, $1\equiv(\mathbf{r}_1,z_1)$ and $\int_{\gamma} \mathrm{d} 1 \equiv \int \mathrm{d}^3 r_1 \int_{\gamma} \mathrm{d} z_1$,
while $\Sigma[G]$ is the \emph{self-energy}, which in itself depends on $G$ and captures all the effects of interactions.

From Dyson's equation on the contour, we can derive an equation of motion for
the lesser Green's function $G^<$, which is needed to compute the current density from Eq.~\eqref{eq:currentdensity}.
This equation is known as one of the \emph{Kadanoff-Baym equations} (KBEs)~\cite{Kadanoff1962}.
In the limit $t_0\to-\infty$,  it reads:
\begin{multline}\label{eq:kbe}
      i\hbar\frac{\partial}{\partial t}G^<(\mathbf{r}_1,\mathbf{r}_2;t,t) \\
    = \left[h_0(\mathbf{r}_1,-i\hbar\nabla_1)-h_0(\mathbf{r}_2,+i\hbar\nabla_2)\right]G^<(\mathbf{r}_1,\mathbf{r}_2;t,t) \\
      +\int \! \mathrm{d}^3 \! r_3 \, \bigg[\Sigma^{\delta}(\mathbf{r}_1,\mathbf{r}_3;t)G^<(\mathbf{r}_3,\mathbf{r}_2;t,t) \\
       \qquad \qquad - G^<(\mathbf{r}_1,\mathbf{r}_3;t,t)\Sigma^{\delta}(\mathbf{r}_3,\mathbf{r}_2;t) \bigg] \\
      +\! \int_{-\infty}^t \!\! \mathrm{d} t' \! \int \mathrm{d}^3 r_3 \, \bigg[\Sigma^>(\mathbf{r}_1,\mathbf{r}_3;t,t')G^<(\mathbf{r}_3,\mathbf{r}_2;t',t) \\
      \qquad +G^<(\mathbf{r}_1,\mathbf{r}_3;t,t')\Sigma^>(\mathbf{r}_3,\mathbf{r}_2;t',t) \\
      \qquad -\Sigma^<(\mathbf{r}_1,\mathbf{r}_3;t,t')G^>(\mathbf{r}_3,\mathbf{r}_2;t',t) \\
       -G^>(\mathbf{r}_1,\mathbf{r}_3;t,t')\Sigma^<(\mathbf{r}_3,\mathbf{r}_2;t',t) \bigg].
\end{multline}
Here, $\Sigma^{\delta}(\mathbf{r}_1,\mathbf{r}_2;t)$ is the part of the self-energy that is local in time following the notation of Stefanucci and van Leeuwen (Ref.~\cite{Stefanucci2013}, Eq.~9.12, p.~252),
while $f^>(t_1,t_2) = f(z_1=t_{1+},z_2=t_{2-})$ and $f^<(t_1,t_2) = f(z_1=t_{1-},z_2=t_{2+})$
are the greater and lesser counterparts of a function $f(z_1,z_2)$ on the contour.
The explicit derivation of Eq.~\eqref{eq:kbe} from Eq.~\eqref{eq:dysoncontour} is given in \ref{appendix}.

The first of the three terms on the right-hand side of this equation describes
the unperturbed time-evolution of the lesser Green's function in a static potential $V(\mathbf{r})$.
The second term involves the local-time self-energy that includes both the screened external potential
as well as the effects of static electron-electron and electron-phonon interactions.
This includes, for example, the difference between the instantaneous Hartree-Fock self-energy and the mean-field potential already included in $h_0$ through $V(\mathbf{r})$.
Finally, the remaining terms, involving various combinations of $G^{<,>}$ and $\Sigma^{<,>}$,
describe the effects of internal dynamical correlations, such as particle collisions and scattering as well as their interaction with the screened external potential.


\subsection{Boltzmann transport equation}\label{Sec2.2}

It is currently not possible to determine exact solutions of Eq.~\eqref{eq:kbe}.
However, with a few approximations this equation can be converted into a computationally accessible problem that yields quantitatively predictive results.

We approximate the term involving $\Sigma^{\delta}$ by neglecting the screening of the external potential.
This approximation can be relaxed by taking into account the response of the electronic density to the external field, but
the final form of the equation would not change; therefore we prefer to omit this detail for the sake of brevity.
We also neglect the difference between the internal instantaneous electron self-energy and the mean-field potential already included in $V(\mathbf{r})$.
In this approximation, the local-time self-energy reads
\begin{equation}
  \Sigma^{\delta}(\mathbf{r}_1,\mathbf{r}_2;t) \approx -e\phi_{\mathrm{ext}}(\mathbf{r}_1,t)\delta^{(3)}(\mathbf{r}_1-\mathbf{r}_2),
\end{equation}
whereupon the corresponding part of Eq.~\eqref{eq:kbe} becomes:
\begin{multline}\label{eq:drive}
     \int \mathrm{d}^3 r_3 \, \Big[\Sigma^{\delta}(\mathbf{r}_1,\mathbf{r}_3;t)G^<(\mathbf{r}_3,\mathbf{r}_2;t,t) \\
       - G^<(\mathbf{r}_1,\mathbf{r}_3;t,t)\Sigma^{\delta}(\mathbf{r}_3,\mathbf{r}_2;t) \Big] \\
     \approx -e\left[\phi_{\mathrm{ext}}(\mathbf{r}_1,t)-\phi_{\mathrm{ext}}(\mathbf{r}_2,t)\right]G^<(\mathbf{r}_1,\mathbf{r}_2;t,t).
\end{multline}
As in a typical experiment, we assume the electric field to be spatially homogeneous, in which case
\begin{equation}\label{eq:drivehomogeneous}
  \phi_{\mathrm{ext}}(\mathbf{r}_1,t)-\phi_{\mathrm{ext}}(\mathbf{r}_2,t)
   = -\mathbf{E}(t)\cdot(\mathbf{r}_1-\mathbf{r}_2).
\end{equation}
We now consider electrons in a solid and we choose the unperturbed Hamiltonian in the position representation as
\begin{equation}
  h_0(\mathbf{r},-i\hbar\nabla) = \frac{-\hbar^2 \nabla^2}{2m} + V_{\mathrm{lat+Hxc}}(\mathbf{r}),
\end{equation}
where $V_{\mathrm{lat+Hxc}}(\mathbf{r})$ is given by the sum of an ionic lattice potential
and the effective mean-field Hartree and exchange and correlation potentials generated by the electrons.
The eigenstates of this Hamiltonian,
\begin{equation}\label{eq:h0Ham}
  h_0(\mathbf{r},-i\hbar\nabla) \varphi_{n\mathbf{k}}(\mathbf{r}) = \varepsilon_{n\mathbf{k}}\varphi_{n\mathbf{k}}(\mathbf{r}),
\end{equation}
can be labeled by a band index $n$ and a crystal wavevector $\mathbf{k}$.

Next, we express both sides of Eq.~\eqref{eq:kbe} in the basis $\{\varphi_{n\mathbf{k}}(\mathbf{r})\}$ and implicitly assume
that the external field $\textbf{E}$ is screened self-consistently by $V_{\mathrm{Hxc}}$.
For simplicity we adopt the commonly used approximation of retaining only
the diagonal matrix elements of the Green's function and self-energy~\cite{Ponce2018}.
This approximation only affects the band indices since the Green's function is diagonal
in $\mathbf{k}$ for a homogeneous electric field, which does not break the translational symmetry of the lattice.
This approximation is expected to be valid in systems where the interactions contained in $\hat{H}_{\mathrm{int}}$
and the external fields in $\hat{H}_{\mathrm{ext}}$ do not mix the bands significantly.
Now we can write the diagonal matrix elements of the lesser and greater Green's function as
\begin{multline}\label{eq:blochgreensfun}
    \mp \frac{i}{\hbar} f^{>,<}_{n\mathbf{k}}(t,t') \equiv \\
    \int \! \mathrm{d}^3 r_1 \int \! \mathrm{d}^3 r_2 \, \varphi^*_{n\mathbf{k}}(\mathbf{r}_1) G^{>,<}(\mathbf{r}_1,\mathbf{r}_2;t,t') \varphi_{n\mathbf{k}}(\mathbf{r}_2),
\end{multline}
where the -(+) sign corresponds to the left (right) symbol $>$($<$).
We then take the diagonal elements of the term involving the electric field, Eq.~\eqref{eq:drivehomogeneous}.
By expressing the Bloch wave functions as $\varphi_{n\mathbf{k}}(\mathbf{r}) = \mathrm{e}^{i\mathbf{k}\cdot\mathbf{r}}u_{n\mathbf{k}}(\mathbf{r})$,
where $u_{n\mathbf{k}}(\mathbf{r})$ is lattice-periodic, we obtain:
\begin{multline}
  \int \! \mathrm{d}^3 r_1 \!\! \int \! \mathrm{d}^3 r_2 \, \varphi^*_{n\mathbf{k}}(\mathbf{r}_1) e \mathbf{E}(t) \cdot (\mathbf{r}_1-\mathbf{r}_2) G^<(\mathbf{r}_1,\mathbf{r}_2;t,t)  \\
  \times \varphi_{n\mathbf{k}}(\mathbf{r}_2)  = -e \mathbf{E}(t) \cdot \frac{1}{\hbar} \frac{\partial f^<_{n\mathbf{k}}}{\partial \mathbf{k}}(t,t).
\end{multline}
Finally, we define the diagonal components of the greater and lesser self-energy as
\begin{multline}
    \mp i\hbar \Gamma^{>,<}_{n\mathbf{k}}(t,t') \equiv \\
    \int \! \mathrm{d}^3 r_1 \int \! \mathrm{d}^3 r_2 \, \varphi^*_{n\mathbf{k}}(\mathbf{r}_1) \Sigma^{>,<}(\mathbf{r}_1,\mathbf{r}_2;t,t') \varphi_{n\mathbf{k}}(\mathbf{r}_2).
\end{multline}
With these definitions and approximations, Eq.~\eqref{eq:kbe} takes on the form
\begin{equation}\label{eq:boltzmann}
  \frac{\partial f^<_{n\mathbf{k}}}{\partial t}(t,t)  -e \mathbf{E}(t) \cdot \frac{1}{\hbar}\frac{\partial f^<_{n\mathbf{k}}}{\partial \mathbf{k}}(t,t)
  = - \Gamma^{(\mathrm{co})}_{n\mathbf{k}}(t),
\end{equation}
where the unperturbed time evolution of $f^<_{n\mathbf{k}}$ vanishes identically due to Eq.~\eqref{eq:h0Ham} and
where we have introduced the \emph{collision rate}
\begin{multline}\label{eq:collrate}
  \! \Gamma^{(\mathrm{co})}_{n\mathbf{k}}(t) \! \equiv \!\!\! \int_{-\infty}^{t} \!\! \!\!\!\!\!\! \mathrm{d} t' \,
  \big[   \Gamma^>_{n\mathbf{k}}(t,t')f^<_{n\mathbf{k}}(t',t) + f^<_{n\mathbf{k}}(t,t')\Gamma^>_{n\mathbf{k}}(t',t) \\
        - \Gamma^<_{n\mathbf{k}}(t,t')f^>_{n\mathbf{k}}(t',t) - f^>_{n\mathbf{k}}(t,t')\Gamma^<_{n\mathbf{k}}(t',t) \big].
\end{multline}
Equation~\eqref{eq:boltzmann} is the quantum equivalent of the Boltzmann transport equation in the approximation
of neglecting the off-diagonal matrix elements of the Green's function and self-energy.

In the case of direct-current (DC) transport, the electric field does not depend on time.
As a result, the total Hamiltonian is time-independent and any two-time function $f(t,t')$ depends only on the time difference $t-t'$, $f(t,t')=f(t-t')$.
In particular, $f^<_{n\mathbf{k}}(t,t)$ becomes time-independent, and the collision integral of Eq.~\eqref{eq:collrate} simplifies to
\begin{equation}
  \Gamma^{(\mathrm{co})}_{n\mathbf{k}} \! = \!\! \int_{-\infty}^{+\infty} \!\!\!\!\!\! \mathrm{d} t' \big[ f^<_{n\mathbf{k}}(t')\Gamma^>_{n\mathbf{k}}(-t')
    \!-\! f^>_{n\mathbf{k}}(t')\Gamma^<_{n\mathbf{k}}(-t') \big],
\end{equation}
where we shifted the integration variable $t'$ by $t$ and let $t' \to -t'$ in the second and fourth terms of Eq.~\eqref{eq:collrate}.
Finally, we can write the Boltzmann equation in the frequency domain by introducing the Fourier transform of a function $F(t)$ as
\begin{equation}
  F(\omega) = \int_{-\infty}^{+\infty} \!\!\! \mathrm{d} t \, \mathrm{e}^{i \omega t} F(t).
\end{equation}
The Boltzmann equation for time-independent electric fields then reads
\begin{multline}\label{eq:boltzmannfourier}
  -e \mathbf{E} \cdot \frac{1}{\hbar} \frac{\partial f_{n\mathbf{k}}}{\partial \mathbf{k}} =
  -\int \! \frac{\mathrm{d} \omega}{2\pi} \big[ f^<_{n\mathbf{k}}(\omega)\Gamma^>_{n\mathbf{k}}(\omega) \\
  - f^>_{n\mathbf{k}}(\omega)\Gamma^<_{n\mathbf{k}}(\omega) \big],
\end{multline}
where we defined the $\mathbf{E}$-field dependent occupation number as
\begin{equation}\label{eq:occupnumber}
  f_{n\mathbf{k}} \equiv \int \frac{\mathrm{d} \omega}{2\pi} f^<_{n\mathbf{k}}(\omega).
\end{equation}
From the definition of $f^<_{n\mathbf{k}}(t,t')$ in Eq.~\eqref{eq:blochgreensfun}, it follows immediately that
$f_{n\mathbf{k}} = \langle \hat{c}^{\dagger}_{n\mathbf{k}}\hat{c}_{n\mathbf{k}} \rangle$, where $\hat{c}^{(\dagger)}_{n\mathbf{k}}$
is the annihilation (creation) operator for a Bloch state $|n\mathbf{k}\rangle$.
Therefore $f_{n\mathbf{k}}$ has the physical meaning of the (fractional) number of electrons in the state $|n\mathbf{k}\rangle$.

We now specialize Eq.~\eqref{eq:boltzmannfourier} to the case of scattering by lattice vibrations.
To this end, the matrix elements $\Gamma^{>,<}_{n\mathbf{k}}(\omega )$ of the greater/lesser self-energy
are commonly approximated as the diagonal matrix elements of the Fan-Migdal self-energy~\cite{Giustino2017}
\begin{multline}\label{eq:FMscatrate}
  \Gamma^{>,<}_{n\mathbf{k}}(\omega)
  \approx \frac{i}{\hbar} \sum_{m,\nu}  \int \frac{\mathrm{d}^3 q}{\Omega_{\mathrm{BZ}}} \int \frac{\mathrm{d} \omega'}{2\pi}
  \left| g^{\mathrm{R}}_{mn\nu}(\mathbf{k},\mathbf{q};\omega') \right|^2 \\
   \times f^{>,<}_{m\mathbf{k}+\mathbf{q}}(\omega+\omega') D^{<,>}_{\mathbf{q}\nu}(\omega'),
\end{multline}
where $D^{>,<}_{\mathbf{q}\nu}(\omega )$ is the Green's function of a phonon of branch $\nu$ and crystal wavevector $\mathbf{q}$
and the summation and integration run over all electronic bands $m$, phonon branches $\nu$,
and crystal wavevectors $\mathbf{q}$ in the first Brillouin zone, whose volume is denoted by $\Omega_{\mathrm{BZ}}$.
The Fourier-transformed $g^{\mathrm{R}}_{mn\nu}(\mathbf{k},\mathbf{q};\omega)$ is the retarded matrix element
for absorption of a phonon of branch $\nu$ with wavevector $\mathbf{q}$ and frequency $\omega$ that scatters
an electron from state $|n\mathbf{k}\rangle$ into state $|m\mathbf{k}+\mathbf{q}\rangle$.

\begin{figure}[t!]
  \centering
  \includegraphics[width=0.99\linewidth]{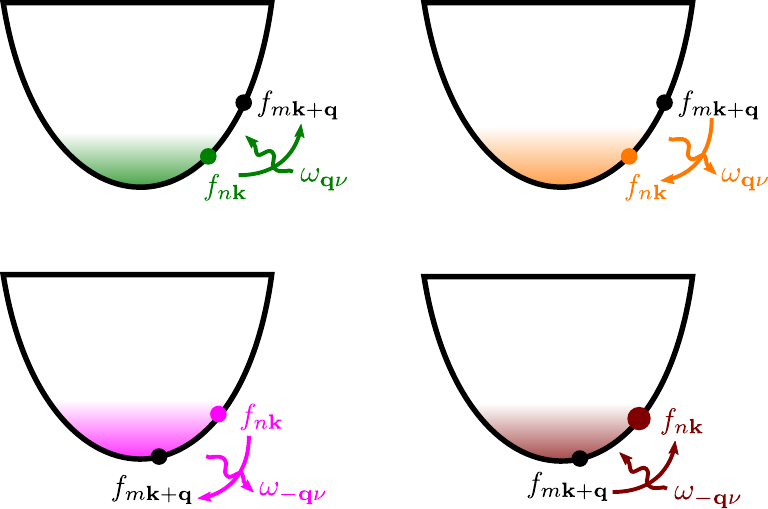}
  \caption{\label{fig:collrate}
  The four processes included in the collision rate in Eq.~\eqref{eq:collrate} derived from the Fan-Migdal self-energy:
  Scattering of an electron out of state $|n\mathbf{k}\rangle$ via phonon absorption
  (green, first term) and emission (purple, second term)
  and scattering of an electron into state $|n\mathbf{k}\rangle$ via phonon absorption (brown, third term)
  and emission (orange, fourth term).}
\end{figure}

In practice, Eq.~\eqref{eq:FMscatrate} is usually evaluated under three further approximations:
(i) the phonon Green's function is written in the Born-Oppenheimer approximation:
\begin{multline}
    D^{>,<}_{\mathbf{q}\nu}(\omega ) \approx \frac{-i}{\hbar} 2\pi \big[ (n_{\mathbf{q}\nu}+1) \delta(\omega\mp\omega_{\mathbf{q}\nu}) \\
                                                                 + n_{\mathbf{q}\nu} \delta(\omega\pm\omega_{\mathbf{q}\nu}) \big],
\end{multline}
with $n_{\mathbf{q}\nu} \equiv 1/[\exp(\hbar\omega_{\mathbf{q}\nu}/k_{\mathrm{B}}T)-1]$ denoting the
Bose-Einstein distribution evaluated at the adiabatic phonon frequency $\omega_{\mathbf{q}\nu}$;
(ii) the electron-phonon matrix elements are approximated as frequency-independent quantities
obtained from the
self-consistent first derivative of the effective potential $V_{\mathrm{lat+Hxc}}(\mathbf{r})$~\cite{Giustino2017}:
\begin{equation}\label{eq:gDFPT}
	g_{mn\nu}(\mathbf{k},\mathbf{q}) = \langle m\mathbf{k}+\mathbf{q}  | \partial_{\mathbf{q}\nu}V_{\mathrm{lat+Hxc}}(\hat{\mathbf{r}}) | n\mathbf{k} \rangle,
\end{equation}
where
\begin{equation}
\partial_{\mathbf{q}\nu} = \sum_{p\kappa\alpha} \sqrt{\frac{\hbar}{2M_{\kappa}\omega_{\mathbf{q}\nu}}} \mathrm{e}^{i\mathbf{q}\cdot \mathbf{R}_p} e_{\kappa\alpha\nu}(\mathbf{q}) \frac{\partial}{\partial R_{p\kappa\alpha}}
\end{equation}
where $R_{p\kappa\alpha}$ denotes the $\alpha$th cartesian component of the equlibrium position of atom $\kappa$ of mass $M_\kappa$ in the $p$th unit cell with origin $\mathbf{R}_p$
and $e_{\kappa\alpha\nu}(\mathbf{q})$ is the $\alpha$th cartesian component of the vibrational eigenmodes with frequency $\omega_{\mathbf{q}\nu}$ for atom $\kappa$;
and (iii) the frequency dependence of the electronic Green's functions is approximated at the level of the unperturbed Hamiltonian:
\begin{align}
  f^{>}_{n\mathbf{k}}(\omega ) & \approx 2\pi (1-f_{n\mathbf{k}}) \delta(\omega - \varepsilon_{n\mathbf{k}}/\hbar), \\
  f^{<}_{n\mathbf{k}}(\omega ) & \approx 2\pi  f_{n\mathbf{k}} \delta(\omega - \varepsilon_{n\mathbf{k}}/\hbar).
\end{align}
Here $f_{n\mathbf{k}}$ still depends on the electric field and needs to be determined from the Boltzmann equation.
Within these three approximations, the collision rate reads
\begin{align}\label{eq:collisionrate}
  \Gamma&^{(\mathrm{co})}_{n\mathbf{k}} = \int \frac{\mathrm{d} \omega }{2\pi} \big[  f^<_{n\mathbf{k}}(\omega )\Gamma^>_{n\mathbf{k}}(\omega )- f^>_{n\mathbf{k}}(\omega )\Gamma^<_{n\mathbf{k}}(\omega ) \big] \nonumber \\
  & \quad \, \, \, \approx \frac{2\pi}{\hbar}\sum_{m,\nu} \int \frac{\mathrm{d}^3 q}{\Omega_{\mathrm{BZ}}} \left| g_{mn\nu}(\mathbf{k},\mathbf{q}) \right|^2  \nonumber\\
        \times \big[& f_{n\mathbf{k}}(1 - f_{m\mathbf{k}+\mathbf{q}})\delta(\Delta \varepsilon_{\mathbf{k},\mathbf{q}}^{nm} + \hbar\omega_{\mathbf{q}\nu}) n_{\mathbf{q}\nu} \nonumber\\
              +& f_{n\mathbf{k}}(1 - f_{m\mathbf{k}+\mathbf{q}})\delta(\Delta \varepsilon_{\mathbf{k},\mathbf{q}}^{nm} - \hbar\omega_{\mathbf{q}\nu}) (n_{\mathbf{q}\nu}\!+\!1) \nonumber\\
              -& (1 - f_{n\mathbf{k}})f_{m\mathbf{k}+\mathbf{q}}\delta(-\Delta \varepsilon_{\mathbf{k},\mathbf{q}}^{nm} + \hbar\omega_{\mathbf{q}\nu}) n_{\mathbf{q}\nu} \nonumber\\
              -& (1 - f_{n\mathbf{k}})f_{m\mathbf{k}+\mathbf{q}}\delta(-\Delta \varepsilon_{\mathbf{k},\mathbf{q}}^{nm} - \hbar\omega_{\mathbf{q}\nu}) (n_{\mathbf{q}\nu}\!+\!1) \big],
\end{align}
where $\Delta \varepsilon_{\mathbf{k},\mathbf{q}}^{nm} \equiv \varepsilon_{n\mathbf{k}}\!-\!\varepsilon_{m\mathbf{k}+\mathbf{q}}$.
Equation~\eqref{eq:collisionrate} represents the difference of the rate
for an electron in state $|n\mathbf{k}\rangle$ to scatter \emph{out} of the state (first two terms)
and the rate for an electron to scatter \emph{into} the state $|n\mathbf{k}\rangle$ (last two terms).
Both processes can be mediated either by phonon \emph{absorption} (first and third term)
or phonon \emph{emisssion} (second and forth term).
We note that we let $\mathbf{q}\to-\mathbf{q}$ in the terms involving phonon emission to write them also in terms
of $f_{m\mathbf{k}+\mathbf{q}}$ instead of $f_{m\mathbf{k}-\mathbf{q}}$, making use of $\omega_{\mathbf{q}\nu}=\omega_{-\mathbf{q}\nu}$
and the fact that the matrix elements for phonon emission and absorption are related by complex conjugation.
The four scattering processes included in $\Gamma^{(\mathrm{co})}_{n\mathbf{k}}$
are illustrated in Fig.~\ref{fig:collrate}.

Equation~\eqref{eq:boltzmannfourier} is solved iteratively to obtain the $\mathbf{E}$-field-dependent occupancies $f_{n\mathbf{k}}$.
Then the experimentally accessible macroscopic average of the current density $\mathbf{J}(\mathbf{r})$ can be obtained via
\begin{align}
    \mathbf{J}_{\mathrm{M}}(\mathbf{E}) &= \frac{1}{V}\int \mathrm{d}^3 r \, \mathbf{J}(\mathbf{r};\mathbf{E}) \\
                            &= \frac{-e}{V_{\mathrm{uc}}} \sum_n \int \frac{\mathrm{d}^3 k}{\Omega_{\mathrm{BZ}}} \, \mathbf{v}_{n\mathbf{k}}f_{n\mathbf{k}}(\mathbf{E}),\label{eq.curr.v}
\end{align}
where we made use of Eqs.~\eqref{eq:currentdensity}, \eqref{eq:blochgreensfun}, and \eqref{eq:occupnumber} and where $V$ and $V_{\mathrm{uc}}$ denote the crystal and unit cell volume, respectively.
In Eq.~\eqref{eq.curr.v} we introduced the diagonal velocity matrix elements $\mathbf{v}_{n\mathbf{k}}=\langle n\mathbf{k} | \hat{\mathbf{p}}/m | n\mathbf{k} \rangle$ and explicitly indicated
 the $\mathbf{E}$-field dependence of all quantities for clarity.
%

In the case of weak electric fields, we can restrict ourselves to the \emph{linear response} of the current density, which defines the \emph{conductivity} tensor:
\begin{equation}\label{eq:conductivitydef}
  \sigma_{\alpha\beta} \equiv \left. \frac{\partial J_{\mathrm{M},\alpha}}{\partial E_{\beta}}\right|_{\mathbf{E}=\mathbf{0}} \!\!\!\! = \frac{-e}{V_{\mathrm{uc}}} \sum_n \! \int \! \frac{\mathrm{d}^3 k}{\Omega_{\mathrm{BZ}}} \, v^{\alpha}_{n\mathbf{k}} \partial_{E_{\beta}} f_{n\mathbf{k}}.
\end{equation}
Here $\alpha,\beta$ run over the three Cartesian directions and we introduced the short-hand notation
$\partial_{E_{\beta}}f_{n\mathbf{k}} = (\partial f_{n\mathbf{k}}/\partial E_{\beta})|_{\mathbf{E}=\mathbf{0}}$.
From Eq.~\eqref{eq:boltzmannfourier}, we can obtain an expression for the linear response coefficients $\partial_{E_{\beta}}f_{n\mathbf{k}}$ by taking derivatives
on both sides with respect to the electric field:
\begin{multline}\label{eq:iterboltzmann}
   -e v^{\beta}_{n\mathbf{k}} \frac{\partial f^0_{n\mathbf{k}}}{\partial \varepsilon_{n\mathbf{k}}}
  = \sum_{m} \int \frac{\mathrm{d}^3 q}{\Omega_{\mathrm{BZ}}} \,
     \big[ \tau^{-1}_{m\mathbf{k}+\mathbf{q} \to n\mathbf{k}} \, \partial_{E_{\beta}}f_{m\mathbf{k}+\mathbf{q}} \\
       - \tau^{-1}_{n\mathbf{k} \to m\mathbf{k}+\mathbf{q}} \, \partial_{E_{\beta}}f_{n\mathbf{k}} \big],
\end{multline}
where we introduced the partial decay rate
\begin{multline}\label{eq.tau.partial}
  \tau^{-1}_{n\mathbf{k}  \to m\mathbf{k}+\mathbf{q}} \equiv \sum_{\nu} \frac{2\pi}{\hbar} \left| g_{mn\nu}(\mathbf{k},\mathbf{q}) \right|^2 \\
  \times \big[ (n_{\mathbf{q}\nu}+1-f^0_{m\mathbf{k}+\mathbf{q}})\delta(\Delta \varepsilon_{\mathbf{k},\mathbf{q}}^{nm} -\hbar\omega_{\mathbf{q}\nu}) \\
  + (n_{\mathbf{q}\nu}+f^0_{m\mathbf{k}+\mathbf{q}})\delta(\Delta \varepsilon_{\mathbf{k},\mathbf{q}}^{nm} +\hbar\omega_{\mathbf{q}\nu}) \big],
\end{multline}
and its analog $\tau^{-1}_{m\mathbf{k}+\mathbf{q} \to n\mathbf{k}}$ with the indices $n\mathbf{k}$ and $m\mathbf{k}+\mathbf{q}$ swapped.
Here, $f^0_{n\mathbf{k}}$ denotes the equilibrium occupancies in the absence of an electric field,
which are given by the Fermi-Dirac distribution evaluated at the band energies,
 $f^0_{n\mathbf{k}} \equiv 1/\{\exp[(\varepsilon_{n\mathbf{k}}-\mu)/k_{{\mathrm{B}}}T]+1\}$, where $\mu$ is the chemical potential.
We also used the fact that, ignoring the Berry curvature~\cite{Xiao2010},
the diagonal matrix elements of the velocity operator
are simply given by $v^{\alpha}_{n\mathbf{k}}=\hbar^{-1}\partial \varepsilon_{n\mathbf{k}}/\partial k_{\alpha}$.

Equation~\eqref{eq:iterboltzmann} is known in the literature~\cite{Ponce2018} as the \emph{Boltzmann transport equation}.
Its solution yields the linear response coefficients $\partial_{E_{\beta}}f_{n\mathbf{k}}$, which are needed in Eq.~\eqref{eq:conductivitydef} to obtain the conductivity tensor.

The electrical conductivity in Eq.~\eqref{eq:conductivitydef} scales with the density of carriers.
This is generally not an issue when studying metals, for which temperature, bias voltage, and defects do not alter the carrier density near the Fermi energy.
However, in semiconductors the carrier density can change by many orders of magnitude with doping, temperature, and applied voltage.
In these cases, in order to single out the intrinsic transport properties of the material, it is convenient to introduce the \emph{carrier drift mobility}, which
is defined as the ratio between conductivity and carrier density:
\begin{equation}\label{eq:definitionmob}
  \mu^{\mathrm{d}}_{\alpha\beta} \equiv \left|\frac{\sigma_{\alpha\beta}}{e n_{\mathrm{c}}}\right|.
\end{equation}
The charge carrier density entering the electron mobility tensor, $n_{\mathrm{c}} = n_{\mathrm{el}}$, is defined as
\begin{equation}
  n_{\mathrm{el}} = \frac{1}{V_{\mathrm{uc}}} \! \sum_{n\in \mathrm{CB}}  \! \int \frac{\mathrm{d}^3 k}{\Omega_{\mathrm{BZ}}}
  \left[ f_{n\mathbf{k}}^0(\mu,T) - f_{n\mathbf{k}}^0(\varepsilon_{\mathrm{F}},0) \right],
\end{equation}
with CB denoting the set of all conduction bands.
In the case of the hole mobility, $n_{\mathrm{c}} = n_{\mathrm{h}}$, the sum is understood to run over all valence bands.


\subsection{Approximations to the Boltzmann equation}\label{Sec2.3}

Besides the BTE given in Eq.~\eqref{eq:iterboltzmann}, there also exist
various simplified versions that can reduce the computational complexity at the cost of further approximations.
In this section we briefly discuss three common approximations, in decreasing order of accuracy:
(i) the momentum relaxation time approximation (MRTA),
(ii) the self-energy relaxation time approximation (SERTA),
and (iii) the lowest-order variational approximation (LOVA).

The main source of complexity in the BTE comes from the dependence of the linear response coefficients
$\partial_{E_{\beta}}f_{n\mathbf{k}}$ of state $|n\mathbf{k} \rangle$  on the linear response coefficients of all other states $|m\mathbf{k}+\mathbf{q}\rangle$.
In the \emph{momentum relaxation time approximation} (MRTA), this obstacle is overcome by using two approximations.
Firstly, the linear response coefficients $\partial_{E_{\beta}}f_{n\mathbf{k}}$
are taken to possess only a component in the direction of the band velocity $v^{\beta}_{n\mathbf{k}}$
\begin{equation}\label{eq:mrtaansatz}
  \partial_{E_{\beta}}f_{n\mathbf{k}} \stackrel{\mathrm{MRTA}}{\approx} e v^{\beta}_{n\mathbf{k}} \frac{\partial f^0_{n\mathbf{k}}}{\partial \varepsilon_{n\mathbf{k}}}  \widetilde{\tau}_{n\mathbf{k}},
\end{equation}
where $\widetilde{\tau}_{n\mathbf{k}}$ is now an unknown \emph{scalar} quantity to solve for.
Using Eq.~\eqref{eq:mrtaansatz}, the identity
\begin{equation}\label{eq:fdderiv}
  \frac{\partial f^0_{n\mathbf{k}}}{\partial \varepsilon_{n\mathbf{k}}} = -\frac{1}{k_{\rm B}T} f^0_{n\mathbf{k}}(1-f^0_{n\mathbf{k}}),
\end{equation}
 and multiplying with $\sum_{\alpha} v^{\alpha}_{n\mathbf{k}}$ on both sides of Eq.~\eqref{eq:iterboltzmann},
we obtain an equation for $\widetilde{\tau}_{n\mathbf{k}}$:
\begin{multline}\label{eq:mrtaansatz3}
  1 = \sum_m \int \frac{\mathrm{d}^3 q}{\Omega_{\mathrm{BZ}}} \, \Big[ \widetilde{\tau}_{n\mathbf{k}}
  \tau^{-1}_{n\mathbf{k} \to m\mathbf{k}+\mathbf{q}}
   - \frac{\mathbf{v}_{n\mathbf{k}} \cdot \mathbf{v}_{m\mathbf{k}+\mathbf{q}}}{|\mathbf{v}_{n\mathbf{k}}|^2} \\
     \times \frac{f^0_{m\mathbf{k}+\mathbf{q}}(1-f^0_{m\mathbf{k}+\mathbf{q}})}{f^0_{n\mathbf{k}}(1-f^0_{n\mathbf{k}})}
      \widetilde{\tau}_{m\mathbf{k+q}}\tau^{-1}_{m\mathbf{k}+\mathbf{q} \to n\mathbf{k}}\Big].
\end{multline}
At this point one can make use of the explicit algebraic forms
of the Fermi-Dirac and Bose-Einstein distribution functions and of the decay rates $\tau^{-1}_{n\mathbf{k} \to m\mathbf{k}+\mathbf{q}}$ to prove the detailed balance condition~\cite{Li2015}:
\begin{multline}\label{eq:detailbalance}
  f^0_{m\mathbf{k}+\mathbf{q}}(1-f^0_{m\mathbf{k}+\mathbf{q}}) \tau^{-1}_{m\mathbf{k}+\mathbf{q} \to n\mathbf{k}} \\
  = f^0_{n\mathbf{k}}(1-f^0_{n\mathbf{k}}) \tau^{-1}_{n\mathbf{k} \to m\mathbf{k}+\mathbf{q}}.
\end{multline}
Secondly, one makes the approximation
\begin{equation}\label{eq:mrtaansatz2}
  \widetilde{\tau}_{n\mathbf{k}}\tau^{-1}_{n\mathbf{k} \to m\mathbf{k}+\mathbf{q}}  \approx \widetilde{\tau}_{m\mathbf{k}+\mathbf{q}}\tau^{-1}_{m\mathbf{k}+\mathbf{q} \to n\mathbf{k}},
\end{equation}
in Eq.~\eqref{eq:mrtaansatz3} and using Eq.~\eqref{eq:detailbalance} one obtains an explicit expression for $\widetilde{\tau}_{n\mathbf{k}}$:
\begin{multline}\label{eq:taumrta}
  \widetilde{\tau}_{n\mathbf{k}}^{-1} = \sum_m \int \frac{\mathrm{d}^3 q}{\Omega_{\mathrm{BZ}}}
  \Big[ 1 - \frac{\mathbf{v}_{n\mathbf{k}} \cdot \mathbf{v}_{m\mathbf{k}+\mathbf{q}}}{|\mathbf{v}_{n\mathbf{k}}|^2} \Big] \tau^{-1}_{n\mathbf{k} \to m\mathbf{k}+\mathbf{q}}.
\end{multline}
%
It partially incorporates the effects of scattering back into the state $|n\mathbf{k}\rangle$
by reducing the rate for scattering out of it by a geometrical factor that involves the scattering angle and favors forward scattering.
The inverse of Eq.~\eqref{eq:taumrta} constitutes an effective, state-dependent, total scattering time.
By inserting $\widetilde{\tau}_{n\mathbf{k}}$ from Eq.~\eqref{eq:taumrta} into Eq.~\eqref{eq:mrtaansatz} and subsequently using the so-obtained
linear response coefficients in Eqs.~\eqref{eq:conductivitydef} and \eqref{eq:definitionmob}, we obtain the electron drift mobility in the MRTA:
\begin{equation}\label{eq:mobilitymrta}
  \mu^{\mathrm{d},\mathrm{MRTA}}_{\alpha\beta} \! = \! \frac{e}{n_{\mathrm{el}} V_{\mathrm{uc}}} \! \sum_n \! \int \! \frac{\mathrm{d}^3 k}{\Omega_{\mathrm{BZ}}}
  \bigg[\!-\frac{\partial f^0_{n\mathbf{k}}}{\partial \varepsilon_{n\mathbf{k}}} \bigg] v^{\alpha}_{n\mathbf{k}} v^{\beta}_{n\mathbf{k}} \widetilde{\tau}_{n\mathbf{k}}.
\end{equation}
%
The MRTA can be simplified even further if the rate for scattering back into the state $|n\mathbf{k}\rangle$ is neglected entirely.
This corresponds to setting the geometric factor in the square bracket of Eq.~\eqref{eq:taumrta} to one,
so that the effective scattering rate becomes equal to the total decay rate
\begin{equation}\label{eq:tauserta}
  \tau_{n\mathbf{k}}^{-1} \equiv \sum_m \int \frac{\mathrm{d}^3 q}{\Omega_{\mathrm{BZ}}} \tau^{-1}_{n\mathbf{k} \to m\mathbf{k}+\mathbf{q}}.
\end{equation}
As the total decay rate is also equal to twice the negative imaginary part of the retarded electron self-energy,
this approximation has been referred to as the \emph{self-energy relaxation time approximation} (SERTA)~\cite{Ponce2018}.
Similar to the case of the MRTA, the linear response coefficients in the SERTA are given by
\begin{equation}
  \partial_{E_{\beta}}f_{n\mathbf{k}} \stackrel{\mathrm{SERTA}}{\approx} e v^{\beta}_{n\mathbf{k}} \frac{\partial f^0_{n\mathbf{k}}}{\partial \varepsilon_{n\mathbf{k}}} \tau_{n\mathbf{k}}
\end{equation}
and the drift mobility reads
\begin{equation}\label{eq:mobilityserta}
  \mu^{\mathrm{d},\mathrm{SERTA}}_{\alpha\beta} \!=\! \frac{e}{n_{\mathrm{el}} V_{\mathrm{uc}}} \! \sum_n \!\! \int \!\! \frac{\mathrm{d}^3 k}{\Omega_{\mathrm{BZ}}}
  \bigg[\!-\frac{\partial f^0_{n\mathbf{k}}}{\partial \varepsilon_{n\mathbf{k}}}\bigg] v^{\alpha}_{n\mathbf{k}} v^{\beta}_{n\mathbf{k}} \tau_{n\mathbf{k}}.
\end{equation}

Lastly, we introduce a further approximation which is used for metals
and is referred to as the \emph{lowest-order variational approximation} (LOVA),
or the Ziman resistivity formula~\cite{Ziman1960,Allen1978,Pinski1981}.
In his original derivation, Ziman started from the Drude formula for the resistivity
of the electron gas, and derived an expression for the average scattering rate using a variational principle~\cite{Ziman1960}.
In order to keep the presentation self-contained, here we follow the alternative derivation by Grimvall~\cite{Grimvall1981},
who linked the isotropic scattering rate $\langle \tau^{-1} \rangle$
to an average of the state- and momentum-resolved total decay rates $\tau^{-1}_{n\mathbf{k}}$.
From Eq.~\eqref{eq:mobilityserta} we see that the conductivity
in the SERTA involves a weighted integral of velocity matrix elements
$v^{\alpha}_{n\mathbf{k}}v^{\beta}_{n\mathbf{k}}$ and the decay times $\tau_{n\mathbf{k}}$,
with the weighting factor being given by minus the derivative of the Fermi-Dirac distribution function.
This suggests evaluating the Drude formula with a scattering rate $\langle \tau^{-1} \rangle$ obtained using the same weighted average
\begin{equation}\label{eq:taulova}
  \left\langle \tau^{-1} \right\rangle \equiv
  \frac{\sum_n \int \frac{\mathrm{d}^3 k}{\Omega_{\mathrm{BZ}}} \big[-\frac{\partial f^0_{n\mathbf{k}}}{\partial \varepsilon_{n\mathbf{k}}}\big] \tau^{-1}_{n\mathbf{k}}}
       {\sum_n \int \frac{\mathrm{d}^3 k}{\Omega_{\mathrm{BZ}}} \big[-\frac{\partial f^0_{n\mathbf{k}}}{\partial \varepsilon_{n\mathbf{k}}}\big]}.
\end{equation}
Using Eqs.~\eqref{eq:tauserta} and \eqref{eq.tau.partial}, we can express the numerator as
\begin{multline}\label{eq:lovanumerator}
  \sum_n \int \frac{\mathrm{d}^3 k}{\Omega_{\mathrm{BZ}}} \bigg[ -\frac{\partial f^0_{n\mathbf{k}}}{\partial \varepsilon_{n\mathbf{k}}}\bigg] \tau^{-1}_{n\mathbf{k}}
  = \int \! \mathrm{d} \varepsilon \! \int \!\! \mathrm{d} \varepsilon ' \! \! \int \! \mathrm{d} \omega \bigg[-\frac{\partial f(\varepsilon )}{\partial \varepsilon }\bigg] \\
  \times \Big\{ \big[ n(\omega ) + 1 - f(\varepsilon') \big] \delta(\varepsilon - \varepsilon ' - \hbar\omega ) \\
  + \big[ n(\omega ) + f(\varepsilon ') \big] \delta(\varepsilon - \varepsilon ' + \hbar\omega ) \bigg\} \gamma(\varepsilon ,\varepsilon ',\omega ),
\end{multline}
where $n(\omega ) \equiv 1/[\exp(\hbar\omega/k_{\mathrm{B}}T) - 1]$ denotes the Bose-Einstein distribution, and
$f(\varepsilon ) \equiv 1/\{\exp[(\varepsilon -\varepsilon_{\mathrm{F}})/k_{\mathrm{B}}T] + 1\}$ is the Fermi-Dirac distribution,
in which we approximated the chemical potential $\mu$ by the Fermi energy $\varepsilon_{\mathrm{F}}$.
In Eq.~\eqref{eq:lovanumerator} we defined the energy-resolved and positive-definite decay function
\begin{multline}\label{eq:alphasquaref}
  \gamma(\varepsilon ,\varepsilon ',\omega ) \equiv
  \frac{2\pi}{\hbar} \sum_{mn\nu} \int \frac{\mathrm{d}^3 k}{\Omega_{\mathrm{BZ}}} \int \frac{\mathrm{d}^3 q}{\Omega_{\mathrm{BZ}}}
  \left| g_{mn\nu}(\mathbf{k},\mathbf{q}) \right|^2 \\
  \times \delta(\varepsilon -\varepsilon_{n\mathbf{k}}) \delta(\varepsilon '-\varepsilon_{m\mathbf{k}+\mathbf{q}}) \delta(\omega -\omega_{\mathbf{q}\nu}).
\end{multline}
Since the weight function $-\partial f(\varepsilon )/\partial \varepsilon$ appearing in Eq.~\eqref{eq:lovanumerator}
is peaked at the Fermi energy, Eq.~\eqref{eq:alphasquaref} only needs to be evaluated with $\varepsilon$ lying within a narrow window around the Fermi energy.
In addition, the Dirac delta function $\delta(\varepsilon -\varepsilon '-\hbar\omega )$ also forces $\varepsilon '$ to be close to the Fermi level,
as the phonon energies $\hbar \omega_{\mathbf{q}\nu}$ are typically of the same order of magnitude as the thermal energy $k_{\mathrm{B}}T$.
Allen~\cite{Allen1976a} noted that the electron-phonon matrix elements $g_{mn\nu}(\mathbf{k},\mathbf{q})$
usually do not vary much within a narrow window around the Fermi level.
%
In this case, $\gamma(\varepsilon ,\varepsilon ',\omega )$ can be approximated by
\begin{equation}
  \gamma(\omega ) \approx \gamma(\varepsilon =\varepsilon_{\mathrm{F}},\varepsilon '=\varepsilon_{\mathrm{F}},\omega ).
\end{equation}
The $\varepsilon$- and $\varepsilon '$-integrals in Eq.~\eqref{eq:lovanumerator} can then be carried out analytically
with the help of Eq.~\eqref{eq:fdderiv}, yielding
\begin{multline}
  \sum_n \int \frac{\mathrm{d}^3 k}{\Omega_{\mathrm{BZ}}} \bigg[\! -\frac{\partial f^0_{n\mathbf{k}}}{\partial \varepsilon_{n\mathbf{k}}}\bigg] \tau^{-1}_{n\mathbf{k}} \\
  \approx \frac{2k_{\mathrm{B}}T}{\hbar} \int_0^{\infty} \frac{\mathrm{d} \omega }{\omega }
  \frac{[\hbar \omega /( 2k_{\mathrm{B}}T)]^2 \gamma(\omega ) }{\sinh^2 [\hbar \omega /( 2k_{\mathrm{B}}T)]}.
\end{multline}
The denominator of Eq.~\eqref{eq:taulova} can be approximated by replacing the derivatives of the Fermi-Dirac distribution as $\delta$-functions
centered at the Fermi level:
$-\partial f^0_{n\mathbf{k}}/\partial \varepsilon_{n\mathbf{k}} \approx \delta(\varepsilon_{\mathrm{F}}-\varepsilon_{n\mathbf{k}})$.
As a result, the denominator of Eq.~\eqref{eq:taulova} becomes the density of states at the Fermi level, $\text{DOS}(\varepsilon_{\mathrm{F}})$.
By combining the resulting expression for the decay rate $\langle \tau^{-1}\rangle$
with Drude's formula, $\rho = m^* \langle \tau^{-1}\rangle/ e^2 n_{\mathrm{c}}$, one arrives at Ziman's resistivity formula:
\begin{equation}\label{eq:Ziman}
  \rho =  \frac{4 \pi m^*}{e^2 \hbar} \frac{2k_{\mathrm{B}}T}{n_{\mathrm{c}} }
   \int_0^{\infty} \!\! \frac{\mathrm{d} \omega }{\omega }  \frac{[\hbar \omega /( 2k_{\mathrm{B}}T)]^2 \alpha^2_{\mathrm{tr}}F(\omega ) }{\sinh^2 [\hbar \omega /( 2k_{\mathrm{B}}T)]},
\end{equation}
where the \emph{transport Eliashberg function}~\cite{Savrasov1996,Grimvall1981}
$\alpha^2_{\mathrm{tr}}F$ is defined as
\begin{equation}
  \alpha^2_{\mathrm{tr}}F(\omega ) \equiv \frac{1}{4 \pi \mathrm{DOS}(\varepsilon_{\mathrm{F}})}\gamma(\omega ).
\end{equation}
We note that Ziman's formula is semi-empirical in nature, since the density of carriers $n_c$ enters as an empirical parameter.


\subsection{Mobility at finite magnetic field}\label{Sec2.4}

While Eq.~\eqref{eq:boltzmannfourier} describes the dynamic equilibrium between a driving electrostatic force
and a restoring force due to carrier scattering, it can also be extended to include a finite magnetic field $\mathbf{B}$.
This extension requires the following replacement
\begin{equation}
  -e \mathbf{E} \to -e \left[ \mathbf{E} + \mathbf{v}_{n\mathbf{k}} \times \mathbf{B} \right]
\end{equation}
inside Eq.~\eqref{eq:boltzmannfourier}.
%
After carrying out the algebra, we obtain
a result similar to Eq.~\eqref{eq:iterboltzmann}:
\begin{multline}\label{eq:iterboltzmannwithb}
  -e v_{n\mathbf{k}}^{\beta} \frac{\partial f^{0}_{n\mathbf{k}}}{\partial \varepsilon_{n\mathbf{k}}}  -  e \left( \mathbf{v}_{n\mathbf{k}} \times \mathbf{B} \right)
  \cdot \frac{1}{\hbar} \frac{\partial}{\partial \mathbf{k}} \partial_{E_{\beta}}f_{n\mathbf{k}} = \sum_{m} \int \frac{\mathrm{d}^3 q}{\Omega_{\mathrm{BZ}}} \\
    \times  \big[ \tau^{-1}_{m\mathbf{k}+\mathbf{q} \to n\mathbf{k}} \, \partial_{E_{\beta}}f_{m\mathbf{k}+\mathbf{q}}  - \tau^{-1}_{n\mathbf{k} \to m\mathbf{k}+\mathbf{q}} \, \partial_{E_{\beta}}f_{n\mathbf{k}} \big],
\end{multline}
%
where we assumed that, to first order, the magnetic field alone does not perturb $f_{n\mathbf{k}}^0$~\cite{Ziman1972,Macheda2018}.
This assumption seems plausible and has been successfully used in the past~\cite{Macheda2018} but we are not aware of a formal proof. 

For practical implementations in first-principles software, it is useful to re-write Eq.~\eqref{eq:iterboltzmannwithb} and isolate
the linear response coefficients $\partial_{E_{\beta}}f_{n\mathbf{k}}$:
\begin{multline}\label{eq:iterwithbimpl}
 \left[ 1 - \frac{e}{\hbar}\tau_{n\mathbf{k}}(\mathbf{v}_{n\mathbf{k}}\times\mathbf{B})\cdot \frac{\partial}{\partial \mathbf{k}} \right]  \partial_{E_{\beta}} f_{n\mathbf{k}} = \\
  e v_{n\mathbf{k}}^\beta \frac{\partial f^0_{n\mathbf{k}}}{\partial \varepsilon_{n\mathbf{k}}} \tau_{n\mathbf{k}} + \frac{2\pi\tau_{n\mathbf{k}}}{\hbar}
  \sum_{m\nu} \!\int\! \frac{\mathrm{d}^3 q}{\Omega_{\mathrm{BZ}}} | g_{mn\nu}(\mathbf{k},\mathbf{q})|^2 \\
 \times \Big[(n_{\mathbf{q}\nu}+1-f_{n\mathbf{k}}^0)\delta(\Delta \varepsilon^{nm}_{\mathbf{k},\mathbf{q}} + \hbar \omega_{\mathbf{q}\nu} )  \\
  +  (n_{\mathbf{q} \nu}+f_{n\mathbf{k}}^0)\delta(\Delta \varepsilon^{nm}_{\mathbf{k},\mathbf{q}} - \hbar \omega_{\mathbf{q}\nu} ) \Big] \partial_{E_{\beta}} f_{m\mathbf{k}+\mathbf{q}},
\end{multline}
where $\tau_{n\mathbf{k}}$ is the total decay rate from Eqs.~\eqref{eq:tauserta} and \eqref{eq.tau.partial}:
\begin{multline}
  \tau_{n\mathbf{k}}^{-1} = \frac{2\pi}{\hbar} \sum_{m\nu} \!\int\! \frac{d\mathbf{q}}{\Omega_{\text{BZ}}} | g_{mn\nu}(\mathbf{k,q})|^2 \big[ (n_{\mathbf{q}\nu} +1 - f_{m\mathbf{k+q}}^0 ) \\
  \times \! \delta( \Delta \varepsilon^{nm}_{\mathbf{k},\mathbf{q}} \!  - \! \hbar \omega_{\mathbf{q}\nu}) \!  +  \! (n_{\mathbf{q}\nu} \! + \!  f_{m\mathbf{k+q}}^0 )\delta( \Delta \varepsilon^{nm}_{\mathbf{k},\mathbf{q}}  \! + \! \hbar \omega_{\mathbf{q}\nu}   ) \big].
\end{multline}

We note that, strictly speaking, electronic Bloch states and band structures are no longer well defined in the presence of a uniform magnetic field.
A rigorous treatment of this problem requires the use of ``magnetic boundary conditions'', which impose that the magnetic flux through a unit cell surface be an integer multiple of the flux quantum~\cite{Brown1964,Cai2004}.
Furthermore, for sufficiently strong magnetic field and low enough temperature, it is important to consider the quantization of the electron orbits into Laudau levels~\cite{Cohen2016}.
Equation~~\eqref{eq:iterwithbimpl} does not take into account these effects and therefore it is only valid for weak magnetic fields that can be treated perturbatively.
A useful approximate criterion to establish the crossover regime between Bloch bands and Landau levels is
the ratio between the cyclotron frequency $\omega_{\mathrm{c}} = eB/m^*$ (with $m^*$ denoting the effective carrier mass) and the scattering rate $\tau^{-1}$.
When the scattering rate is much larger than the cyclotron frequency, electrons are effectively hindered from remaining in stable cyclotron orbits.
Using the Drude formula $\mu = e\tau/m^*$, the crossover criterion $\omega_{\mathrm{c}} \sim \tau^{-1}$
can be written as $B \sim 1/\mu$. This simple expression provides a useful rule of thumb for
estimating the magnetic field at which the effects of Landau quantization become important.
For example, in a material with a mobility of 1,000~\cmVs{}, a magnetic field of $\sim$10~T would be required before the electronic density of states condenses into Landau levels~\cite{Lundstrom2009}.

An alternative to solving the BTE for weak magnetic fields is to consider the exact bilinear response coefficients $\partial_{E_{\beta}}\partial_{B_{\gamma}}f_{n\mathbf{k}}$.
Taking derivatives on both sides of Eq.~\eqref{eq:iterwithbimpl} with respect to $B_{\gamma}$ at zero field yields an iterative equation for the linear response coefficients
$\partial_{E_{\beta}}f_{n\mathbf{k}}$ and $\partial_{E_{\beta}}\partial_{B_{\gamma}}f_{n\mathbf{k}}$
\begin{multline}\label{eq:secondderivofB}
  \partial_{E_\beta}\partial_{B_\gamma} f_{n\mathbf{k}} = -\frac{e}{\hbar}\tau_{n\mathbf{k}} \Big( \mathbf{v}_{n\mathbf{k}} \times \frac{\partial}{\partial \mathbf{k}}\Big)_{\gamma} \partial_{E_\beta} f_{n\mathbf{k}} + \frac{2\pi\tau_{n\mathbf{k}}}{\hbar}  \\
 \times  \sum_{m\nu} \!\int\! \frac{d\mathbf{q}}{\Omega_{\text{BZ}}} | g_{mn\nu}(\mathbf{k,q})|^2 \big[ (n_{\mathbf{q}\nu} +1 - f_{n\mathbf{k}}^{0} ) \delta( \Delta \varepsilon^{nm}_{\mathbf{k},\mathbf{q}}  + \hbar \omega_{\mathbf{q}\nu} )\\
    +  (n_{\mathbf{q}\nu} + f_{n\mathbf{k}}^{0} )\delta( \Delta \varepsilon^{nm}_{\mathbf{k},\mathbf{q}} - \hbar \omega_{\mathbf{q}\nu} )   \big] \partial_{E_\beta} \partial_{B_\gamma} f_{m\mathbf{k+q}}.
\end{multline}
%

\begin{figure}[t!]
  \centering
  \includegraphics[width=0.99\linewidth]{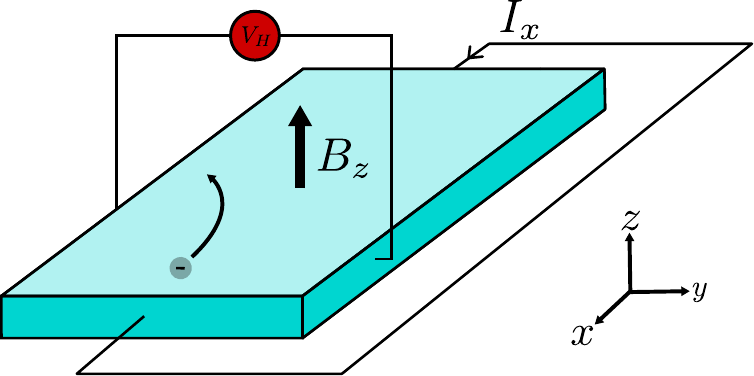}
  \caption{\label{fig:halleffect} Schematic setup for Hall mobility measurement.
  A current is flowing in the $+x$ direction and a magnetic field is applied in the $+z$ direction.
  The resulting Lorentz force will accelerate electrons in the $-y$ direction, resulting in a measurable Hall voltage $V^\mathrm{H}$.}
\end{figure}

The Hall conductivity tensor is obtained from the second derivatives $\partial_{E_{\beta}}\partial_{B_{\gamma}}f_{n\mathbf{k}}$ using
\begin{align}
  \sigma^{\mathrm{H}}_{\alpha\beta\gamma} \equiv& \left. \frac{\partial^2 J_{\mathrm{M},\alpha}}{\partial E_{\beta} \partial B_{\gamma}}
  \right|_{\substack{\mathbf{E}=\mathbf{0} \\ \mathbf{B}=\mathbf{0}}} \\
  =& \frac{-e}{V_{\mathrm{uc}}} \sum_n \int \frac{\mathrm{d}^3 k}{\Omega_{\mathrm{BZ}}} \, v^{\alpha}_{n\mathbf{k}} \partial_{E_{\beta}}\partial_{B_{\gamma}}f_{n\mathbf{k}}. \label{eq:conductivityhall}
\end{align}
A schematic setup for a Hall mobility measurement is shown in Fig.~\ref{fig:halleffect}.
Besides the drift and Hall conductivities and their mobility analogs, a commonly reported quantity is the dimensionless Hall factor (tensor), which is
defined as the ratio between the Hall conductivity (or mobility) and the drift conductivity (or mobility):
\begin{equation}\label{eq:hallfactor}
  r^{\mathrm{H}}_{\alpha\beta\gamma} \equiv \frac{\sigma^{\mathrm{H}}_{\alpha\beta\gamma}}{\sigma^{\mathrm{d}}_{\alpha\beta}}
  = \frac{\mu^{\mathrm{H}}_{\alpha\beta\gamma}}{\mu^{\mathrm{d}}_{\alpha\beta}}.
\end{equation}
A popular approximation to Eq.~\eqref{eq:hallfactor} consists of assuming a parabolic and non-degenerate band extremum,
following Ref.~\cite{Wiley1975}, p.~118 and Ref.~\cite{Price1957}, Eq.~3.12.
Within this approximation, the isotropic and temperature-dependent Hall factor is given by~\cite{Beer1963}
\begin{equation}\label{eq65}
  r^{\mathrm{H}} = \frac{\langle \tau^2 \rangle}{\langle \tau \rangle^2},
\end{equation}
with
\begin{equation}\label{eq66}
  \langle \tau^n \rangle \equiv \frac{\int_0^\infty \tau^n(x k_{\mathrm{B}}T) x^{3/2} e^{-x} \mathrm{d} x }{\int_0^\infty x^{3/2} e^{-x} \mathrm{d} x}.
\end{equation}
Here,  $x = \varepsilon/(k_{\mathrm{B}}T)$ and we introduced the distribution function of the total decay rate,
$\tau(\varepsilon ) = \sum_n \int \frac{\mathrm{d}^3 k}{\Omega_{\mathrm{BZ}}} \delta(\varepsilon -\varepsilon_{n\mathbf{k}}) \tau_{n\mathbf{k}}$.
In addition, the anisotropy present in band structures has been described by including a correction factor $K$~\cite{Herring1956}:
\begin{equation}\label{eq67}
  r^{\mathrm{H}} = \frac{ \langle \tau^2 \rangle }{\langle \tau \rangle^2} \frac{3K(K+2)}{(2K+1)^2}.
\end{equation}
This always results in a lowering of $r^{\mathrm{H}}$ compared to the fully isotropic ($K$=1) case.


\subsection{Kubo formalism}\label{Sec2.5}

The BTE formalism provides an efficient framework for dealing with time-independent electric fields in a self-consistent way.
However, the case of time-dependent fields is more conveniently dealt with
by directly evaluating the linear response of the current density instead of solving an equation of motion iteratively.
This approach has been developed by Kubo~\cite{Kubo1966} and the corresponding formula for the linear response of an observable
is known as the \emph{Kubo formula}.
The Kubo formula is especially convenient to study the linear response to time-dependent external perturbations as found in AC transport.
%

For time-dependent fields in the linear-response regime,
it is convenient to adopt a gauge in which the external electric field is expressed in terms of a vector potential:
\begin{equation}
  \mathbf{E}(t) = -\frac{\partial}{\partial t} \mathbf{A}(t),
\end{equation}
where we treat the electric field as being spatially homogeneous.
We then take the external Hamiltonian on the Keldysh-Schwinger contour as
\begin{multline}\label{eq:hextkubo}
  \hat{H}_{\mathrm{ext}}(z) = - \mathbf{A}(z) \cdot \int \mathrm{d}^3 r \, \hat{\mathbf{J}}(\mathbf{r},z) \\
                     + \frac{e}{2 m} \mathbf{A}^2(z) \int \mathrm{d}^3 r \, \hat{\varrho}(\mathbf{r}),
\end{multline}
where $\hat{\varrho}(\mathbf{r}) \equiv (-e) \hat{\psi}^{\dagger}(\mathbf{r})\hat{\psi}(\mathbf{r})$ is the electronic charge density operator and where we have introduced the
gauge-invariant current density operator:
\begin{align}\label{eq:currentdefkubo}
  \hat{\mathbf{J}}(\mathbf{r},z) &= \frac{-e}{m} \hat{\psi}^{\dagger}(\mathbf{r})\Big[-i\hbar\nabla\hat{\psi}(\mathbf{r})\Big] + \frac{e}{m} \mathbf{A}(z) \hat{\varrho}(\mathbf{r}) \\
                                 &\equiv \hat{\mathbf{J}}^{(\mathrm{p})}(\mathbf{r}) + \hat{\mathbf{J}}^{(\mathrm{d})}(\mathbf{r},z)[\mathbf{A}]. \label{eq:currentdefkubo2}
\end{align}
Here we have identified the two contributions as the \emph{paramagnetic} current density $\hat{\mathbf{J}}^{(\mathrm{p})}(\mathbf{r})$, corresponding to Eq.~\eqref{eq:currentdef},
and the \emph{diamagnetic} current density $\hat{\mathbf{J}}^{(\mathrm{d})}(\mathbf{r},z)$.
The external Hamiltonian given in Eq.~\eqref{eq:hextkubo} can be obtained by
applying the minimal coupling substitution $\hat{\mathbf{p}} \to \hat{\mathbf{p}} - (-e) \mathbf{A}(z)$ to the equilibrium Hamiltonian $\hat{H}_{\mathrm{eq}} \equiv \hat{H}_0 + \hat{H}_{\mathrm{int}}$.
We note that in Eq.~\eqref{eq:currentdefkubo} we chose for convenience not to symmetrize $\mathbf{J}^{(\mathrm{p})}(\mathbf{r})$ in the gradient as was done in Eq.~\eqref{eq:currentdef}.
For a spatially constant vector potential or more generally in the Coulomb gauge, corresponding to $\nabla \cdot \mathbf{A}(\mathbf{r},z) = 0$, the two forms of $\mathbf{J}^{(\mathrm{p})}(\mathbf{r})$ are equivalent.

As detailed in \ref{appendix2}, we use the Keldysh-Schwinger contour formalism to obtain the expectation value of the current density at time $t$.
We then define the conductivity tensor in the time domain as the functional derivative of the macroscopic current density with respect to electric field:
\begin{equation}
  \sigma_{\alpha\beta}(t,t') \equiv  \frac{1}{V} \int \mathrm{d}^3 r \,\left.\frac{\delta J_{\alpha}(\mathbf{r},t)}{\delta E_{\beta}(t')}\right|_{\mathbf{E}=\mathbf{0}}.
\end{equation}
Using the chain rule
\begin{align}
  \frac{\delta}{\delta E_{\beta}(t')} =& \sum_{\gamma} \int_{-\infty}^{+\infty} \! \mathrm{d} t''
  \frac{\delta A_{\gamma}(t'')}{\delta E_{\beta}(t')} \frac{\delta}{\delta A_{\gamma}(t'')} \\
                                      =& \int_{-\infty}^{+\infty} \! \mathrm{d} t'' \! \int \frac{\mathrm{d} \omega }{2 \pi} \, \frac{1}{i \omega } \mathrm{e}^{i\omega (t'-t'')} \frac{\delta}{\delta A_{\beta}(t'')},
\end{align}
we can write the time-domain conductivity as
\begin{multline}
  \sigma_{\alpha\beta}(t,t') = \int \frac{\mathrm{d} \omega }{2\pi} \frac{e}{i m \omega } \delta_{\alpha\beta} \mathrm{e}^{i\omega (t'-t)}
                        \frac{1}{V} \int \mathrm{d}^3 r \, \varrho_0(\mathbf{r},t) \\
                      + \int \frac{\mathrm{d} \omega }{2\pi} \frac{1}{\hbar \omega } \int_{-\infty}^{+\infty} \mathrm{d} t'' \, \mathrm{e}^{i\omega (t'-t'')} \\
                        \times \frac{1}{V} \int \mathrm{d}^3 r \int \mathrm{d}^3 r' \, \mathcal{J}_{\alpha,\beta}^{(\textrm{p}),\textrm{R}}(\mathbf{r},\mathbf{r}';t,t'').
\end{multline}
Here  $\mathcal{J}_{\alpha,\beta}^{(\textrm{p}),\textrm{R}}$ is the retarded component of the paramagnetic current-current correlation function, defined explicitly in \ref{appendix2},
while $\varrho_0(\mathbf{r},t)$ denotes the equilibrium charge density.
Note that all expectation values and correlation functions in the expression above
are defined with respect to the time-independent Hamiltonian $\hat{H}_{\mathrm{eq}}$ and hence can only depend on time differences
or, in the case of one-time functions, are time-independent.
We can then define the Fourier transform of the real-time conductivity tensor,
\begin{equation}
  \sigma_{\alpha\beta}(\omega ) \equiv \int_{-\infty}^{+\infty} \mathrm{d} (t-t') \mathrm{e}^{i \omega (t-t')} \sigma_{\alpha\beta}(t,t'),
\end{equation}
which is commonly referred to as the \emph{optical conductivity}.
Defining the Fourier transform of the retarded paramagnetic current-current correlation function in the same way,
the optical conductivity tensor takes on the compact form:
\begin{multline}\label{eq:conductivityac}
  \sigma_{\alpha\beta}(\omega) = i \frac{e^2}{m \omega } \delta_{\alpha\beta} n_{\mathrm{el}} \\
  + \frac{1}{\hbar \omega } \frac{1}{V} \int \! \mathrm{d}^3 r \! \int \! \mathrm{d}^3 r' \mathcal{J}_{\alpha,\beta}^{(\textrm{p}),\textrm{R}}(\mathbf{r},\mathbf{r}';\omega).
\end{multline}

In practice, the current-current correlation function is seldom evaluated exactly.
Instead, it is common to work in the \emph{independent-particle approximation} (IPA).
%
In this approximation, the conductivity tensor reads
\begin{multline}\label{eq:sigmaIPA}
  \sigma_{\alpha\beta}^{\rm IPA}(\omega ) = i \frac{e^2}{m \omega } \delta_{\alpha\beta} n_{\mathrm{el}}
  + i \frac{e^2}{V_{\mathrm{uc}}} \sum_{mn} \int \frac{\mathrm{d}^3 k}{\Omega_{\mathrm{BZ}}} v^{\alpha}_{mn\mathbf{k}} v^{\beta}_{nm\mathbf{k}} \\
  \! \times \! \! \int \!\! \mathrm{d} \omega ' \!\! \int \! \mathrm{d} \omega'' \! \frac{f(\hbar\omega'')-f(\hbar\omega')}{\hbar \omega} \frac{\mathcal{A}_{n\mathbf{k}}(\omega')\mathcal{A}_{m\mathbf{k}}(\omega'')}{\omega+\omega''-\omega'+i\eta},
\end{multline}
where $\mathcal{A}_{n\mathbf{k}}$ is the electronic spectral function of state $|n\mathbf{k}\rangle$.
The spectral function can be written in terms of the unperturbed eigenvalues of $\hat{H}_0$, $\varepsilon_{n\mathbf{k}}$,
and the retarded electron self-energy $\Sigma^{\mathrm{R}}_{n\mathbf{k}}(\omega )$ as
\begin{multline}\label{eq:spectralfct}
  \mathcal{A}_{n\mathbf{k}}(\omega) = \\
  \frac{\hbar}{\pi}\frac{-\mathrm{Im} [\Sigma^{\mathrm{R}}_{n\mathbf{k}}(\omega )]}
  { \{\hbar\omega - \varepsilon_{n\mathbf{k}} - \mathrm{Re} [\Sigma^{\mathrm{R}}_{n\mathbf{k}}(\omega ) ] \}^2
  + \{\mathrm{Im} [\Sigma^{\mathrm{R}}_{n\mathbf{k}}(\omega) ] \}^2},
\end{multline}
where we used the fact that $\mathcal{A}_{n\mathbf{k}}(\omega ) = -\frac{\hbar}{\pi}\mathrm{Im}\left[G^{\mathrm{R}}_{n\mathbf{k}}(\omega )\right]$
and expressed the retarded electron Green's function using Dyson's equation.
The derivation of Eq.~\eqref{eq:sigmaIPA} is given in \ref{indepparaapprox}.
The AC conductivity can be obtained by taking the real part of $\sigma_{\alpha\beta}^{\rm IPA}(\omega )$.

For completeness, we remark that in the limit $\omega \to 0$, the Kubo formula simplifies to
\begin{multline}\label{eq:sigma0IPA}
  \mathrm{Re} \sigma_{\alpha\beta}^{\rm IPA}(\omega \to 0) = \frac{\pi e^2}{V_{\mathrm{uc}}} \sum_{mn} \int \frac{\mathrm{d}^3 k}{\Omega_{\mathrm{BZ}}} \,
  \mathrm{Re} \left[ v^{\alpha}_{mn\mathbf{k}} v^{\beta}_{nm\mathbf{k}} \right] \\
  \times \int \mathrm{d} \omega' \Big[-\frac{\partial f}{\partial \varepsilon }(\hbar\omega') \Big] \mathcal{A}_{n\mathbf{k}}(\omega') \mathcal{A}_{m\mathbf{k}}(\omega'),
\end{multline}
where we made use of the fact that the $\omega \to 0$ limit of the term involving $\mathrm{Im}[v^{\alpha}_{mn\mathbf{k}}v^{\beta}_{nm\mathbf{k}}]$ vanishes identically.
This expression has a similar algebraic structure as Eq.~\eqref{eq:mobilityserta} for the mobility in the SERTA of the Boltzmann formalism.
We note that the effects of carrier scattering enter in the Kubo formalism through the spectral function.

Compared to the BTE, the Kubo formula has the advantage that it allows the incorporation
of higher-order electronic correlation effects on the electronic structure through the spectral function.
In addition, it also presents a simple way to calculate the AC conductivity.
On the downside, since the spectral function is almost invariably evaluated non-selfconsistently, it is difficult to achieve the same accuracy
as in the iterative BTE method for DC transport.

Finally, we note that a link between Eq.~\eqref{eq:sigma0IPA} and the conductivity equivalent of Eq.~\eqref{eq:mobilityserta} can be established
by neglecting the real part of the self-energy in Eq.~\eqref{eq:spectralfct} and evaluating the imaginary part at $\omega =\varepsilon_{n\mathbf{k}}/\hbar$.
Using $-2\mathrm{Im} \Sigma^{\mathrm{R}}(\varepsilon_{n\mathbf{k}}/\hbar)\equiv\hbar\tau^{-1}_{n\mathbf{k}}$, the spectral function in this approximation becomes
\begin{equation}
  \mathcal{A}_{n\mathbf{k}}(\omega ) \approx \frac{1}{\pi} \frac{\tau^{-1}_{n\mathbf{k}}/2}{(\omega-\varepsilon_{n\mathbf{k}}/\hbar)^2+(\tau^{-1}_{n\mathbf{k}}/2)^2}.
\end{equation}
If we further neglect the off-diagonal velocity matrix elements in Eq.~\eqref{eq:sigma0IPA}, the conductivity tensor reads
\begin{multline}\label{eq:conductivityipadc}
  \sigma_{\alpha\beta}^{\textrm{IPA-DC}} = \frac{\pi e^2}{V_{\mathrm{uc}}} \sum_{n} \int \frac{\mathrm{d}^3 k}{\Omega_{\mathrm{BZ}}} \, v^{\alpha}_{n\mathbf{k}} v^{\beta}_{n\mathbf{k}} \\
  \times \int \mathrm{d} \omega' \Big[ -\frac{\partial f}{\partial \varepsilon }(\hbar\omega') \Big] \mathcal{A}_{n\mathbf{k}}^2(\omega').
\end{multline}
Lastly, we make the approximation
\begin{equation}\label{eq:kubosertaapprox}
  \frac{\partial f}{\partial \varepsilon }(\hbar\omega')\mathcal{A}^2_{n\mathbf{k}}(\omega ') \approx \frac{\partial f}{\partial \varepsilon }(\varepsilon_{n\mathbf{k}})\mathcal{A}^2_{n\mathbf{k}}(\omega '),
\end{equation}
which is justified when the total decay rate is much smaller than the thermal energy $k_{\mathrm{B}}T$.
This approximation is reasonable for tetrahedral semiconductors at room temperature, where
the quasiparticle approximation is valid, but it breaks down for correlated narrow-gap semiconductors~\cite{Tomczak2018}.
Within the approximation of Eq.~\eqref{eq:kubosertaapprox}, the frequency integral involving the spectral function can be carried out analytically, yielding
\begin{equation}
  \int \mathrm{d} \omega \, \mathcal{A}_{n\mathbf{k}}^2(\omega ) = \frac{\tau_{n\mathbf{k}}}{\pi}.
\end{equation}
In this case, Eq.~\eqref{eq:sigma0IPA} reduces to the conductivity in the SERTA of the Boltzmann formalism, Eq.~\eqref{eq:mobilityserta}.

\subsection{Summary of available theoretical approaches}\label{Sec2.6}

\begin{figure}[b]
  \centering
  \includegraphics[width=\columnwidth]{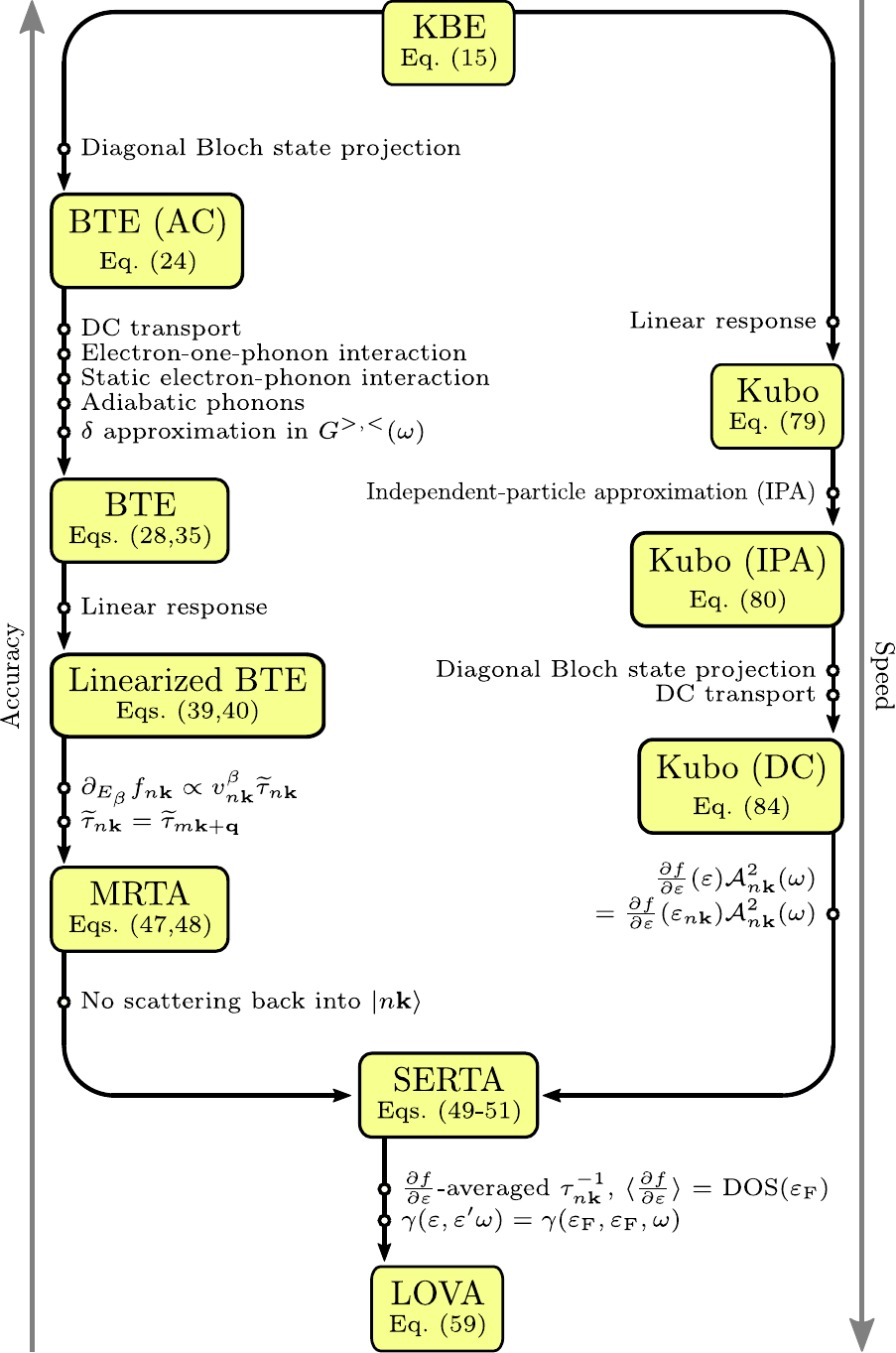}
  \caption{\label{fig:overview}
  Overview of the key theoretical equations and approximations presented in this review.
  KBE: the Kadanoff-Baym equation of motion for the lesser Green's function.
  BTE (AC): the Boltzmann transport equation for time-dependent electric fields.
  BTE: the Boltzmann transport equation for time-independent electric fields,
  with only electron-one-phonon scattering processes considered in the collision rate.
  Linearized BTE: the Boltzmann transport equation for the linear response coefficients $\partial_{E_{\beta}}f_{n\mathbf{k}}$.
  MRTA: the momentum relaxation time approximation to the linearized BTE.
  SERTA: the self-energy relaxation time approximation to the linearized BTE.
  LOVA: the lowest-order variational approximation to the linearized BTE.
  Kubo: the Kubo formula for the exact linear response of the lesser Green's function to a time-dependent electric field.
  Kubo (IPA): the Kubo formula in the independent-particle approximation.
  Kubo (DC): the Kubo formula in the IPA without off-diagonal Bloch state matrix elements in the DC transport limit.
  }
\end{figure}

In this section, we provide a concise overview of the theoretical approaches described so far.
A graphical summary is presented in Fig.~\ref{fig:overview}.
The central ingredient in calculations of charge transport is the current density, which can be obtained from
the lesser one-electron Green's function $G^<$, Eq.~\eqref{eq:greenslesser}.
The Green's function obeys one of the Kadanoff-Baym equations of motion, Eq.~\eqref{eq:kbe}, which are equivalent to Dyson's equation on the Keldysh-Schwinger contour.
The KBE for $G^<$ can be written in the basis of Bloch states of a reference Hamiltonian for a crystalline solid.
If we retain only the diagonal matrix elements, it takes the form of the Boltzmann transport equation for a homogeneous, time-dependent electric field [BTE (AC)], Eq.~\eqref{eq:boltzmann}.
From this point, we can further simplify the formalism by considering time-independent fields (DC transport) and retaining only one-phonon scattering processes in the collision term, using frequency-independent electron-phonon coupling matrix elements and adiabatic phonons.
Similarly, the Green's functions in the collision term can be approximated using the non-interacting, single-particle Hamiltonian.
Using these approximations, we arrive at the steady-state version of the Boltzmann transport equation (BTE), Eqs.~\eqref{eq:boltzmannfourier} and \eqref{eq:collisionrate}.
To obtain the conductivity and the mobility tensor, we consider weak fields and linearize the BTE, Eq.~\eqref{eq:iterboltzmann}.

From the BTE one can then identify a hierarchy of three further approximations, namely
the momentum relaxation-time approximation, the self-energy relaxation time approximation, and the lowest-order variational approximation.
In the MRTA, Eqs.~\eqref{eq:taumrta} and \eqref{eq:mobilitymrta}, the change of the occupation number $f_{n\mathbf{k}}$ with electric field is taken
to only have a component in the direction of the band velocity $\mathbf{v}_{n\mathbf{k}}$, with its magnitude being proportional to an effective scattering time $\widetilde{\tau}_{n\mathbf{k}}$.
The latter is further taken to be independent of the electron wavevector in the collision rate.
Starting from the MRTA, one can make the further approximation of considering only scattering processes out of the state $|n\mathbf{k}\rangle$, while neglecting the scattering into this state.
This approximation leads to the self-energy relaxation time approximation, Eqs.~(\ref{eq:tauserta}-\ref{eq:mobilityserta}), and it is equivalent to a non-iterative solution of the BTE.
The central quantity in the SERTA is the total decay rate $\tau^{-1}_{n\mathbf{k}}$, which can be obtained from the imaginary part of the retarded electron self-energy.
As a further simplification one can consider the lowest-order variational approximation, which leads to the Ziman resistivity formula, Eq.~\eqref{eq:Ziman}.
In the LOVA, one considers metals, the state- and momentum-resolved decay rates $\tau^{-1}_{n\mathbf{k}}$ are approximated by their weighted average
in a small window around the Fermi energy, and the scattering function $\gamma(\varepsilon ,\varepsilon ',\omega)$ is simplified by considering only electrons at the Fermi level.

A different approach to the transport problem is obtained by considering
the Kubo formula, Eq.~\eqref{eq:conductivityac}.
While in deriving the linearized BTE one first approximates the equation of motion for $G^<$ and then linearizes in the electric field, in the derivation of the Kubo formula one directly considers the linear response of $G^<$ in perturbation theory.
This procedure directly yields the AC conductivity.
%
%
In practice the Kubo formula is employed within the independent-particle approximation [Kubo (IPA)], Eq.~\eqref{eq:sigmaIPA}.
%
A further approximation consists of neglecting the off-diagonal matrix elements of the velocity operator and of the spectral function.
In the case of time-independent electric fields this leads to the SERTA of the Boltzmann formalism [Kubo (DC)], Eq.~\eqref{eq:sigma0IPA}.
Therefore there is a clear connection between the BTE and the Kubo approach under well-defined approximations.
The relation between the Kadanoff-Baym approach, the Boltzmann formalism and its
various approximations, and the Kubo formalism is schematically illustrated in Fig.~\ref{fig:overview}.

\section{Implementation in modern electronic structure codes}\label{Sec3}

While transport properties have been studied with analytical approaches for decades,
first-principles-based calculations have made their appearance much more recently due to the numerical complexity involved and the lack of adequate software infrastructure.
Even though these calculations are not very streamlined and still require large
high-performance computing (HPC) facilities to be performed, various computer codes to perform these calculations have appeared in the past fifteen years.
A non-exhaustive list is given in Table~\ref{table1}.
\begin{table*}
\begin{small}
\begin{center}
  \begin{tabular}{r r r r r l}
  \hline
Method             & Approximation           & Software                              & License & Size       &  Notes \\
                   &                         &                                       &         & (\# atoms) &        \\
\hline
DFT-NEGF           & local GF                & \texttt{TRANSIESTA}~\cite{Papior2017} & GPL &  3000  & LCAO (DFT or TB) \\
                   &                         & \texttt{SMEAGOL}~\cite{Rocha2005}     & SAL &   100  & LCAO with DFT, supercell  \\
                   &                         & \texttt{AITRANSS}~\cite{Bagrets2013}  & COM &   100  & MO with no $\mathbf{k}$-points    \\
                   &                         & \texttt{GIPAW}~\cite{Chen2012}        & GPL &   100  & AO     \\
                   &                         & \texttt{OMEN}~\cite{Luisier2009}      & COM & 10000  & dissipative transport (TB) \\
Kubo               & linear response         & \texttt{KGEC}~\cite{Calderin2017}     & GPL &   100  & Kubo-Greenwood with PW  \\
                   &                         & \texttt{ABINIT}~\cite{Gonze2002}      & GPL &   100  & Kubo-Greenwood with PW  \\
                   &                         & no name~\cite{Roche1997}              & PRI &  1000  & Kubo-Greenwood with  TB \\
BTE                & linear response         & \texttt{EPW}~\cite{Ponce2016a}        & GPL &    50  & PW and Wannier interpolation \\
                   &                         & no name~\cite{Li2015}                 & PRI &     5  & PW and linear interpolation \\
                   &                         & no name~\cite{Sohier2018}             & PRI &     5  & PW and linear interpolation \\
BTE-SERTA          & no ``in''-scattering    & no name~\cite{Restrepo2009}           & PRI &     5  & PW no interpolation \\
                   &                         & \texttt{ATK}~\cite{Gunst2016}         & COM &     5  & localized basis set, supercell \\
                   &                         & \texttt{PERTURBO}~\cite{Agapito2018}  & PRI &    10  & Atomic orbital interpolation  \\
BTE-cSERTA         & constant scattering     & \texttt{ShengBTE}~\cite{Li2014b}      & GPL &   100  & iterative method \\
                   &                         & \texttt{BOLTZTRAP}~\cite{Madsen2006}  & GPL &   100  & smoothed Fourier interpolation \\
                   &                         & \texttt{BOLTZWANN}~\cite{Pizzi2014}   & GPL &   100  & Wannier interpolation of bands \\
BTE-models         & model scattering        & Rode iteration~\cite{Rode1970}        & PRI &  1000  & model EP interaction \\
                   &                         & variational~\cite{Howarth1953}        & PRI &  1000  & model EP interaction \\
                   &                         & Monte Carlo~\cite{Jacoboni1983}       & PRI &  1000  & model EP interaction \\
  \hline
  \end{tabular}
    \caption{\label{table1}
 Various methods for the calculation of carrier mobility,
 ranging from non-equilibrium Green's function DFT-NEGF, Kubo, and the Boltzmann transport equation (BTE).
 BTE calculations in various flavors are possible: self-energy relaxation time approximation (SERTA),
 constant self-energy relaxation time approximation (cSERTA),
 and BTE with analytical models for the relaxation time.
 The DFT-NEGF usually relies
 on linear combinations of atomic orbitals (LCAO), molecular orbitals (MO), or atomic orbitals (AO)
 computed with DFT or tight-binding (TB) methods.
 GPL denotes the GNU general public license, SAL stands for the Smeagol academic license,
 COM stands for commercial license, and PRI denotes a private code not available to the community.}
\end{center}
\end{small}
\end{table*}
Existing codes can broadly be grouped into three categories:
(i) non-equilibrium Green's function (NEGF) methods coupled with DFT or tight-binding methods,
to describe ballistic transport between leads and atomic wires, molecules, or surfaces
~\cite{Taylor2001,Nardelli2001,Brandbyge2002,Palacios2002,Wortmann2002,Rocha2005,Pecchia2008,Saha2009,Ozaki2010,Chen2012,Bagrets2013,Papior2017};
(ii) codes that solve the linearized BTE relying on \textit{ab initio} band structures and velocities and employ empirical relaxation times~\cite{Madsen2006,Li2014b,Pizzi2014};
and (iii) implementations in which the linearized BTE is solved from first principles, without empirical parameters~\cite{Restrepo2009,Li2015,Gunst2016,Ponce2016a,Agapito2018,Sohier2018}.

\begin{figure}[t!]
  \centering
  \includegraphics[width=0.75\linewidth]{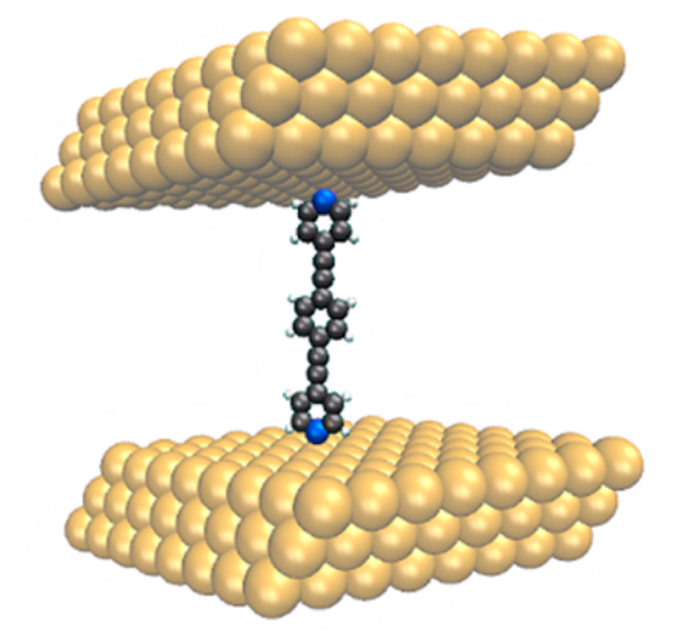}
  \caption{\label{fig:junction}
  Structural model employed in first-principles calculations of
  charge transport in 1,2,-bis(2-phenylethynyl)benzene
  bonded to Au electrodes.
  Different anchoring groups including pyridine, thiolate, and isocyanide have been studied.
 Adapted from Ref.~\cite{Obersteiner2017}; copyright (2017) by the American Chemical Society.}
\end{figure}

In the first category, we find software dedicated to ballistic transport,
where typically a molecule is placed in-between two leads (e.g., a C$_{60}$ molecule placed between two semi-infinite copper leads),
which is modeled \textit{ab initio} or using tight-binding methods.
Although in principle one would like to solve for the fully interacting non-equilibrium Green's function,
this has not been achieved yet; instead one usually works in the ballistic regime, wherein the scattering of carriers in the conduction region between the leads is neglected.
NEGF calculations are based on the same formalism presented in Sec.~\ref{Sec2.1}; the calculations rely on DFT
to compute the electronic structure and the unperturbed Hamiltonian and the Schwinger-Keldysh formalism to obtain the non-equilibrium density matrix (DFT-NEGF).
Within this approach, various basis sets have been employed to describe the Green's function, ranging from density-functional-based tight-binding~\cite{Pecchia2008} to numerical atomic orbitals,
which feature an efficient linear scaling~\cite{Brandbyge2002,Rocha2005,Papior2017};
Gaussian orbitals~\cite{Palacios2002,Bagrets2013}, pseudoatomic orbitals (PAOs)~\cite{Ozaki2010},
real-space-optimized orbitals~\cite{Nardelli2001,Saha2009},
linearized augmented plane waves (LAPW)~\cite{Wortmann2002},
projector-augmented plane waves (PAW)~\cite{Chen2012},
and linear combinations of atomic orbitals (LCAO)~\cite{Taylor2001}.
Some of these codes can also be used to model multi-lead junctions.
Recent codes can easily cope with over 10,000 orbitals for DFT-NEGF calculations, and over 1,000,000 orbitals for tight-binding NEGF-type calculations, which makes it possible to study nanoscale systems such as flakes of two-dimensional materials and molecular junctions.
For example, Fig.~\ref{fig:junction} shows the structure of a molecular junction for a calculation involving clusters of up to 16 molecules and approximately 3,000 atoms.
In the linear-response limit (Kubo formula), it is possible to compute the conductivity of metals
and a few codes have been developed for this purpose~\cite{Gonze2002,Calderin2017}.
Furthermore, codes have been developed to tackle large models of 2D materials, including defects, using the tight-binding, real-space, $\mathcal{O}(N)$ Kubo-Greenwood method with parameters derived from {\it ab initio} calculations~\cite{Mayou1988,Roche1997,Roche1999}.
%

In the second category, the solution to the linearized BTE can be computed with Rode's iterative approach~\cite{Budd1967,Rode1970,Rode1970a,Rode1975},
variational principles~\cite{Howarth1953,Ehrenreich1960,Ehrenreich1961},
or by Monte-Carlo sampling~\cite{Jacoboni1983,Fischetti1988,Fischetti1993},
where the electron-phonon interaction is modeled using various semiempirical models (BTE-models in Table~\ref{table1}).
Clearly the use of simplified models to describe the electron-phonon coupling reduces the range of applicability of the methods; typical simplifications include the study of a single phonon branch (Debye or Einstein models), of a single parabolic band, and the neglect of anisotropy.
For materials where those approximations hold, earlier methods are very affordable and have found widespread application in the 1970s and 1980s; several models
are still successfully being used nowadays~\cite{Frost2017b}.
To go beyond isotropic materials with multiple and non-parabolic electron bands,
various methods based on the efficient calculation of DFT band structures have been developed,
including the use of Fourier interpolation of the bands~\cite{Madsen2006} or the use of maximally localized Wannier functions (MLWF) to interpolate the eigenstates and velocities~\cite{Souza2001,Pizzi2014}.
Another, more time-consuming approach is to use real-space supercells to evaluate the interatomic force constants (IFC) with the added advantage of including anharmonic effects~\cite{Li2014b}.
In all cases, the codes rely on constant scattering rates (BTE-cSERTA).
The key challenge when solving the BTE is that the momentum integral over $\mathbf{k}$ converges extremely slowly due to the sharpness of the Fermi-Dirac distribution~\cite{Pizzi2014}.
For this reason, these codes rely on efficient interpolation techniques~\cite{Madsen2006,Pizzi2014} to evaluate the eigenvalues and carrier velocities obtained from first principles
on ultra-dense grids, for example using grids of 200$\times$200$\times$200 $\mathbf{k}$-points or denser.

In the third category we find codes where also the electron-phonon matrix elements and scattering rates are evaluated \textit{ab initio} using DFPT.
The main challenge when computing \textit{ab initio}
electron-phonon scattering rates is the requirement of ultra-dense momentum grids close to the band edges~\cite{Ponce2018}.
Furthermore, the problem is exacerbated by the fact that transport properties require at the same time ultra-dense grids for the phonon momenta ($\mathbf{q}$-points)
as well as the electron momenta ($\mathbf{k}$-points).
In the case of bulk three-dimensional crystals, this requirement leads to a challenging $\mathcal{O}(N^6)$ scaling of the computational workload, if $N$ is the number of grid points in the Brillouin zone along a reciprocal lattice vector.
Various approaches have been attempted for tackling this task, including the direct evaluation of electron-phonon matrix elements using DFPT~\cite{Restrepo2009},
the linear interpolation of the \textit{ab initio} scattering rates~\cite{Li2015,Sohier2018},
the use of local orbital implementations~\cite{Gunst2016,Agapito2018}, and the use of MLWFs~\cite{Giustino2007,Ponce2016a}.

The interpolation method based on MLWFs is viewed by many as the most accurate and computationally affordable approach.
In this method
the electron-phonon matrix elements on dense momentum grids can be obtained from~\cite{Giustino2007}:
\begin{multline}\label{eq:gdefinition}
  g_{mn\nu}(\mathbf{k},\mathbf{q}) = \sum_{pp'\kappa}\sqrt{\frac{\hbar}{2M_{\kappa} \omega_{\mathbf{q}\nu}}}
  \mathrm{e}^{i(\mathbf{k} \cdot \mathbf{R}_p + \mathbf{q} \cdot \mathbf{R}_{p'})} \\
  \times \! \sum_{m'n'\alpha} \! e_{\kappa\alpha\nu}(\mathbf{q})
  U_{mm'\mathbf{k}+\mathbf{q}} g_{m'n'\kappa\alpha}(\mathbf{R}_p,\mathbf{R}_{p'}) U^{\dagger}_{n'n\mathbf{k}},
\end{multline}
where $U_{nm\mathbf{k}}$ is a unitary transformation matrix that converts the periodic part of the electronic wave function into real-space Wannier functions that are maximally localized~\cite{Marzari1997}.
The $g_{mn\kappa\alpha}(\mathbf{R}_p,\mathbf{R}_{p'})$ are the real-space Wannier electron-phonon matrix elements, which decay rapidly in real space as a function of
$|\mathbf{R}_p|$ and $|\mathbf{R}_{p'}|$. This property enables the efficient interpolation of the matrix elements to arbitrary points of the Brillouin zone using Eq.~\eqref{eq:gdefinition}.

In the case of polar materials, the interaction of electrons with longitudinal optical modes is long-ranged.
As a consequence, the electron-phonon matrix elements $g_{mn\nu}(\mathbf{k},\mathbf{q})$ diverge as $1/|\mathbf{q}|$ for $|\mathbf{q}|\rightarrow 0$~\cite{Frohlich1954,Ponce2015}.
The correct treatment of this divergence when performing Wannier interpolation has recently been proposed~\cite{Verdi2015,Sjakste2015} and consists of splitting
the electron-phonon matrix elements into a short- ($\mathcal{S}$) and a long-range ($\mathcal{L}$) contribution~\cite{Verdi2015}
\begin{equation}\label{eq:gspliting}
  g_{mn\nu}(\mathbf{k},\mathbf{q}) = g_{mn\nu}^{\mathcal{S}}(\mathbf{k},\mathbf{q}) + g_{mn\nu}^{\mathcal{L}}(\mathbf{k},\mathbf{q}),
\end{equation}
where
\begin{multline}\label{eq:glongrange}
  g_{mn\nu}^{\mathcal{L}}(\mathbf{k},\mathbf{q}) = i \sum_\kappa
   \sqrt{ \frac{\hbar}{2 M_\kappa \omega_{\mathbf{q}\nu} } }  \sum_{\mathbf{G}\neq -\mathbf{q}} \\
  \times \frac{(\mathbf{q}+\mathbf{G}) \mathbf{Z}^*_{\kappa}  \mathbf{e}_{\kappa,\nu}(\mathbf{q})}{(\mathbf{q}+\mathbf{G})
  \boldsymbol{\varepsilon }^{\infty}  (\mathbf{q}+\mathbf{G})} \big\langle  m\mathbf{k}+\mathbf{q}
  \big| \mathrm{e}^{i(\mathbf{q}+\mathbf{G}) \cdot {\mathbf{r}}} \big| n\mathbf{k} \big\rangle.
\end{multline}
Here, $\mathbf{Z}^*_\kappa$ is the Born effective charge tensor, $\mathbf{e}_{\kappa,\nu}(\mathbf{q})$ is the vibrational eigendisplacement vector,
and $\boldsymbol{\varepsilon }^{\infty}$ is the macroscopic high-frequency dielectric constant tensor, evaluated at clamped ions.
The overlap matrices in Eq.~\eqref{eq:glongrange} can be computed in the approximation of small $\mathbf{q}+\mathbf{G}$~\cite{Verdi2015} as
\begin{equation}
  \big\langle m\mathbf{k}+\mathbf{q} \big| \mathrm{e}^{i(\mathbf{q}+\mathbf{G}) \cdot {\mathbf{r}}}
  \big| n\mathbf{k} \big\rangle = \big[ U_{\mathbf{k}+\mathbf{q}+\mathbf{G}} U_{\mathbf{k}}^{\dagger} \big]_{mn},
\end{equation}
where the periodic gauge $|n\mathbf{k}\rangle$ = $|n\mathbf{k+G}\rangle$ is implied. 
The Wannier rotation matrices $U_{nm\mathbf{k}}$ can be obtained at arbitrary $\mathbf{k}$-points and $\mathbf{q}$-points through the interpolation of the electronic Hamiltonian~\cite{Souza2001}.
One can therefore accurately interpolate electron-phonon matrix elements using the following four-step procedure:
(i) the matrix elements $g_{mn\nu}(\mathbf{k},\mathbf{q})$ are computed on a coarse grid using DFPT;
(ii) the long-range part $g_{mn\nu}^{\mathcal{L}}(\mathbf{k},\mathbf{q})$ is subtracted
to obtain the short-range component $g_{mn\nu}^{\mathcal{S}}(\mathbf{k},\mathbf{q})$;
(iii) the standard Wannier electron-phonon interpolation of Ref.~\cite{Giustino2007}
is applied to the short-range component only;
and (iv) the long-range component is added back to the interpolated
short-range part for each arbitrary $\mathbf{k}$-point and $\mathbf{q}$-point.

Most codes today rely on the self-energy relaxation time approximation (Table~\ref{table1}, BTE-SERTA) due to the simplicity of its implementation.
Indeed, within this approximation, the scattering rate is directly related to the imaginary part
of the retarded electron-phonon (Fan-Migdal) self-energy and can therefore be computed easily.
However, this approximation is not reliable in materials with strong band structure anisotropy and in polar materials~\cite{Liu2017,Ma2018}.
In order to go beyond SERTA, it is possible to solve the BTE iteratively, with a small computational overhead.
For example, in simple semiconductors it takes approximately 20 iterations to reach convergence~\cite{Fiorentini2016}.

The calculation of the electron mobility in GaAs is representative of the computational requirements in terms of Brillouin zone sampling and of the importance of iterative solutions.
In this case, converged calculations required grids as dense as 400$\times$400$\times$400 $\mathbf{k}$-points and the iterative BTE solution yielded mobility values approximately 50\% higher than in the SERTA~\cite{Ma2018}.
Moreover, significantly denser grids are needed to obtain converged mobility results at lower temperatures
and this may explain why many authors present calculation results at relatively high temperatures.
Low temperatures pose a challenge because the product of Fermi-Dirac occupations and the electronic density of states is peaked close to the band edges and the width of the peaks decreases with decreasing temperature.
Adaptive broadening strategies such as proposed by Li~\cite{Li2015} can be used to address this challenge
by using a smaller Gaussian broadening at lower temperatures and closer to the band edges.
We also note that at very low temperature, the mobility is no longer limited by phonon-induced scattering and other mechanisms need to be taken into account~\cite{Lundstrom2009}.
Finally, the importance of including spin-orbit coupling (SOC) was recently highlighted even for materials like silicon, where the spin-orbit splitting is small~\cite{Ma2018,Ponce2018}.

At present, for simple tetrahedral semiconductors, \textit{ab initio} mobility calculations typically agree with experiments to within 20\%,
but the agreement worsens for narrow-gap semiconductors as a result of the poor description of the effective masses in DFT.
Transport calculations based on accurate band structures obtained from many-body $GW$ perturbation theory have recently been demonstrated~\cite{Ponce2018}, but
in order to go forward, it will be important to also include many-body corrections to the electron-phonon matrix elements~\cite{Faber2011,Yin2013,Antonius2014,Monserrat2016a,Li2019a}.


\section{Recent \textit{ab initio} calculations of carrier mobilities}\label{Sec4}
\subsection{Bulk materials}\label{Sec4.1}

\begin{figure*}[t]
\begin{centering}
\includegraphics[width=1.01\textwidth]{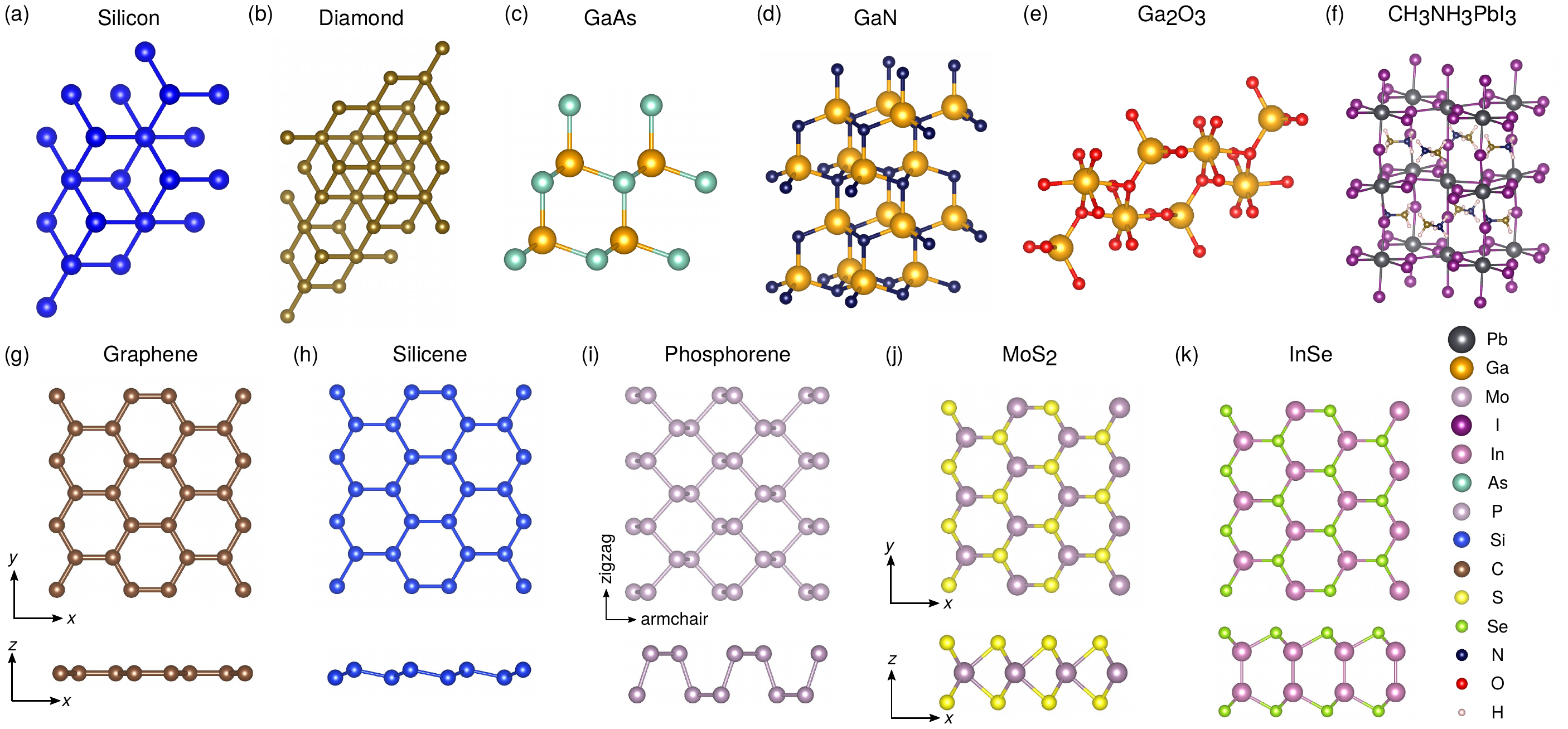}
\caption{
 Ball-stick representations of the semiconductors reviewed in this work:
 (a) silicon, (b) diamond, (c) GaAs, (d) wurtzite GaN, (e) $\beta$-Ga$_2$O$_3$,
 (f) CH$_3$NH$_3$PbI$_3$, (g) graphene, (h) silicene, (i) phosphorene, (j) MoS$_2$, and (k) InSe.}
\label{fig:structures}
\end{centering}
\end{figure*}

In this section, we review some of the key efforts toward developing predictive
methods for calculating carrier transport properties of bulk materials.
We discuss two nonpolar semiconductors, namely silicon and diamond, and four polar semiconductors,
namely GaAs, GaN, Ga$_2$O$_3$, and halide perovskites.
These compounds find application in semiconductor research and technology, including electronics, optoelectronics, lighting, and energy research.
Ball-stick representations of these compounds are shown in Fig.~\ref{fig:structures}(a)-(f).

\subsubsection{Silicon}

\begin{figure*}[t!]
  \centering
  \includegraphics[width=0.99\linewidth]{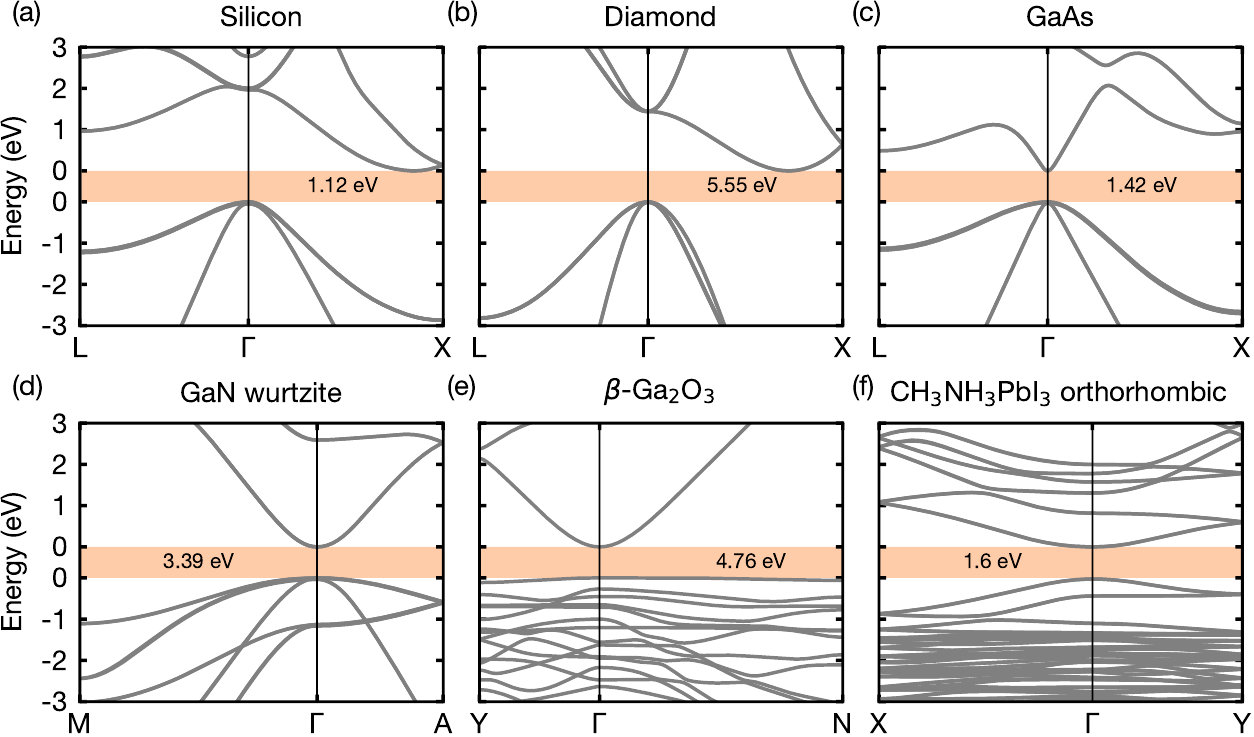}
  \caption{\label{fig:bs}
  Electronic band structure of (a) silicon, (b) diamond, (c) GaAs, (d) wurtzite GaN,
  (e) $\beta$-Ga$_2$O$_3$, and (f) the low-temperature orthorhombic phase of CH$_3$NH$_3$PbI$_3$.
  The calculated band structures are obtained from DFT and include spin-orbit coupling.
  The energy axis are aligned with the band edges, and the experimental bandgaps are indicated in the shaded areas.
  }
\end{figure*}

Under ambient conditions, silicon crystallizes in the diamond structure.
Bulk silicon has an indirect bandgap of 1.12~eV, with a valence band top composed
of a degenerate heavy and light hole band and a band splitting of 8~meV due to spin-orbit interaction.
The conduction band minima are located 0.85~$2\pi/a$ away from the Brillouin zone center, along the $\Gamma$-$X$ directions leading to six elongated electron pockets~\cite{Sze2007}.
The electronic band structure of silicon is shown in Fig.~\ref{fig:bs}(a).

In the 1950s, Smith measured the piezoresistance of $n$-doped silicon, i.e., the dependence of resistivity on strain, and relied on early band structure calculations to understand the deformation of the six conduction band valleys under strain~\cite{Smith1954}.
In the 1960s, empirical models were developed to rationalize experimental observations whereby the carrier velocity increases with field strength and the mobility decreases with doping~\cite{Caughey1967}.
The prevalent semi-empirical model to account for impurity scattering and the reduction of the mobility with carrier concentration is due to Brooks and Herring~\cite{Brooks1951,Li1977}.
This model relies on a static, single-site description of carrier-impurity interaction and on the Born approximation~\cite{Kosina1998}.
According to this model the hole mobility is given by:
\begin{equation}\label{eq92}
  \mu = \frac{2^{7/2} \epsilon_s^2 (k_B T)^{3/2} }{100 \pi^{3/2} e^3 \sqrt{m_d^*} \,n_{\rm i}G(b)}.
\end{equation}
Here, $G(b) = \ln(b+1) - b/(b+1)$, $b = 24\pi m_d^* \epsilon_s (k_B T)^{2}/(10^6 e^2 h^2 n')$,
$n'=n_{\rm h}(2-n_{\rm h}/n_{\rm i})$, $m_d^*=0.55 m_0$
is the density-of-state effective mass for holes~\cite{Balkanski2000}, $n_{\rm h}$ and $n_{\rm i}$
are the hole densities and the density of ionized impurities, respectively, and $\epsilon_s=11.9\epsilon_0$ is the static dielectric constant of silicon.
The extension of this model to describe the anisotropic electron effective masses of silicon was developed by Long and Norton~\cite{Norton1973,Li1977,Arora1982}.

Subsequently, refined analytical models that include the screening of the Coulomb potential of impurities
by charge carriers, electron-hole scattering, and clustering of impurities were developed for device simulations~\cite{Klaassen1992}.
After it was reported that strained silicon on a (100) Si$_{1-x}$Ge$_x$ substrate could significantly
increase the carrier mobility~\cite{Manku1992,Manku1993}, Nayak and Chun employed $\mathbf{k}\cdot \mathbf{p}$ perturbation theory
to calculate the low-field hole mobility of strained (100) Si$_{1-x}$Ge$_x$ and they obtained a 2.4-6-fold increase for $x=0.1$-$0.2$.
These findings were later confirmed by Fischetti \textit{et al.}~\cite{Fischetti2003}.
These authors attributed the mobility improvement to the increased energy splitting between
the occupied light-hole band and the empty heavy-hole band, which results in a much smaller effective mass~\cite{Nayak1994}.
In 1995 Schenk calculated the mobility of silicon under both low and high electric fields~\cite{Schenk1996}.
In his work the BTE was solved by relying on Kohler's variational method~\cite{Kohler1948}, which was also
subsequently implemented in the device simulator DESSIS$_{\text{ISE}}$~\cite{Mueller1994}.

First-principles calculations of the mobility started appearing in the late 2000s.
In 2007 Dziekan~\textit{et al.}~\cite{Dziekan2007} studied the mobility enhancement of
strained silicon by combining first-principles band structures, the \textit{ab initio} deformation potential method~\cite{Wang2011a}, and the BTE-SERTA.
The following year, Murphy-Armando and Fahy performed similar calculations for a SiGe alloy~\cite{Murphy-Armando2008}, while Yu \textit{et al.} showed that the SERTA is a good approximation for the electron mobility
but not for the hole mobility under strain, due to the large strain-induced suppression of scattering~\cite{Yu2008}.

A complete first-principles calculation of the electron mobility of silicon was reported by Restrepo \textit{et al.} in 2009~\cite{Restrepo2009}.
They relied on DFPT to compute the electron-phonon matrix elements and solved the BTE within the SERTA.
They reported a phonon-limited room-temperature mobility of 1970~\cmVs{}.
In 2013, Rhyner and Luisier~\cite{Rhyner2013} compared on an equal footing the low-field mobility of bulk silicon using the BTE and the NEGF method within a full-band, nearest-neighbor tight-binding model~\cite{Luisier2009}.
Using the BTE, they obtained an electron and a hole mobility at room temperature of 1080~\cmVs{} and 400~\cmVs{}, respectively.
These values underestimate the measured mobilities of 1350-1450~\cmVs{}~\cite{Ludwig1956,Cronemeyer1957,Li1977,Jacoboni1977}
and 445-510~\cmVs{}~\cite{Ludwig1956,Cronemeyer1957,Jacoboni1977,Dorkel1981}, respectively.
On the other hand, the NEGF calculations of Rhyner and Luisier~\cite{Rhyner2013} yielded room-temperature mobilities of 1550~\cmVs{} and 640~\cmVs{} for electrons and holes, respectively, which overestimate the experimental data.
In their work, the discrepancy between BTE and experiment was attributed to the limitations of Fermi's golden rule in the calculation of the scattering rates.
However, as we discuss below, it is more likely that the tight-binding parameterization and the relatively coarse momentum grid ($40^3$ points) employed in this work might be at the origin of the discrepancy.

In 2015 Li~\cite{Li2015} reported a complete \textit{ab initio} calculation of the BTE electron mobility of silicon,
including an extensive convergence study of the scattering rates.
He relied on a linear interpolation of the DFPT electron-phonon matrix elements
from a 16$\times$16$\times$16 $\mathbf{k}$-/$\mathbf{q}$-point grid to 96$\times$96$\times$96-point fine grids.
Li~\cite{Li2015} obtained a room-temperature electron mobility of 1860~\cmVs{} and found that the iterative Boltzmann solution yields similar results as the SERTA.
This finding was rationalized by noting that, in silicon, forward and backward carrier scattering balance each other, so that the collision integral in the BTE due to incoming electrons [first term in the square bracket of Eq.~\eqref{eq:iterboltzmann}] is strongly suppressed; this is precisely the term which is neglected in the SERTA.
Shortly after, Fiorentini and Bonini~\cite{Fiorentini2016} also reported calculations on silicon; in this case the authors interpolated
the electron-phonon matrix element using MLWFs~\cite{Giustino2007}, which allowed them to use ultra-dense 110$\times$110$\times$110-point fine grids.
Fiorentini and Bonini~\cite{Fiorentini2016} also developed an efficient conjugate gradient algorithm to solve the BTE, and obtained an electron mobility of 1750~\cmVs{}.

\begin{figure}[t!]
  \centering
  \includegraphics[width=0.99\linewidth]{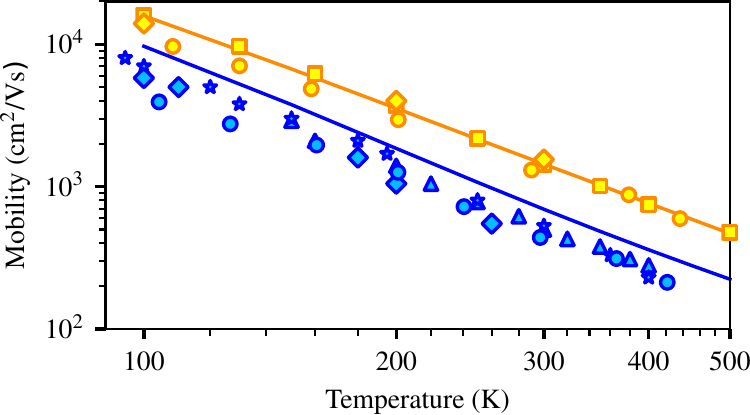}
  \caption{\label{fig:mobsilicon}
  Comparison between calculated and measured intrinsic
  electron and hole mobilities of silicon as a function of temperature.
  The intrinsic material was modeled using carrier concentrations below
  $10^{15}$ cm$^{-3}$.
  The blue line is for holes and the orange line is for electrons.
  Experiments are from~\cite{Morin1954} ($\triangle$), \cite{Logan1960} ($\Diamond$),
  \cite{Ludwig1956} (\ding{73}), \cite{Jacoboni1977} ($\circ$), and \cite{Norton1973} ($\Box$).
  Adapted from Ref.~\cite{Ponce2018}; copyright (2018) by the American Physical Society.}
\end{figure}

In 2018 Ma~\textit{et al.}~\cite{Ma2018} and Ponc\'e \textit{et al.}~\cite{Ponce2018} studied the electron and hole mobility of silicon using MLWFs.
Both teams found that it is important to include SOC in the calculation of the hole mobility,
since the split-off hole is removed from the valence band top and the available scattering channels are reduced; on the other hand it was found that the electron mobility is largely unaffected by SOC effects.
Furthermore, Ponc\'e \textit{et al.}~\cite{Ponce2018} showed that numerical convergence of the Brillouin-zone integrals could be achieved using as little as $10^5$ grid points when using quasi-random Sobol grids~\cite{Zacharias2015}.
In this work the authors also quantified the effect on the calculated mobility of many-body quasiparticle corrections (5\%), many-body corrections to the DFPT electron-phonon matrix elements (14\%), and phonon interaction-induced renormalization of the band gap (5\%)~\cite{Ponce2018}.
The most accurate mobility values reported in this work at room temperature are 1366~\cmVs{} and 658~\cmVs{} for electrons and holes, respectively, and the temperature dependence of the mobility was found to be in good agreement with experiments, see  Fig.~\ref{fig:mobsilicon}.
Finally, these \textit{ab initio} calculations~\cite{Ponce2018,Fischetti2019} revealed that acoustic-phonon scattering in silicon is much more important than previously thought~\cite{Kent1998}.


\subsubsection{Diamond}

Diamond is a superhard material with a wide indirect bandgap of 5.55~eV [Fig.~\ref{fig:bs}(b)],
high thermal conductivity, high breakdown field, and high carrier mobility~\cite{Wort2008,Carrier2004}.
%
Despite the importance of diamond, there are still significant uncertainties about its carrier mobility and dependence on doping, temperature, and magnetic field.
For example, measured room-temperature hole mobilities range from 2000 to 3800~\cmVs{}, and electron mobilities range from 1800 to 4500~\cmVs{}~\cite{Reggiani1981,Isberg2002,Isberg2005,Pernegger2005,Nesladek2008,Jansen2013};
the highest reported electron mobilities~\cite{Isberg2002} have not been confirmed~\cite{Nesladek2008}.
%

There are only a few theoretical investigations of the carrier mobility in diamond.
In 2010, Pernot~\textit{et al.}~\cite{Pernot2010} studied $p$-doped diamond by considering four scattering  mechanisms in a semi-empirical model:
neutral and ionized-impurity scattering, and acoustic and nonpolar optical phonon scattering.
They computed the intrinsic hole mobility and found a value of 1830~\cmVs{}, significantly larger than that of any group-IV semiconductor.
One of the reasons why diamond should outperform other semiconductors at room temperature and above is the high energy of its optical phonons, 165~meV.
Indeed, in the case of silicon, the mobility is partially limited by optical-phonon scattering
at room temperature, whereas this mechanism becomes important only at significantly higher temperatures in the case of diamond.

In 2012, Restrepo and Windl~\cite{Restrepo2012} studied for the first time the electronic spin relaxation rate of diamond from first principles~\cite{Fabian1999,Zutic2004}.
Their study is discussed in more detail in Section~\ref{Sec5.1} on spintronics, but it should be mentioned here that they obtained a very low electron mobility of 130~\cmVs{} at room temperature.
This work marks the first \textit{ab initio} calculation of the intrinsic mobility of diamond.
Shortly after, L\"of\oldr{a}s~\textit{et al.}~\cite{Loefaas2013} studied hole transport in diamond using the BTE-cSERTA and included SOC using the BolzTrap code~\cite{Madsen2006}.
They found that acoustic-phonon scattering is the dominant scattering mechanism at room temperature.
\begin{figure}[t!]
  \centering
  \includegraphics[width=0.99\linewidth]{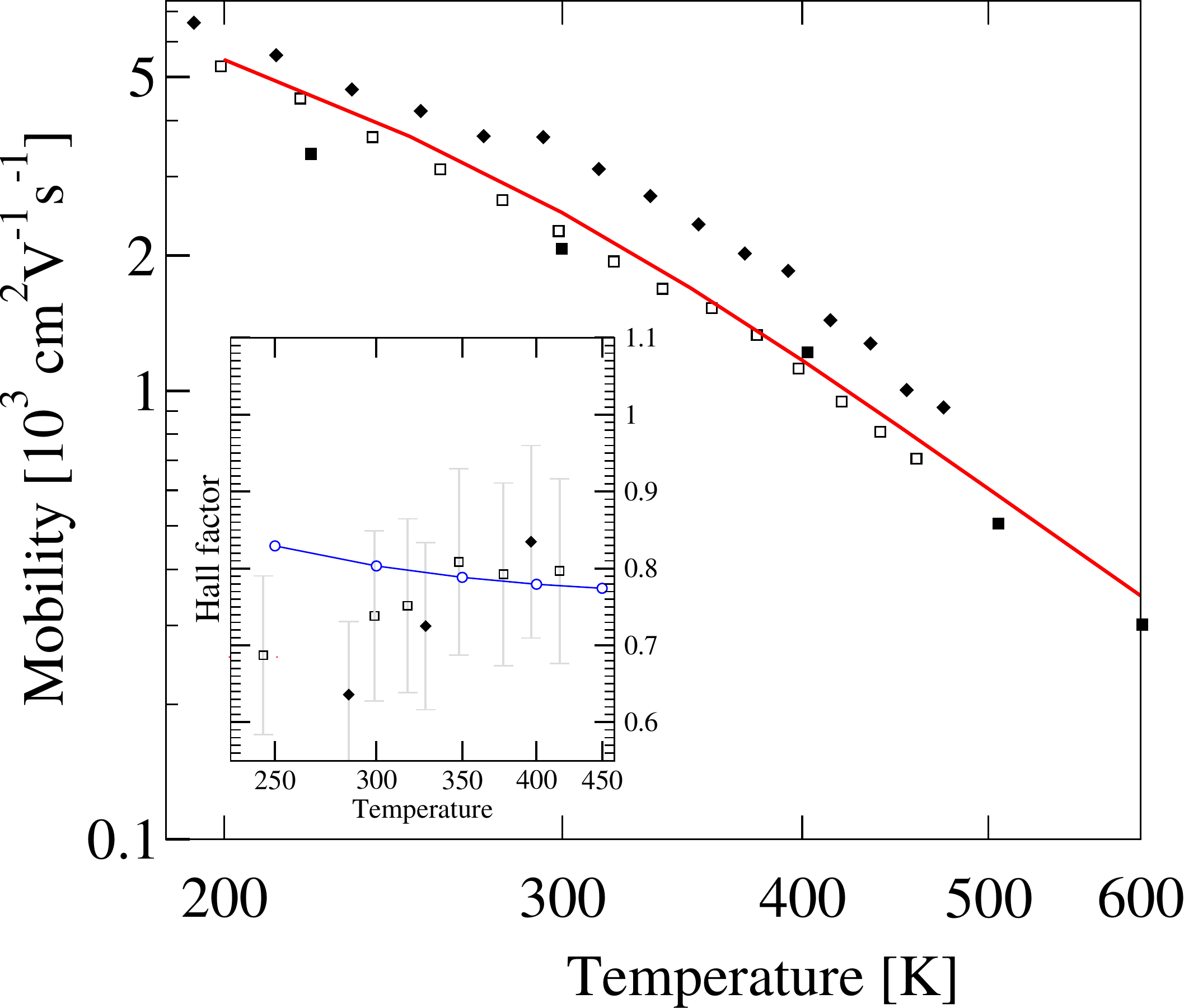}
  \caption{\label{fig:mobdiamond}
  Temperature dependence of the hole drift mobility of $p$-doped diamond
  with a hole density of 10$^{15}$cm$^{-3}$ (red solid line).
  Experiments are from~\cite{Reggiani1981} ($\blacksquare$), \cite{Gabrysch2011} ($\Box$) and
  \cite{Isberg2005} ($\blacklozenge$).
  Inset: The corresponding Hall factor as a function of temperature;
  experimental data are from~\cite{Dean1965} ($\Box$) and~\cite{Konorova1967} ($\blacklozenge$).
  Adapted from Ref.~\cite{Macheda2018}; copyright (2018) by the American Physical Society. }
\end{figure}
More recently, Macheda and Bonini~\cite{Macheda2018} solved the \textit{ab initio} BTE including the effect of a finite magnetic field using Eq.~\eqref{eq:iterboltzmannwithb}, and determined the drift mobility, Hall mobility, and Hall factor.
As shown in Fig.~\ref{fig:mobdiamond}, they obtained a room-temperature hole mobility of 2500~\cmVs{} and a Hall factor $r_{\rm H}=0.81$, using a dense 100$\times$100$\times$100 $\mathbf{k}$-/$\mathbf{q}$-point mesh
interpolated using EPW~\cite{Ponce2016a}.
This calculation does not include SOC.
Therefore slightly lower hole mobilities are expected upon inclusion of SOC.

\subsubsection{Gallium arsenide}\label{sec:gaas}

In the 1960s Ehrenreich~\cite{Ehrenreich1960} was among the first authors to theoretically investigate the transport properties of GaAs.
He found that a combination of ionized-impurity and polar optical-phonon scattering gives qualitative agreement between theory and experiment for high-purity GaAs.
Later Wolfe~\textit{et al.}~\cite{Wolfe1970} refined the model by adding the effect of piezoacoustic scattering, acoustic-deformation potential scattering,
and neutral-impurity scattering.
He obtained a mobility of 240,000~\cmVs{} at 77~K, in good agreement with experiments.

In the case of GaAs, the usual approximation that the drift mobility and the Hall mobility are similar does not hold.
There is still considerable uncertainty in measurements of the Hall factor $r^{\mathrm{H}}$, which has been found to range between 0.8 and 4~\cite{Stillman1970,Lee1983,Look1996,Wenzel1997,Wenzel1998}.
Neumann and Van Nam~\cite{Neumann1978} theoretically investigated the drift and Hall mobilities in GaAs and found that the Hall factor should be in the range $r^{\mathrm{H}}=1.1$-2.5.
They also observed that the Brooks-Herring formula~\cite{Brooks1955} is inadequate for describing ionized-impurity scattering in $p$-doped GaAs.
They postulated that this shortcoming may be due to the existence of two degenerate bands, so that interband scattering should also be taken into account.

In 1994, Scholz~\cite{Scholz1995} studied the hole mobility of GaAs by approximating the Fermi-Dirac distribution with the Maxwell-Boltzmann distribution and obtained a room-temperature mobility of 400~\cmVs{}.
In contrast to earlier studies~\cite{Lee1983}, he unambiguously determined that polar LO-phonon scattering was the dominant mechanism for low-field mobility.
More recently Arabshahi~\cite{Arabshahi2008} studied the electron Hall mobility, using the BTE and considering various models to describe each scattering mechanism.
He obtained a mobility of 8300~\cmVs{} at room temperature and concluded that the mobility was limited by longitudinal optical-phonon scattering at high temperature, while neutral-impurity scattering dominates at low temperature.

In 2016, Zhou and Bernardi~\cite{Zhou2016} studied for the first time the mobility of GaAs within the SERTA using the EPW code~\cite{Ponce2016a}.
They relied on the recently proposed method to analytically obtain the long-range electron-phonon matrix elements of the LO modes~\cite{Sjakste2015,Verdi2015} using Eq.~\eqref{eq:gspliting}.
They computed the short- and long-range part of the matrix elements separately, in order to achieve a much denser sampling of the analytic part, for example, using Brillouin-zone grids of 600$\times$600$\times$600 points.
Using this procedure they succeeded in obtaining the electron mobility of GaAs between 200~K and 500~K from first principles.
They obtained a room-temperature mobility of 8900~\cmVs{}, in good agreement with the experimental value of 7200-9000~\cmVs{}
~\cite{Hicks1969,Rode1970,Stillman1970,Rode1971,Nichols1980,Blakemore1982,Wenzel1998,Sotoodeh2000,Madelung2003,Beaton2010}.
Based on the good agreement with experiment, they concluded that the SERTA should be a reasonable approximation in the case of GaAs.
However, one year later, Liu~\textit{et al.}~\cite{Liu2017} used a similar approach in combination with $GW$ band structures~\cite{Deslippe2012}, $600^3$ $\mathbf{k}$-points and $100^3$ $\mathbf{q}$-points, and the iterative BTE.
They obtained a mobility of 7050~\cmVs{} in the SERTA and a mobility of 8340~\cmVs{} using the iterative BTE.
Therefore the SERTA underestimates the more accurate BTE result by approximately 20\%.
Liu~\textit{et al.}~\cite{Liu2017} also computed the mobility using earlier semi-empirical models and obtained a value of 4930~\cmVs{}.
The underestimation of the mobility in the semi-empirical calculation was attributed to (i) the lack of non-parabolicity of the conduction band and (ii) the lack of intervalley scattering.
These authors also emphasized the importance of piezoacoustic and acoustic-deformation potential scattering, which account for a 15\% reduction of the mobility at 300~K.
They also showed that intervalley scattering plays an important role.
These findings explain why a model based solely on deformation potential scattering or Fr\"ohlich scattering is not accurate enough for GaAs, as both mechanisms are almost equally important.
Altogether, these recent investigations~\cite{Zhou2016,Liu2017}
highlight the necessity of using parameter-free \textit{ab initio} calculations for achieving
an accurate description of carrier transport in GaAs.
\begin{figure}[t!]
  \centering
  \includegraphics[width=0.99\linewidth]{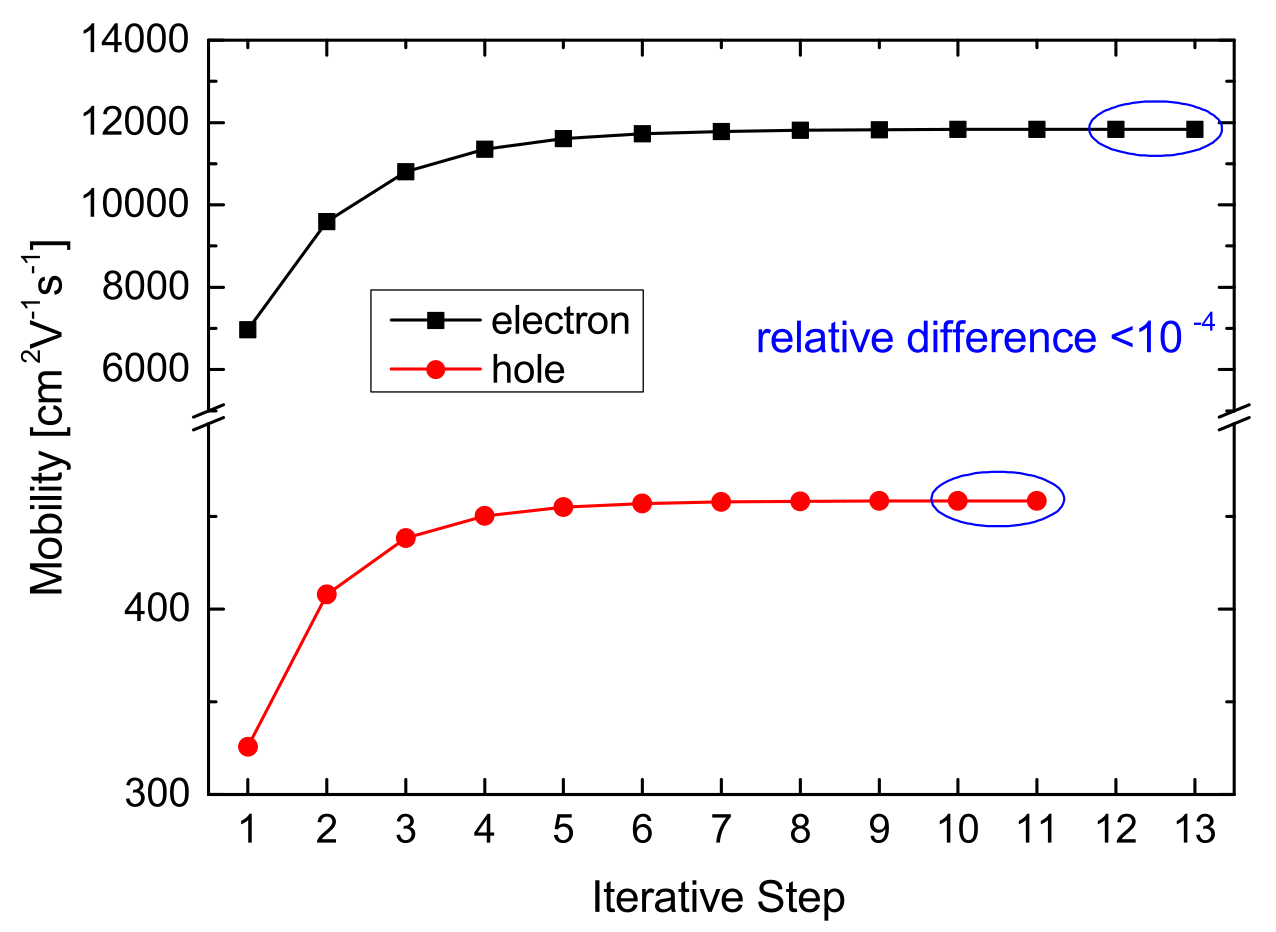}
  \caption{\label{fig:mobgaas}
   Calculated intrinsic electron and hole mobilities of GaAs at room temperature, plotted
   as a function of the number of iterations performed in the solution of the BTE.
   The first iteration corresponds to the SERTA result, the asymptotic value is the BTE mobility.
   Adapted from Ref.~\cite{Ma2018}; copyright (2018) by the American Physical Society.
  }
\end{figure}

Recently, Ma~\textit{et al.}~\cite{Ma2018} clarified the differences between the calculated electron mobilities of Ref.~\cite{Zhou2016} and Ref.~\cite{Liu2017}
and computed the hole mobility including SOC using the EPW code~\cite{Ponce2016a}, see Fig.~\ref{fig:mobgaas}.
By using the same lattice constant and pseudopotentials, they reproduced the SERTA result of Zhou and Bernardi~\cite{Zhou2016}.
When using the same pseudopotentials and $GW$ band structure as in Liu \textit{et al.}~\cite{Liu2017}, they obtained a smaller SERTA result but a similar BTE mobility.
By analyzing the scattering rates, Ma~\textit{et al.}~\cite{Ma2018} concluded that the mobility is very sensitive to the effective mass and the $\Gamma$-$L$ energy gap
and suggested that the discrepancy between the two previous studies is to be ascribed to the different band structures, as opposed to the electron-phonon matrix elements.
They confirmed the findings of Liu~\textit{et al.}~\cite{Liu2017} that the iterative solution of the BTE yields larger mobilities than SERTA, but they found that the difference is even larger than previously thought
(40\% versus 18\%), as is shown in Fig.~\ref{fig:mobgaas}.
Moreover, Ma~\textit{et al.}~\cite{Ma2018} obtained a hole mobility of 459~\cmVs{} at room temperature within the BTE, which is about 30\% larger than the SERTA result.
Their results are in good agreement with experimental values, ranging from 188~\cmVs{} to 460~\cmVs{}~\cite{Hill1970,Mears1971,Kim1991,Wenzel1998,Sotoodeh2000,Beaton2010}.
As in the case of silicon, the hole mobility was found to be strongly affected by SOC: neglecting SOC underestimates the calculated (BTE) hole mobility of GaAs by as much as 50\%.


\subsubsection{Gallium nitride}

\begin{figure*}[th]
  \centering
  \includegraphics[width=0.95\linewidth]{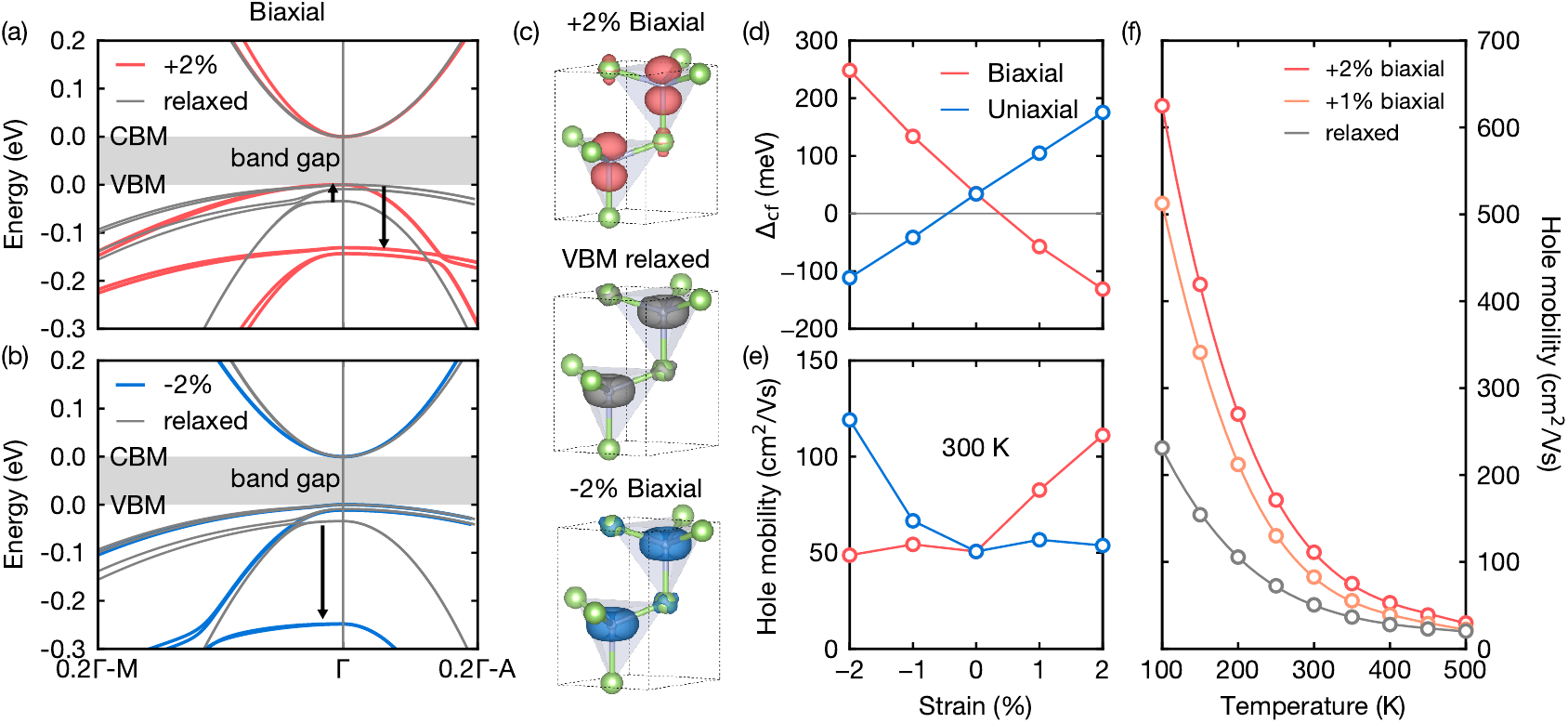}
  \caption{\label{fig:mobgan}
  Crystal field engineering of the band structure and mobility of GaN.
  (a), (b) Change in the GW quasiparticle band structure of GaN upon biaxial dilation and compression, respectively.
  The energy levels are aligned to the conduction and valence band edges.
  (c) Electron wavefunction at the valence band maximum at $\Gamma$ for the undistorted wurtzite GaN structure
  as well as for 2\% biaxial dilation and 2\% biaxial compression, respectively.
  (d) Crystal field splitting $\Delta_{\rm cf}$ versus  strain and (e) corresponding hole Hall mobility at 300~K.
  (f) Predicted   temperature-dependent hole mobility in wurtzite GaN as a function of biaxial strain.
  Adapted from Ref.~\cite{Ponce2019a}; copyright (2018) by the American Physical Society.}
\end{figure*}
The first calculation of the carrier mobility of GaN was performed by Ilegems and Montgomery in 1973~\cite{Ilegems1973},
taking into account the nonparabolicity of the conduction band as well as deformation potential scattering, piezoacoustic scattering, and polar optical-phonon scattering.

More recently, Mnatsakanov \textit{et al.}~\cite{Mnatsakanov2003} developed a simple analytical model
based on experimental results to accurately describe low-field carrier mobilities in a wide temperature and doping range:
\begin{equation}\label{eq93}
  \mu(N,T) = \mu^{\text{max}}(T_0) \frac{B(N)\big[T/T_0 \big]^\beta}{1+B(N)\big[T/T_0 \big]^{\beta+\alpha}},
\end{equation}
where
\begin{equation}\label{eq94}
  B(N) = \frac{\mu^{\text{min}}+\mu^{\text{max}}\big[ N_{\mathrm{g}}/N \big]^{\gamma}}{\mu^{\text{max}}-\mu^{\text{min}}},
\end{equation}
with $T_0$=300~K, $N_{\mathrm{g}}$=2(3)$\times$10$^{17}$~cm$^{-3}$, $\gamma$=1(2),
$\alpha$=2(5) for electrons (holes), and $\beta$=0.7 for electrons (not given for holes due to a lack of experimental data).
The model of Eq.~\eqref{eq93} works well below room temperature and at low fields.
Farahmand~\textit{et al.}~\cite{Farahmand2001} subsequently developed a model based on Monte Carlo results that also includes high-field mobility, but could not reproduce experimental data accurately.

In 2005, Schwierz~\cite{Schwierz2005} proposed an improved model which included recently published mobility data.
This model was shown to describe the temperature dependence of both the low- and high-field mobility above room temperature more accurately.
Shortly after, Arabshahi~\cite{Arabshahi2008} obtained an electron mobility in GaN of 1300~\cmVs{} at room temperature
by solving the BTE iteratively, using models to describe ionized-impurity as well as acoustic-, piezoelectric-, and polar optical-phonon scattering.

Recently, Jhalani~\textit{et al.}~\cite{Jhalani2017} computed the electron-phonon scattering rates of GaN from first principles using EPW~\cite{Ponce2016a}.
They also computed the time-resolved hot carrier relaxation~\cite{Sjakste2018} by solving the time-dependent BTE.
They found a large asymmetry between the hot electron and hot hole dynamics, with the holes relaxing to the band edge in $\sim$80~fs, while the electron cooling required longer times of $\sim$200~fs.

The first \textit{ab initio} calculation of the mobility of wurtzite GaN was reported by Ponc\'e~\textit{et al.} in 2019~\cite{Ponce2019a}.
This calculation included SOC and $GW$ quasiparticle corrections obtained from the Yambo code~\cite{Marini2009}.
The DFPT matrix elements were computed with Quantum ESPRESSO~\cite{Giannozzi2017} and interpolated using EPW~\cite{Ponce2016a}
on dense grids of 100$\times$100$\times$100 $\mathbf{k}$- and $\mathbf{q}$-points.
Using the BTE, they calculated room-temperature electron and hole drift mobilities of 905~\cmVs{} and 44~\cmVs{}, respectively.
They also found that the SERTA strongly underestimates these values, yielding 457~\cmVs{} and 18~\cmVs{} for electrons and holes, respectively.
To compare with Hall mobility experiments, Ponc\'e~\textit{et al.}~\cite{Ponce2019a} computed the Hall factor using the isotropic approximation of Eq.~\eqref{eq65}
and determined Hall mobilities of 1030~\cmVs{} and 50~\cmVs{} for electrons and holes, respectively.
These values are in reasonable agreement with recent measurements yielding 1265~\cmVs{}~\cite{Kyle2014} and 31~\cmVs{}~\cite{Horita2017}, respectively.
The origin of the low hole mobility in GaN (as compared to the electron mobility) was ascribed to a combination of heavy carrier effective masses
and a high density of final electronic states available for hole scattering via low-energy acoustic phonons.
In fact, it was found that acoustic-phonon scattering accounts for approximately 80\% of the total scattering rates for both electrons and holes,
while the remaining contribution stems from long-wavelength polar longitudinal-optical phonons.
Ponc\'e~\textit{et al.}~\cite{Ponce2019a} also predicted that the hole mobility of GaN could be increased by reversing the sign of the crystal field splitting~\cite{Wagner2002,Dreyer2013}, so as to lift the split-off hole states
above the light and heavy holes, as shown in Fig.~\ref{fig:mobgan}(a-d).
This reversal of crystal-field splitting might be achieved by applying a biaxial tensile strain or a uniaxial compressive strain, as shown in Fig.~\ref{fig:mobgan}(e-f).

\subsubsection{Gallium oxide}
The monoclinic $\beta$-phase of gallium oxide ($\beta$-Ga$_2$O$_3$) has been identified as a promising alternative to GaN and SiC for power electronics, due to its wide bandgap
and high breakdown field~\cite{Higashiwaki2018}.
However, since $\beta$-Ga$_2$O$_3$ has a 10-atom primitive cell and a 20-atom conventional cell,
calculations of transport properties in this material are more challenging than for tetrahedral semiconductors.
Two important questions related to the electron mobility of $\beta$-Ga$_2$O$_3$ have been resolved only recently.
The first one is linked to the shape of the conduction band: Ueda~\textit{et al.}~\cite{Ueda1997} measured a strong anisotropy of the conduction-band effective mass.
However, since then many experiments and theoretical studies have shown that the conduction band is nearly isotropic~\cite{Yamaguchi2004,Irmscher2011,Villora2004,Varley2011,Peelaers2015,Wong2016}.
The second question concerns the relative importance of nonpolar optical-phonon, polar optical-phonon, and ionized-impurity scattering at room temperature.
In 2016, Parisini and Fornari~\cite{Parisini2016} performed a detailed theoretical analysis of the drift and Hall mobilities.
Based on empirical fitting of experimental data, they concluded that the dominant scattering mechanism in $\beta$-Ga$_2$O$_3$ is due to nonpolar optical phonons and reported a large deformation potential of 4$\times$10$^9$~eV/cm.
\begin{figure}[t]
  \centering
  \includegraphics[width=0.99\linewidth]{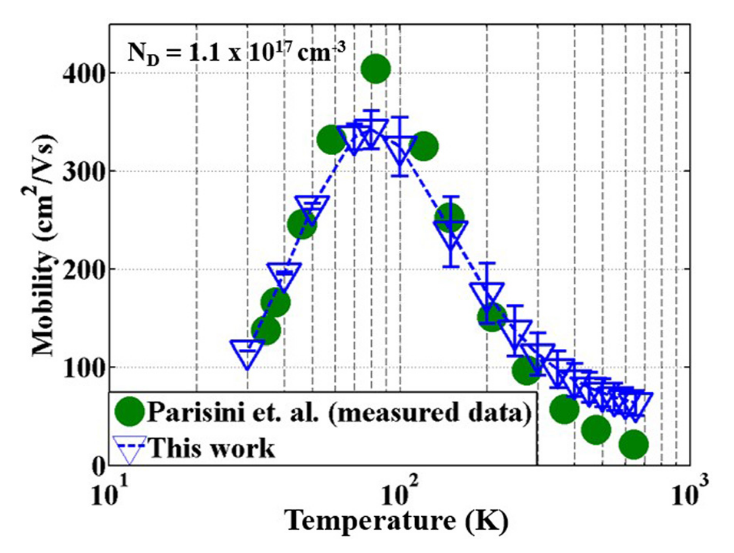}
  \caption{\label{fig:mobgao}
  Calculated electron mobility (triangles) of $\beta$-Ga$_2$O$_3$ compared to the experimental data from Ref.~\cite{Parisini2016} (disks).
  The error bars are determined by assuming 10\% accuracy in the dielectric constant and short-range matrix elements.
  Reproduced from Ref.~\cite{Ghosh2016}; copyright (2016) by AIP Publishing.}
\end{figure}

Later in 2016, Ghosh and Singisetti~\cite{Ghosh2016} performed the first \textit{ab initio} calculation of the electron-phonon coupling and transport properties of $\beta$-Ga$_2$O$_3$.
In their procedure the authors obtained the electron-phonon matrix elements on a dense 40$\times$40$\times$40-point Brillouin zone grid via Wannier interpolation~\cite{Ponce2016a} and then
employed Rode's method~\cite{Rode1975} to iteratively solve the BTE including impurity scattering in the relaxation time approximation.
They obtained a room-temperature mobility of 115~\cmVs{} at a carrier concentration of $10^{17}$~cm$^{-3}$ and a temperature dependence in good agreement with experiment (see Fig.~\ref{fig:mobgao}).
Unlike in Ref.~\cite{Parisini2016}, Ghosh and Singisetti~\cite{Ghosh2016} identified a longitudinal-optical phonon mode with energy around 21~meV as the dominant mechanism in the mobility of $\beta$-Ga$_2$O$_3$.
Shortly after Ma~\textit{et al.}~\cite{Ma2016}, Kang~\textit{et al.}~\cite{Kang2017}, and Mengle and Kioupakis~\cite{Mengle2019} confirmed this finding.

Ma~\textit{et al.}~\cite{Ma2016} used $\mathbf{k}\cdot\mathbf{p}$ perturbation theory to estimate an upper bound of 200~\cmVs{} for the room-temperature electron mobility of $\beta$-Ga$_2$O$_3$
at carrier densities below $10^{18}$~cm$^{-3}$.
They also showed that, despite having an effective mass similar to GaN, the electron mobility of $\beta$-Ga$_2$O$_3$ is almost an order of magnitude smaller due to strong Fr\"ohlich interactions.
Kang~\textit{et al.}~\cite{Kang2017} used a Vogl model in conjunction with Fermi's golden rule to obtain an electron mobility of 155~\cmVs{} at room temperature at a carrier concentration of 10$^{17}$~cm$^{-3}$.
Interestingly, they extracted the deformation potential for nonpolar optical phonons from their first-principles calculations, and obtained a value of  3$\times$10$^8$~eV/cm,
one order of magnitude smaller than Parisini and Fornari~\cite{Parisini2016}.
Mengle and Kioupakis~\cite{Mengle2019} assigned the mobility bottleneck to a polar optical mode around 29~meV, in agreement with the result obtained by Ghosh and Singisetti~\cite{Ghosh2016}.
These calculations agree well with the highest measured room-temperature mobility in bulk $\beta$-Ga$_2$O$_3$, 180~\cmVs{}~\cite{Zhang2018}.
First-principles calculations of hole mobilities are yet to be reported.

\subsubsection{Methylammonium lead triiodide perovskites}

\begin{figure}[t]
  \centering
  \includegraphics[width=0.9\linewidth]{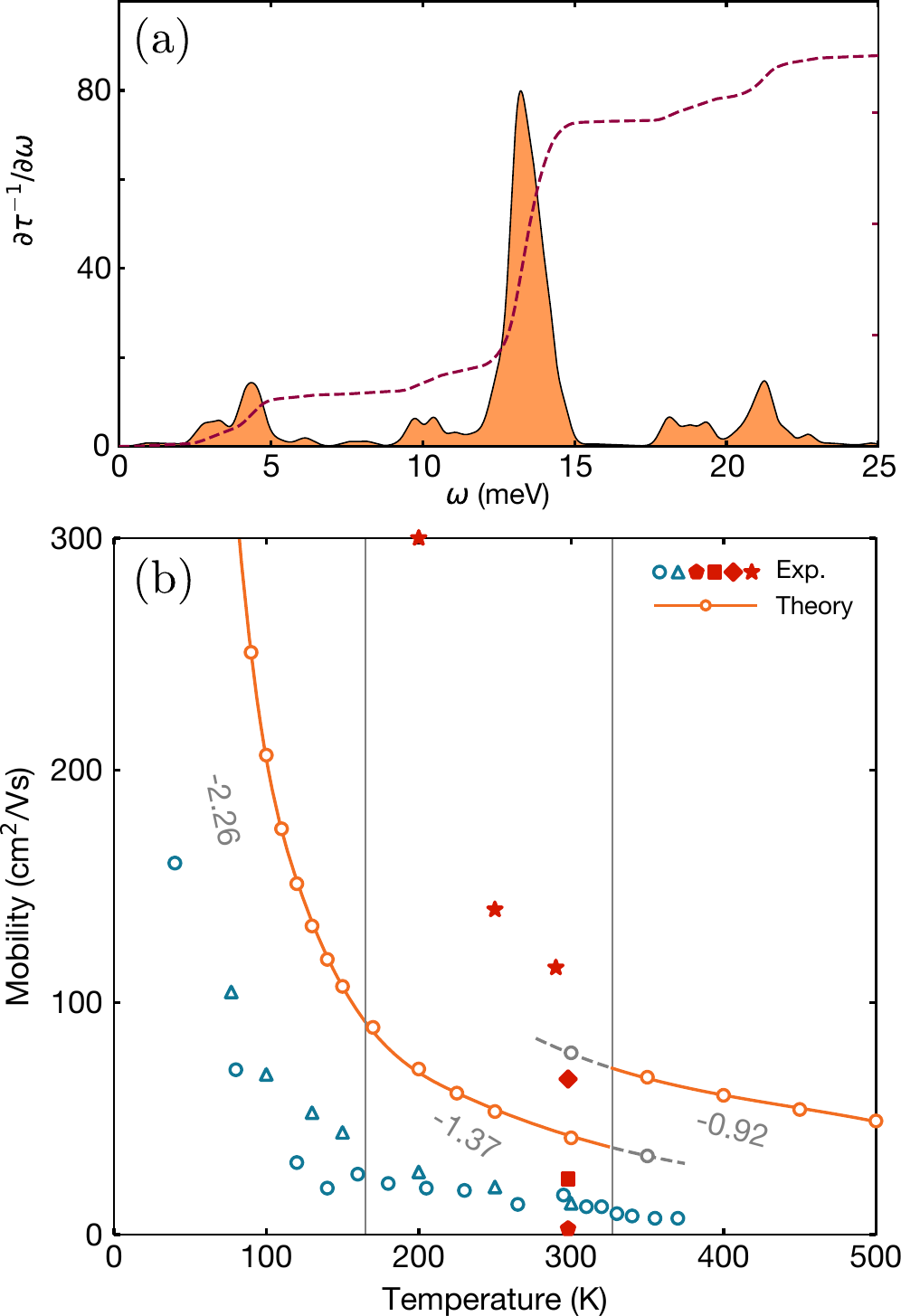}
  \caption{\label{fig:mobmapi}
  (a) Spectral decomposition of the contribution of all phonons with energy $\hbar\omega$
  to the electron scattering rate $\tau^{-1}_{n\mathbf{k}}$ in MAPbI$_3$.
  The height of the peaks indicates the contribution of each phonon and the dashed line indicates
  the cumulative integral of the differential scattering rate.
  Acoustic scattering is negligible in MAPbI$_3$, while the dominant contributions are from vibrations
  corresponding to LO phonons at 4.3~meV, 14.4~meV, and 21~meV.
  (b) Average electron and hole phonon-limited mobilities calculated from the \textit{ab initio} many-body Boltzmann
  equation for MAPbI$_3$ in the $Pnma$ structure (circles and solid, orange lines between 0-350~K)
  and CsPbI$_3$ in the $Pm\overline{3}m$ structure, at the lattice parameter of MAPbI$_3$, between 250-500~K.
  The experimental data are from Ref.~\cite{Milot2015} (circles, optical pump-THz probe) and
  Ref.~\cite{Karakus2015} (triangles, ultra-broadband THz photoconductivity) taken on thin films,
  whereas the filled red symbols are taken on single crystals and are from Ref.~\cite{Shi2015a} (pentagon, space-charge-limited current),
  Ref.~\cite{Dong2015} (square, time of flight), Ref.~\cite{Saidaminov2015} (diamond, dark current-voltage),
  and Ref.~\cite{Shrestha2018} (star, time of flight).
  The vertical bars indicate the boundaries of the orthorhombic, tetragonal, and cubic phases of MAPbI$_3$,
  while the gray numbers indicate the exponent for the temperature dependence of the three phases.
  Adapted from Ref.~\cite{Ponce2019}; copyright (2019) by the American Chemical Society.}
\end{figure}
Organic-inorganic lead halide perovskites~\cite{Kojima2009,Lee2012} attracted considerable attention as promising new materials for photovoltaics and lighting technology~\cite{Green2014,Stranks2015}.
The prototypical compound of this family, methylammonium lead triiodide CH$_3$NH$_3$PbI$_3$ or MAPbI$_3$
(MA = CH$_3$NH$_3$), exhibits three stable phases: orthorhombic $Pnma$ ($T<165$~K), tetragonal $I4/mcm$ (165~K$<T<327$~K),
and cubic $Pm\bar{3}m$ ($T>327$~K).
The electronic band structure of the low-temperature orthorhombic phase is shown in Fig.~\ref{fig:bs}(f).

The first theoretical study of the conductivity and carrier mobility of hybride perovskites was done by Motta~\textit{et al.}~\cite{Motta2015}.
These authors employed the BoltzTrap code~\cite{Madsen2006} and found a room-temperature hole mobility of MAPbBr$_3$ ranging from 5 to 12~\cmVs{} and an electron mobility ranging from 2.5 to 10~\cmVs{} for temperatures spanning the three structural phases.
In these calculations, SOC was not taken into account and the carrier relaxation time was taken to be a constant value of 1~ps from experiments~\cite{Poglitsch1987}.
Motta~\textit{et al.}~\cite{Motta2015} considered two possible orientations of the MA cations in the tetragonal phase and found that the Pb states in the conduction band were strongly affected,
yielding a strong dependence of the electron mobility on the orientation of the cations.

Shorty thereafter, Zhao~\textit{et al.}~\cite{Zhao2016} as well as Lee~\textit{et al.}~\cite{Lee2018} investigated intrinsic and extrinsic charge transport in MAPbI$_3$.
In the intrinsic case, they only considered the coupling to acoustic phonons and obtained a very large deformation potential (5~eV), comparable to that of graphene.
As a result, they obtained an intrinsic mobility of a few thousand \cmVs{}.
By introducing extrinsic effects through charge impurity scattering using the Brooks-Herring formula, they obtained electron and hole mobilities of 101 and 72~\cmVs{}, respectively.

At the same time, Filippetti~\textit{et al.}~\cite{Filippetti2016} showed that polar optical phonons represent the dominant mobility-limiting scattering mechanism at room temperature.
Using the Boltztrap code~\cite{Madsen2006} and the Fr\"ohlich model to evaluate the polar scattering rates, they obtained carrier mobilities of 60 and 40~\cmVs{} for electrons and holes in MAPbI$_3$, respectively.
Shortly after, these findinges were confirmed by Frost~\cite{Frost2017b}, who obtained electron and hole mobilities of 133 and 94~\cmVs{}, respectively, using the more refined Hellwarth model~\cite{Hellwarth1999}.

In 2018, Schlipf~\textit{et al.}~\cite{Schlipf2018} studied the scattering rates and electron lifetimes in the low-temperature orthorhombic phase of MAPbI$_3$ from first principles.
They found that the electron-phonon coupling was dominated by three groups of longitudinal-optical modes, clustering at energies around 4.3~meV, 14.4~meV, and 21~meV, as shown in Fig.~\ref{fig:mobmapi}(a).
These three groups of modes correspond to the Pb-I-Pb bending, Pb-I stretching, and libration-translation of the CH$_3$NH$_3$ cation, respectively.
These findings are supported by photoluminescence measurements by Fu~\textit{et al.}~\cite{Fu2018}, who
observed three phonon replicas in the low-temperature photoluminescence spectra of the closely-related compound CH(NH$_2$)$_2$PbI$_3$ nanocrystals, red-shifted by 3-4~meV, 10-12~meV, and 14-16~meV with respect to the zero-phonon peak.

More recently, Ponc\'e~\textit{et al.}~\cite{Ponce2019} reported the first \textit{ab initio} study of the intrinsic mobility of MAPbI$_3$.
They computed a phonon-limited average electron and hole mobility of 80~\cmVs{} at room temperature.
These authors also obtained a temperature dependence in good agreement with the experimental values determined from pump-probe spectroscopy~\cite{Herz2018}, as shown in Fig.~\ref{fig:mobmapi}(b).
The same figure also shows that the temperature-dependence of the mobility is relatively complex, with power laws changing from $T^{-2.26}$ at low temperature to
$T^{-0.92}$ above room temperature; these changes reflect the relative contributions of the three important groups of phonons at different temperatures.
Further work will be required to include anharmonic effects in the study of carrier mobilities in hybrid perovskites~\cite{Whalley2016}.

\subsection{Two-dimensional materials}\label{Sec4.2}

Two-dimensional (2D) materials~\cite{Novoselov2005,Novoselov2016}, were discovered to possess a wide range of unique properties not found in their bulk counterparts, for example the existence of Dirac fermions in graphene~\cite{CastroNeto2009} and a strong spin-valley coupling in monolayer molybdenum disulfide (MoS$_2$)~\cite{Xu2014}.
Owning to their extraordinary mechanical strength and flexibility~\cite{Lee2008,Bertolazzi2011}, 2D materials attracted considerable attention for potential applications in the next generation of
flexible and energy-efficient electronic and optoelectronic devices~\cite{Akinwande2014,Chhowalla2016}.
However, in order to be competitive with silicon for conventional applications, the carrier mobility of 2D materials needs to be sufficiently high
and should be comparable at least to that of commercially available silicon-based devices, $\sim$500~\cmVs{}~\cite{Schwierz2010}.
So far, the main candidate materials proposed for 2D electronic or optoelectronic applications include graphene, silicene, phosphorene, MoS$_2$, and InSe.
In this section, we present an overview of the key \textit{ab initio} studies that have contributed to improving our understanding of the intrinsic carrier mobility
and charge transport properties of these 2D materials.

\subsubsection{Graphene}

Graphene is a single layer of $sp^2$-bonded carbon atoms arranged in a honeycomb lattice, as shown in Fig.~\ref{fig:structures}(g).
The $p_z$ valence and conduction bands touch each other at the six corners ($K$ and $K'$) of the hexagonal first Brillouin zone, with a linear energy-momentum dispersion around the Fermi level~\cite{CastroNeto2009}.
Graphene is therefore a semimetal with zero bandgap.

The transport properties of graphene have been studied extensively~\cite{DasSarma2011}.
However, early theoretical studies of the intrinsic carrier transport of graphene were mainly based on
simplified assumptions about the electron-phonon interaction in this system, such as the use of deformation potential models~\cite{Hwang2008}.
Despite such simplifications, the smooth crossover from the high-temperature $\rho(T) \sim T$ to the low-temperature $\rho(T) \sim T^4$ dependence were predicted~\cite{Hwang2008}
and subsequently verified experimentally~\cite{Efetov2010}.
Later theoretical works addressed the detailed carrier scattering mechanisms in graphene using partial or fully \textit{ab initio} approaches~\cite{Borysenko2010,Kaasbjerg2012b,Park2014,Sohier2014,Gunst2016}.
Thanks to these studies~\cite{Chen2008,Efetov2010}, we now have a detailed understanding of the temperature and doping dependence of the resistivity of graphene and of the key processes that limit its carrier mobility.

Hwang and Das Sarma~\cite{Hwang2008} presented the first calculation of the intrinsic phonon-limited carrier mobility of graphene as a function of temperature and carrier density,
using the BTE and the acoustic-deformation potential approximation.
A room-temperature mobility exceeding 10$^5$~\cmVs{} was predicted for carrier densities below 10$^{12}$~cm$^{-2}$.
In Hwang and Das Sarma's work~\cite{Hwang2008}, only scattering from longitudinal acoustic (LA) phonons was considered.
The complex dependence of the EPI matrix elements on the electron and phonon wavevectors was condensed into a single effective deformation potential and used as a fitting parameter.
This work was important for its prediction of the temperature and density dependence of the intrinsic resistivity, which was subsequently verified experimentally~\cite{Efetov2010}.

Shishir and Ferry~\cite{Shishir2009} later calculated the intrinsic mobility of graphene by solving the linearized BTE using Rode's iterative method~\cite{Rode1975}.
In addition to considering carrier scattering by acoustic phonons using the model of Hwang and Das Sarma~\cite{Hwang2008}, these authors
also included carrier scattering arising from the optical phonons at the $K$ and $K'$ points, which is responsible for the intervalley scattering between the two valleys $K$ and $K'$ in the conduction band of graphene.
The carrier scattering rates from both acoustic and optical phonons were treated using deformation potential models.
The deformation potential parameters were obtained by fitting the calculated mobility curve to the experimental mobility data by Chen \textit{et al.}~\cite{Chen2008}.
The largest carrier mobility obtained by Shishir and Ferry exceeded 4$\times$10$^5$~\cmVs{}.

Borysenko \textit{et al.}~\cite{Borysenko2010} performed the first \textit{ab initio} study of the intrinsic carrier mobility of graphene that treated the electron-phonon coupling strength
using parameter-free DFT and DFPT calculations.
They were able to obtain the carrier scattering rate due to different phonon modes using Fermi's golden rule.
In contrast to earlier models of carrier scattering based on acoustic deformation potentials, Borysenko~\textit{et al.}~\cite{Borysenko2010} found that all in-plane phonons
play an important role in carrier scattering at room temperature and thus need to be considered simultaneously.
They further calculated the intrinsic resistivity using full-band Monte Carlo simulations~\cite{Jacoboni1983}.
The calculations yielded a low-field carrier mobility of approximately 5$\times$10$^6$~\cmVs{} at 50~K and a room-temperature mobility approaching 10$^6$~\cmVs{}.

Kaasbjerg~\textit{et al.}~\cite{Kaasbjerg2012b,Kaasbjerg2012a} later calculated the EPI matrix elements for acoustic phonons in graphene, using first-principles calculations and the finite-displacement method.
By analyzing the EPI matrix elements using the group theoretical analysis of Ma\~nes \cite{Manes2007},
these authors derived analytical forms for the acoustic EPI in the long-wavelength limit, for both the LA and TA phonons.
The associated model parameters were obtained by fitting to first-principles results.
They calculated the intrinsic acoustic phonon-limited mobility of graphene as a function of temperature and carrier density using the BTE approach of Hwang and Das Sarma~\cite{Hwang2008,Hwang2009}.
They considered temperatures of up to 200~K, as this is the range where their assumption of acoustic phonon-dominated intravelley carrier scattering is considered to be valid.
Indeed, the aforementioned calculations by Borysenko~\textit{et al.}~\cite{Borysenko2010} indicated that both optical phonons and intervalley scattering by the TA and LA phonons start
to dominate the carrier scattering rate at temperatures above 200~K.
Kaasbjerg~\textit{et al.}~\cite{Kaasbjerg2012a} found that in the temperature range considered, carrier relaxation is dominated by TA phonons,
in contrast to the assumption in Hwang and Das Sarma's pioneering work~\cite{Hwang2008}, where coupling to LA phonons was considered to be the principal carrier relaxation mechanism.
They further demonstrated that the inclusion of the complete electron-acoustic phonon matrix elements has qualitative effects on the scaling of the resistivity with temperature in the low-temperature regime, as in this range the backscattering of carriers at the Fermi surface is frozen out.
Kaasbjerg~\textit{et al.}~\cite{Kaasbjerg2012a} found that the EPI causes the temperature dependence of the mobility to become stronger than the $\mu \sim T^{-4}$ scaling law
even when carrier screening is not considered~\cite{Min2011}.

\begin{figure}[t]
\begin{centering}
\includegraphics[width=0.99\linewidth]{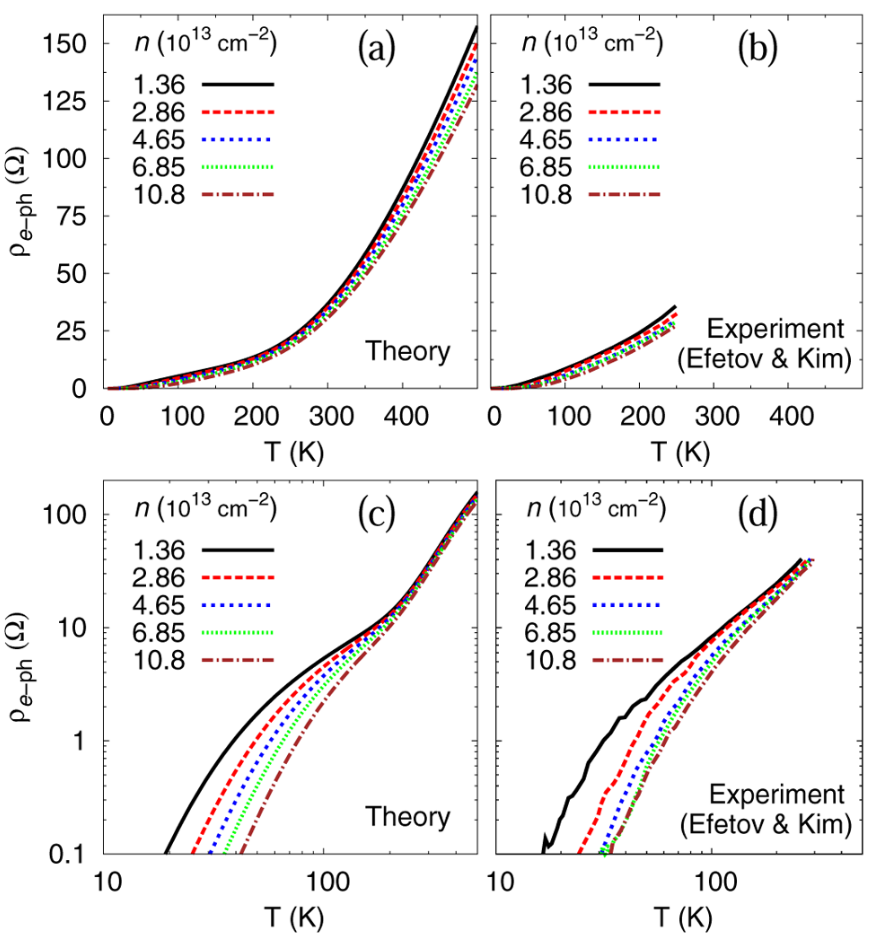}
\caption{(a, c) Calculated intrinsic electrical resistivity of $n$-doped graphene
  as a function of temperature, for different carrier densities. The calculations
  were performed
  using analytical models for the electron-phonon interaction
  and Allen's variational solution~\cite{Allen1978} to the Boltzmann transport equation.
  Effects of the electron-electron interaction, such as the renormalization of the carrier velocity
  and the electron-phonon matrix elements, were incorporated into the analytical models.
  (b,d) The measured temperature dependence of the resistivity of graphene
  at the same carrier densities and for temperatures up to 250~K~\cite{Efetov2010}.
  Reprinted from from~\cite{Park2014}; copyright (2014) by the American Chemical Society.}
\label{fig:mobgraphene}
\end{centering}
\end{figure}

Park~\textit{et al.}~\cite{Park2014} subsequently calculated the intrinsic resistivity of graphene as a function of temperature and carrier density using DFT, DFPT, and the BTE approach.
In their work, the intrinsic resistivity of graphene was computed using the LOVA due to Allen~\cite{Allen1978}.
This approximation was discussed in Sec.~\ref{Sec2.3} and the resistivity is given by Eq.~\eqref{eq:Ziman}.
The electron-phonon matrix elements $g_{mn\nu}(\mathbf{k},\mathbf{q})$ entering Eq.~\eqref{eq:alphasquaref} were obtained at the level of DFPT and interpolated using EPW~\cite{Giustino2007}.
They also focused on the intrinsic resistivity of $n$-doped graphene and analyzed the contribution of different phonon branches to the resistivity.
They found that for temperatures below 200~K, the resistivity is dominated by scattering with acoustic phonons~\cite{Kaasbjerg2012b},
with the contribution from TA phonons being 2.5 times higher than that of LA phonons.
When the temperature is above 200~K, the contribution from high-energy optical phonons from the zone boundary
increases significantly and becomes the dominant mechanism.
Accordingly, the slope of the calculated resistivity curve exhibits a significant increase around 200~K, as shown in Fig.~\ref{fig:mobgraphene}.
Park~\textit{et al.}~\cite{Park2014} further developed an effective tight-binding model of the EPI in graphene.
The model parameters entering the EPI were obtained by calculating the derivative of the nearest-neighbor hopping parameter with respect to the bond length using DFT.
Accounting for the renormalization of the carrier velocity and phonon frequencies, they studied the intrinsic resistivity of graphene within the $GW$ quasiparticle approximation
~\cite{Aryasetiawan1998}, as shown in Fig.~\ref{fig:mobgraphene}.
The effects of electron-electron interactions were found to be more important for optical zone-boundary phonons.
As a result, the relative contribution of the latter to the resistivity is significantly increased, accounting for around 50\% of the total resistivity even at room temperature.
Still, the calculated theoretical resistivity values were 35\% lower than the experimentally data.

The same authors subsequently refined their calculation by considering the EPI at the $GW$ level for both acoustic and optical phonons
and calculated the resistivity through a complete numerical solution of the BTE instead of using Allen's variational solution.
Specifically, they analyzed analytical expressions for the electron-phonon matrix elements~\cite{Manes2007} and their corrections at the $GW$ level~\cite{Lazzeri2008,Attaccalite2010}.

Based on their analysis, these authors concluded that the contribution of acoustic phonons to the resistivity mainly comes from the so-called gauge field~\cite{Vozmediano2010},
which is equivalent to a fictitious strain field that shifts the Dirac point in the first Brillouin zone without changing its energy.
The gauge field contribution was found to be essentially independent of doping and screening.
However, the contribution from the acoustic deformation potential, which is equivalent to a fictitious scalar potential that shifts the energy of the Dirac point
without changing its position in the first Brillouin zone, was negligible and strongly screened.
Overall, the contribution of acoustic phonons to the resistivity was found to be independent of doping and the dielectric environment, in agreement with experimental observations.

By comparing the resistivity from the complete numerical solution of the BTE with their previous calculation~\cite{Park2014}, the authors found that the use of the LOVA
 overestimated the resistivity of graphene, in particular at high temperature.
When compared to the experimental results in the low-temperature regime by Chen~\textit{et al.}~\cite{Chen2008} and Efetov~\textit{et al.}~\cite{Efetov2010},
they found that their calculations underestimated the experimental resistivity by 30\%.
This can be corrected if the acoustic gauge field parameter, computed at the $GW$ level, is increased by 15\%.
In the high-temperature regime, the strong EPI involving the intervalley TO phonons accounts for the strong increase of the resistivity at around 270 K.
However, the underestimation of the high-temperature resistivity by first-principles calculations is even more significant at low temperatures.
This discrepancy had previously been attributed to remote phonon scattering from the substrate in experimental measurements~\cite{Chen2008,Efetov2010}.
At variance with this interpretation, the authors argued that the disagreement could also be explained by the doping-dependent renormalization of the EPI.
However, the required renormalization is much stronger than what was estimated at the $GW$ level~\cite{Attaccalite2010}.

Subsequent studies of carrier transport in graphene focused on the development of first-principles numerical schemes that solve the BTE
without resorting to semi-analytical treatments of the EPI.
Along this line, Restrepo~\textit{et~al.}~\cite{Restrepo2014} calculated the carrier mobility of graphene by solving the BTE in the SERTA and obtained an intrinsic room-temperature mobility of 2$\times$10$^5$~\cmVs{}.
Gunst~\textit{et al.}~\cite{Gunst2016} calculated the phonon-limited mobility of $n$-doped graphene by numerically solving the BTE using the MRTA.
The EPI was computed using a finite-differences supercell method and exhibited a strong dependence on carrier density.
For carrier densities close to typical values in experimental measurements ($\sim$10$^{13}$~cm$^{-2}$), the calculated mobility values ranged from 10$^5$ to 10$^6$~\cmVs{}.
These values are consistent with the results obtained in earlier works~\cite{Hwang2008,Borysenko2010}.

It is noteworthy that Restrepo~\textit{et al.}~\cite{Restrepo2014} also developed a quantum-mechanical
and parameter-free approach to calculate the mobility of graphene as a function of impurity concentration.
Specifically, the Coulomb scattering rate arising from impurities or defects was expressed as
\begin{multline}
  \frac{1}{\tau_{n\mathbf{k}}} =  n_{\mathrm{d}} A_{\mathrm{uc}} \frac{2\pi}{\hbar}
  \sum_m \int \frac{\mathrm{d}^2 q}{\Omega_{\mathrm{BZ}}} \, |T_{mn}(\mathbf{k},\mathbf{q})|^2 \\
  \times \left(1 - \frac{\mathbf{v}_{n\mathbf{k}} \cdot \mathbf{v}_{m\mathbf{k}+\mathbf{q}}}{|\mathbf{v}_{n\mathbf{k}}||\mathbf{v}_{m\mathbf{k}+\mathbf{q}}|}\right)
  \delta(\varepsilon_{n\mathbf{k}} - \varepsilon_{m\mathbf{k}+\mathbf{q}}),
\end{multline}
where $n_{\mathrm{d}}$ is the defect area density, $A_{\mathrm{uc}}$ and $\Omega_{\mathrm{BZ}}$ are the areas of the crystalline unit cell and of the first Brillouin zone, respectively,
$\varepsilon_{n\mathbf{k}}$ and $\varepsilon_{m\mathbf{k}+\mathbf{q}}$ are the band energies, and $\mathbf{v}_{n\mathbf{k}}$ and $\mathbf{v}_{m\mathbf{k}+\mathbf{q}}$ denote the corresponding band velocities.
The scattering matrix $T_{mn}(\mathbf{k},\mathbf{q})$ was treated within the Born approximation as $T_{mn}(\mathbf{k},\mathbf{q}) = \langle m\mathbf{k}+\mathbf{q} | \partial \hat{V} | n\mathbf{k} \rangle$,
where $\partial \hat{V}$ is the self-consistent scattering potential that accounts for the difference between the potential of an unperturbed system and the potential of the system in the presence of defects and impurities.
Restrepo~\textit{et al.}~\cite{Restrepo2014} found that the impurity-limited graphene mobility at a defect density of 10$^{12}$~cm$^{-2}$ ranges from 950~\cmVs{} for Au adatoms to 34,300~\cmVs{} for hydrogen adatoms.
The large difference in mobility originates from the different magnitude of the scattering potentials for the different adatoms.

\subsubsection{Silicene}

\begin{figure}[t]
\begin{centering}
\includegraphics[width=0.99\linewidth]{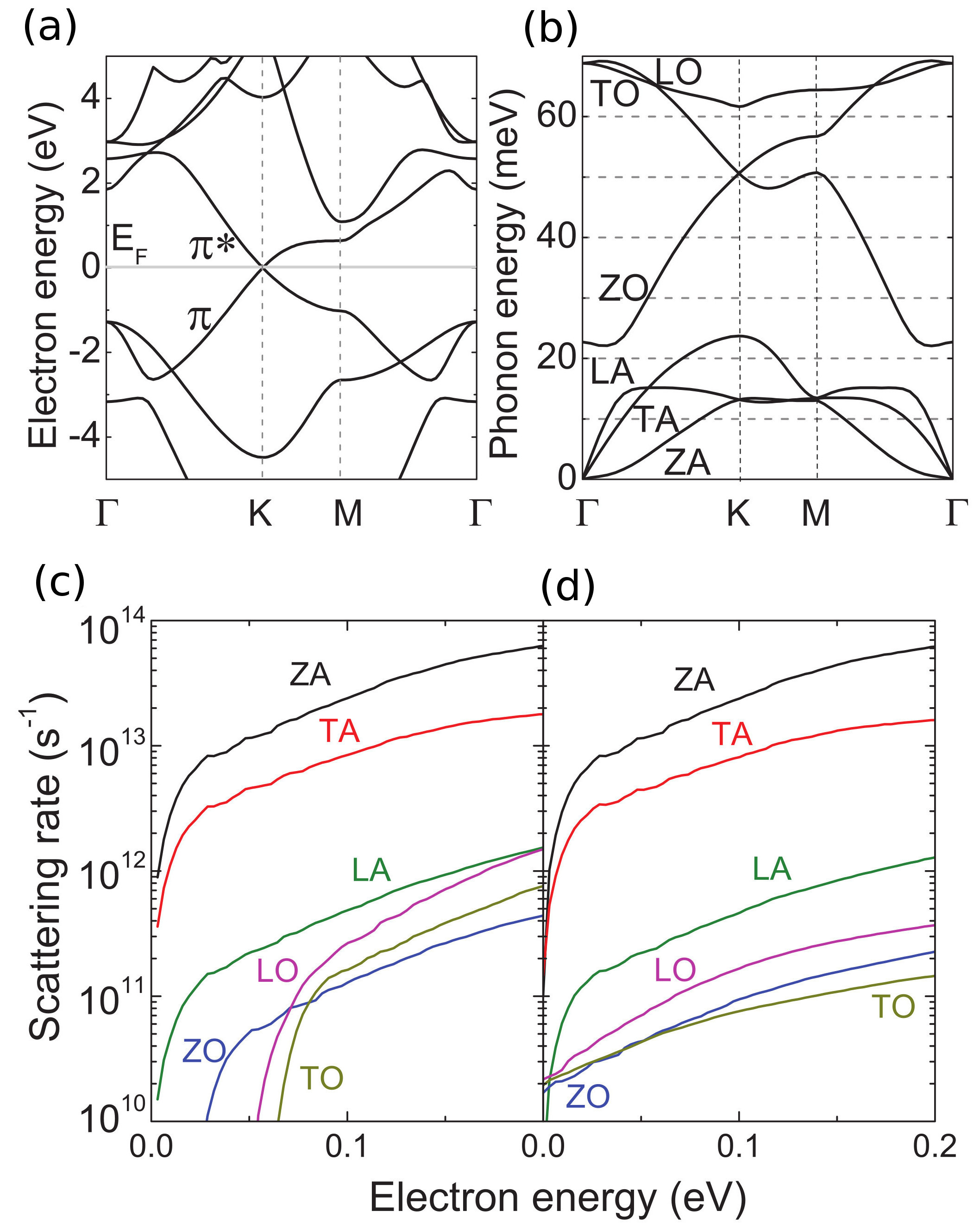}
\caption{The electron (a) and phonon (b) band structures of monolayer
  silicene, as obtained from the generalized gradient approximation to DFT, from Ref.~\cite{Li2013}.
  (c,d) Electron scattering rates in silicene via emission (c) and absorption (d) of phonons
  at room temperature, calculated using \textit{ab initio} electron-phonon matrix elements and Fermi's golden rule.
  In the calculations of the scattering rate, the electron wavevector is assumed to be along the $K$-$\Gamma$ line.
  Adapted from Ref.~\cite{Li2013}; copyright (2013) by the American Physical Society.}
\label{fig:mobsilicene}
\end{centering}
\end{figure}

Silicene is the silicon-based analog of graphene.
Unlike graphene, it exhibits a buckled structure~\cite{Molle2017}, as shown in Fig.~\ref{fig:structures}(h).
The buckled geometry of silicene originates from the weaker $\pi$-bonds as compared to graphene, resulting in a larger bond length.
Therefore, the $p_z$ orbitals have decreased overlap, which renders the planar $sp^2$ hybridization less energetically favorable~\cite{Sahin2009}.
%
%
Despite the buckling, the electronic band structure of silicene still shares many similarities with that of graphene.
In particular, when SOC is not included, silicene is also a zero-gap semimetal with a Dirac-like electronic dispersion
around the Fermi level~\cite{Cahangirov2017}, as shown in Fig.~\ref{fig:mobsilicene}(a).
One of the most appealing aspects of silicene is that, being made of silicon, it should be compatible with existing silicon electronics.
However, silicene is not stable in isolation and requires a supporting substrate.
It has been demonstrated experimentally that silicene can be grown on a number of different substrates using a bottom-up epitaxy method~\cite{Vogt2012,Feng2012,Fleurence12,Chen2012a}.
A single-layer silicene transistor with a mobility of 100~\cmVs{} has been fabricated by Tao~\textit{et al.}~\cite{Tao2015}.

Only a few~\textit{ab initio} studies of the intrinsic transport properties of silicene exist in the literature.
Shao~\textit{et al.}~\cite{Shao2013} calculated the intrinsic carrier mobility of silicene based on the deformation potential approximation.
In their calculation, the deformation potential was determined by calculating the shift of the Fermi level as a function of the lattice constant.
This approach effectively assumes isotropic and intravalley scattering by LA phonons only.
Carrier scattering from intervalley momentum transfer and from other phonon modes were not considered, resulting in a large electron and hole mobility, on the order of 2$\times$10$^5$~\cmVs{}.

Subsequently, Li~\textit{et al.}~\cite{Li2013} calculated the EPI in silicene for all phonon branches using DFPT and employed full-band Monte Carlo simulations~\cite{Jacoboni1983}
to calculate the electric-field dependence of the carrier drift velocity, from which the intrinsic carrier mobility was extracted.
While the coupling of electrons to optical phonons was indeed found to be relatively weak in silicene,
Li~\textit{et al.}~\cite{Li2013} observed that carrier scattering from TA phonons is an order of magnitude higher than from LA phonons.
More importantly, these authors found that the dominant contribution to carrier scattering comes from the out-of-plane acoustic (ZA) phonons, namely the flexural modes [Fig.~\ref{fig:mobsilicene}(c)].
In graphene and other 2D materials with in-plane mirror symmetry, only the in-plane phonons can couple to the charge carriers to first order in the atomic displacements~\cite{Manes2007}.
In silicene, however, the in-plane mirror symmetry is broken and the ZA phonons can couple to the charge carriers already at first order in perturbation theory.
Li~\textit{et al.}~\cite{Li2013} attributed the large carrier scattering rate from ZA phonons to the large values of the calculated EPI and the small phonon energy near the Brillouin zone center.
They found the intrinsic carrier mobility of silicene to be around 1,200~\cmVs{} for free-standing silicene.
When the ZA-phonon contribution is not included, the carrier mobility tripled to 3,900~\cmVs{}.
They argued that since the ZA phonons play the dominant role in the EPI of silicene,
suppressing the ZA phonons by controlling the interaction between silicene and the substrate will be crucial for improving the transport characteristics of silicene-based electronic devices.

It should be noted that in the calculations of Li~\textit{et al.}~\cite{Li2013}, the ZA phonons were regularized to have a frequency versus momentum dispersion relation which is linear in $q$ near the zone center.
The reduction of the carrier mobility is even more severe when the actual $q^2$ dispersion of the ZA phonons is taken into account, as pointed out by Gunst~\textit{et al.}~\cite{Gunst2016}.
The latter authors calculated the carrier mobility of silicene using the full-band MRTA for the BTE, where the EPI matrix elements entering
the calculation of the carrier relaxation time were obtained using a DFT-based supercell method.
Gunst~\textit{et al.}~\cite{Gunst2016} found that the carrier scattering rate originating from the ZA mode is two to three orders of magnitude higher than that of the remaining modes.
In fact, if no long-wavelength cutoff is enforced, the carrier scattering rate arising from the ZA phonons diverges, leading to a vanishing mobility in silicene.
The divergent contribution of the ZA phonons to carrier scattering was shown by Fischetti~\textit{et al.}~\cite{Fischetti2016}
to be a general feature of ideal, freestanding, and infinite 2D crystals lacking horizontal mirror symmetry and results from the divergence of the thermal population of long-wavelength ZA phonons,
in the same spirit as the Mermin-Wagner theorem~\cite{Mermin1966}.
They also showed that the effect of ZA phonons on a 2D crystal lacking horizontal mirror symmetry is particularly severe when the material also exhibits a Dirac-like electron dispersion
at the symmetry point $K$, owing to an increased strength of electron-ZA phonon coupling originating from the band degeneracy at $K$.

In the hypothetical scenario where the coupling to the ZA phonons could be suppressed completely, Gunst~\textit{et al.}~\cite{Gunst2016} found that the mobility of silicene at 300~K and a carrier density of 3$\times$10$^{12}$~cm$^{-2}$
should be approximately 2,100~\cmVs{} when intervalley carrier scattering between the $K$ and $K'$ valleys is not considered.
Including intervalley scattering can decrease the carrier mobility by as much as an order of magnitude, bringing the calculated value of the carrier mobility close to the experimental value of Tao~\textit{et al.} ($\sim$100~\cmVs{})~\cite{Tao2015}.

\subsubsection{Phosphorene}\label{phosphorene}

Phosphorene is the name given to a single layer of black phosphorus [Fig.~\ref{fig:structures}(i)].
Phosphorene has a puckered structure that endows the system with anisotropic mechanical, electronic, optical, and transport properties~\cite{Carvalho2016}, and has a band gap
in the range 0.3-1.8~eV~\cite{Li2016}.
Experimentally, the carrier mobility of multilayer black phosphorus was found to be anisotropic and thickness-dependent~\cite{Li2014,Xia2014,Liu2014,Cao2015,Doganov2015},
with the general trend being that the mobility sharply decreases with decreasing thickness~\cite{Li2014,Cao2015}.
The highest measured room-temperature mobility for hole carriers in multilayer phosphorene of thickness $\sim$10~nm is around 1,000~\cmVs{}~\cite{Li2014},
which is close to the bulk mobility value~\cite{Akahama1983,Morita1986}.
Measurements of carrier mobilities in the monolayer limit are scarce.
Only Cao~\textit{et al.}~\cite{Cao2015} reported room-temperature hole mobilities of 1, 80, and 1,200~\cmVs{} for monolayer, bilayer, and trilayer phosphorene, respectively.

On the theoretical side, there have been several studies aimed at calculating the intrinsic carrier mobility of monolayer
and few-layer phosphorene~\cite{Qiao2014,Liao2015b,Rudenko2016,Trushkov2017,Jin2016,Gaddemane2018,Sohier2018}.
However, the calculated mobility values exhibit significant variations among different groups.
Qiao~\textit{et al.}~\cite{Qiao2014} carried out the first calculations of the phonon-limited carrier mobility in phosphorene multilayers using the so-called Takagi formula~\cite{Takagi1994a,Takagi1994b}:
\begin{equation}\label{eq:Takagi}
  \mu_i = \frac{e \hbar^3 C_i}{k_{\mathrm{B}} T m_i^* m_{\mathrm{d}}^* D_i^2},
\end{equation}
where $i$ refers to the transport direction, $m_i^*$ is the carrier effective mass in the corresponding direction,
and $m_{\mathrm{d}}^* = \sqrt{m_x^* m_y^*}$ is the density-of-state effective mass for an anisotropic electronic band.
The 2D elastic modulus along the transport direction $C_i$ is calculated using $\Delta u = \frac{1}{2}C_i \varepsilon_{ii}^2$, where $\Delta u$ is the change of the total energy
per in-plane area of the 2D crystal in response to the elastic strain $\varepsilon_{ii}$.
The deformation potential constant $D_i$ in Eq.~\eqref{eq:Takagi} is defined as
\begin{equation}
  D_i = a_{0,i}\frac{\Delta E_{c,v}}{\Delta a_i},
\end{equation}
where $\Delta E_{c,v}$ is the energy shift of the band edges under the relative change of lattice constant $\Delta a_i/a_{0,i}$ along the transport direction $i$.
The underlying assumptions in the Takagi formula are~\cite{Takagi1994b,Gaddemane2018}:
(i) if subbands are formed in multilayers, it is assumed that the charge carriers only occupy the lowest subband;
(ii) the carrier relaxation is dominated by intravalley acoustic-phonon scattering from only one in-plane mode;
and (iii) the EPI involving the acoustic mode is isotropic and the matrix elements $g_{mn\nu}(\mathbf{k},\mathbf{q})$ only depend linearly on the magnitude of $\mathbf{q}$.
Any directional dependence of the EPI is ignored.

Using the Takagi formula, Qiao~\textit{et al.}~\cite{Qiao2014} calculated the electron and hole mobilities of phosphorene of up to five layers, along both the armchair and zigzag directions.
They found that the mobilities of both electron and hole carriers are anisotropic and layer dependent.
The calculated mobility values were high, on the order of hundreds to thousands of \cmVs{}.
In multilayer systems, the mobilities of the hole carriers are in general several times larger than those of the electron carriers; furthermore, both electrons and holes were found to be more mobile along the armchair direction.
However, in monolayer phosphorene, the trends of the electron-hole asymmetry and directional anisotropy are reversed.
The electron mobility of monolayer phosphorene was found to be 1,100-1,400~\cmVs{} and 80~\cmVs{} along the armchair and zigzag directions, respectively, whereas for the hole carriers,
the mobility values along the armchair and zigzag directions were found to be 640-700 and 10,000-26,000~\cmVs{}, respectively.
Qiao~\textit{et al.}~\cite{Qiao2014} attributed the exceptionally large hole mobility along the zigzag direction to the extremely small deformation potential along this direction.
The conclusion in the study of Qiao~\textit{et al.}~\cite{Qiao2014} that transport is hole-dominated and anisotropic was subsequently confirmed experimentally in multilayer systems~\cite{Liu2014,Li2015}.

Rudenko~\textit{et al.}~\cite{Rudenko2016} developed a more sophisticated theory of carrier scattering by acoustic phonons in anisotropic 2D systems, starting from the deformation potential method.
Unlike the above-mentioned Takagi formula, in which only a single deformation potential parameter and elastic constant appear,
they constructed electron-phonon scattering matrices of monolayer phosphorene that incorporate the complete elastic and deformation potential tensors.
The scattering matrices proposed by Rudenko~\textit{et al.}~\cite{Rudenko2016} therefore exhibit a sophisticated dependence on both the direction and the magnitude of the scattering wavevector $\mathbf{q}$.
The elastic constants and the deformation potentials corresponding to both in- and out-of-plane deformations were determined from first-principles calculations.
They applied their theory to calculate the direction-dependent carrier mobility of monolayer phosphorene as a function of both temperature and carrier density.
The calculated intrinsic carrier mobility was found to be higher along the armchair direction, and dominated by in-plane phonon scattering for carrier densities above 10$^{13}$~cm$^{-2}$.
At room temperature and a carrier density of 10$^{13}$~cm$^{-2}$, the electron and hole mobilities along the armchair direction were determined to be $\sim$700 and $\sim$250~\cmVs{}, respectively.

Trushkov~\textit{et al.}~\cite{Trushkov2017} calculated the phonon-limited carrier mobility of monolayer phosphorene using a tight-binding treatment of the EPI.
They obtained the electronic band structure using a nearest-neighbour tight-binding model and the phonon spectrum using a valence force model fitted to $GW$ and DFT calculations, respectively.
The EPI was treated using the Su-Shrieffer-Heeger Hamiltonian~\cite{Su1979} with distance-dependent hopping parameters.
Due to the electron-hole symmetry in their model, the calculated electron and hole mobilities were identical.
The authors found the room-temperature mobilities of monolayer phosphorene to be 625~\cmVs{} and 82~\cmVs{} along the armchair and zigzag directions, respectively.
In the above investigations~\cite{Qiao2014,Rudenko2016,Trushkov2017}, the EPI was described using the deformation-potential model.
The complex dependence of the EPI on the electron wavevector $\mathbf{k}$ and phonon wavevector $\mathbf{q}$ was either neglected or treated approximately.
As a result of these approximations, these studies generally found substantially higher carrier mobilities than in first-principles calculations.

Liao~\textit{et al.}~\cite{Liao2015b} performed an \textit{ab initio} investigation of the EPI in monolayer black phosphorus using DFPT.
In their study, the EPI on dense electron and phonon momentum grids was interpolated from DFPT calculations on coarse grids using Wannier interpolation~\cite{Giustino2007}.
They calculated the carrier relaxation time and the intrinsic mobility using the BTE within the MRTA.
The authors compared their calculated carrier scattering rates with those obtained via deformation potential calculations.
They found that for an anisotropic system such as phosphorene, the carrier scattering rates obtained using the deformation potential approximation of Qiao~\textit{et al.}~\cite{Qiao2014}
can be orders of magnitude smaller than the \textit{ab initio} results.
Indeed, the carrier scattering rates along the direction with small deformation potential was significantly underestimated.
This could explain why the extraordinarily large hole mobility of 26,000~\cmVs{} along the zigzag direction of phosphorene was not reproduced by subsequent calculations
that took into account the coupled EPI along different transport directions.
Liao~\textit{et al.}~\cite{Liao2015b} also found that optical phonons, not included in earlier deformation potential models, contribute non-negligibly to carrier scattering.
The phonon-limited carrier mobility calculated by Liao~\textit{et al.}~\cite{Liao2015b} is $\sim$170~\cmVs{} for both electron and hole carriers along the armchair direction,
whereas the calculated carrier mobilities along the zigzag direction are around 50~\cmVs{} and 35~\cmVs{} for electron and hole carriers, respectively.

Jin~\textit{et al.}~\cite{Jin2016} later calculated the intrinsic electron and hole mobilities of both monolayer and bilayer phosphorene, using a full-band Monte Carlo method for the solution of the BTE~\cite{Jacoboni1983}.
Their study confirmed earlier findings that the anisotropic crystal structure of phosphorene imparts anisotropic transport properties via the anisotropic band structure and scattering rates.
They found that, in monolayer phosphorene, the hole mobility in the armchair direction is approximately five times higher than in the zigzag direction at room temperature (460 vs. 90~\cmVs{}).
Transport in bilayer phosphorene, on the other hand, exhibits a more modest anisotropy with substantially higher mobilities (1,610 and 760~\cmVs{}, respectively).

As can be seen from the above, the numerical values of the intrinsic mobilities of phosphorene calculated by different research groups using different approaches exhibit considerable variations.
This prompted Gaddemane~\textit{et al.}~\cite{Gaddemane2018} to critically review the physical models employed in earlier studies.
These authors showed that the assumption of isotropic EPI, the use of deformation potentials instead of electron-phonon matrix elements,
and the neglect of the dependence of the electron-phonon matrix elements on the electron wavevector $\mathbf{k}$ are the main sources of discrepancy among earlier works.
The importance of anisotropic EPI was demonstrated by carrying out calculations of the acoustic-phonon-limited electron mobility of monolayer phosphorene using the Kubo-Greenwood method.
The inclusion of angle-dependent deformation potentials in the Kubo-Greenwood calculations resulted in low electron mobility values of $\sim$25~\cmVs{} and 5~\cmVs{} along the armchair and zigzag directions, respectively.
They also employed a full-band Monte Carlo method~\cite{Jacoboni1983} to numerically solve the BTE and obtained the carrier mobilities of monolayer and bilayer phosphorene.
For their Monte Carlo calculations, the authors calculated the carrier scattering rates using \textit{ab initio} EPI, obtained both from finite differences and from DFPT, which led to very similar results for the computed carrier mobilities.
As in their Kubo-Greenwood calculations, Gaddemane~\textit{et al.}~\cite{Gaddemane2018} obtained rather low carrier mobilities, not exceeding 25~\cmVs{} for electrons and holes in both mono- and bilayer phosphorene.
In contrast to the measurements by Cao~\cite{Cao2015} and the \textit{ab initio} calculations by Jin~\textit{et al.}~\cite{Jin2016},
the bilayer mobilities obtained by Gaddemane~\textit{et al.}~\cite{Gaddemane2018} were almost identical or even smaller than in the monolayer.
This effect was ascribed to the presence of low-energy interlayer optical phonons.
Despite the similar methodology adopted by Jin~\textit{et al.}~\cite{Jin2016} and Gaddemane \textit{et al.}~\cite{Gaddemane2018}, the origin of the large mobility
difference between the two studies remains unclear at the time of writing.

Finally, we mention the recent work by Sohier~\textit{et al.}~\cite{Sohier2018}, in which the authors carried out \textit{ab initio} BTE calculations of the carrier mobilities
of monolayer phosphorene at a high carrier density of $n$=5$\times$10$^{13}$~cm$^{-13}$.
Since a high carrier density was considered, the authors took into account the doping effects on the EPI by employing DFPT for gated 2D materials~\cite{Sohier2017a}.
This method is capable of self-consistently including the static screening effects from the charge carriers on the electron-phonon matrix elements within a linear response approach.
The authors obtained room-temperature hole mobilities of 586 and 44~\cmVs{} along the armchair and zigzag directions, respectively.
The electron mobilities along the armchair and zigzag directions were found to be 302 and 35~\cmVs{}, respectively.
However, as the authors noted in their work, caution is required when comparing their results to earlier works that focused on the intrinsic carrier mobility in the low-density limit,
as the effects of doping on carrier mobility becomes non-trivial at high carrier density.

\subsubsection{Molybdenum disulfide}

Molybdenum disulfide (MoS$_2$) is a representative example of layered transition metal dichalcogenides that can be exfoliated or grown in monolayers.
The atomistic structure of monolayer MoS$_2$ is shown in Fig.~\ref{fig:structures}(j).
Soon after experimental techniques for isolating 2D crystals had been developed, the carrier mobility of monolayer MoS$_2$ was measured~\cite{Novoselov2005}.
However, the reported room-temperature mobility was in the range of 0.5-3~\cmVs{}, much smaller than the early studies of the bulk carrier mobility of 100-260~\cmVs{}~\cite{Fivaz1967}.
Interest in MoS$_2$ for electronic and optoelectronic device applications grew rapidly after the experimental discovery that MoS$_2$ becomes a direct-bandgap semiconductor in its monolayer form~\cite{Mak2010,Splendiani2010}.
Efforts to enhance the carrier mobility of 2D MoS$_2$ ensued and the report by Radisavljevic~\textit{et al.}~\cite{Radisavljevic2011} that room-temperature mobilities as large as 200~\cmVs{} can be achieved by depositing a high-$\kappa$ top-gate dielectric in a double-gated device drew considerable attention.
However, this mobility value was subsequently questioned on the grounds that the methodology employed to extract the field-effect mobility from the experimental current-voltage data
could result in a significant overestimation of the mobility for devices in a double-gate geometry~\cite{Radisavljevic2011}.
Subsequent Hall mobility measurements for single-layer MoS$_2$ devices yielded room-temperature values in the range of 15-60~\cmVs{}~\cite{Radisavljevic2013}.
Cui~\textit{et al.}~\cite{Cui2015} later carried out multi-terminal transport measurements of the Hall mobility of 1-6-layer MoS$_2$ in a van der Waals heterostructure,
obtaining a room-temperature mobility between 40-120~\cmVs{}, with the mobility of monolayer MoS$_2$ located near the lower end of the range.
More recently, Yu~\textit{et al.}~\cite{Yu2014} reported the realization of room-temperature phonon-limited carrier transport in monolayer MoS$_2$ devices, by combining
improvements in sample and interface quality from chemical treatment with the suppression of charged impurity scattering through dielectric and carrier screening~\cite{Yu2016}.
The highest room-temperature mobility of monolayer MoS$_2$ devices measured by the authors was around 150~\cmVs{}.

On the theoretical side, there have been significant efforts to determine the intrinsic carrier mobility of monolayer MoS$_2$.
Kaasbjerg~\textit{et al.}~\cite{Kaasbjerg2012a} carried out the first study of the intrinsic phonon-limited electron mobility of $n$-doped monolayer MoS$_2$,
using a combined first-principles and semi-analytical formalism.
The BTE was solved by using an iterative scheme, in which the scattering rate integral was calculated by summing the quasi-elastic scattering from acoustic phonons
and the inelastic scattering from optical phonons.
Both the acoustic- and the optical-phonon scattering amplitudes were determined from deformation potentials, which were obtained by fitting first-principles electron-phonon matrix elements.
An analytical formula for the Fr\"ohlich interaction in 2D was also derived and its contribution to the phonon collision integral was evaluated numerically.
Kaasbjerg~\textit{et al.}~\cite{Kaasbjerg2012a} considered carrier scattering both within and between the two main valleys ($K$ and $K'$) of monolayer MoS$_2$.
They calculated a room-temperature mobility value of 410~\cmVs{}, which was found to depend only weakly on the carrier density.
They concluded that the room-temperature mobility was largely dominated by optical-deformation potential scattering from intravalley, homopolar phonons, intervalley LO phonons, and by intravalley,
long-wavelength polar LO phonons via Fr\"ohlich interactions.
\begin{figure}[t]
\begin{centering}
\includegraphics[width=0.99\linewidth]{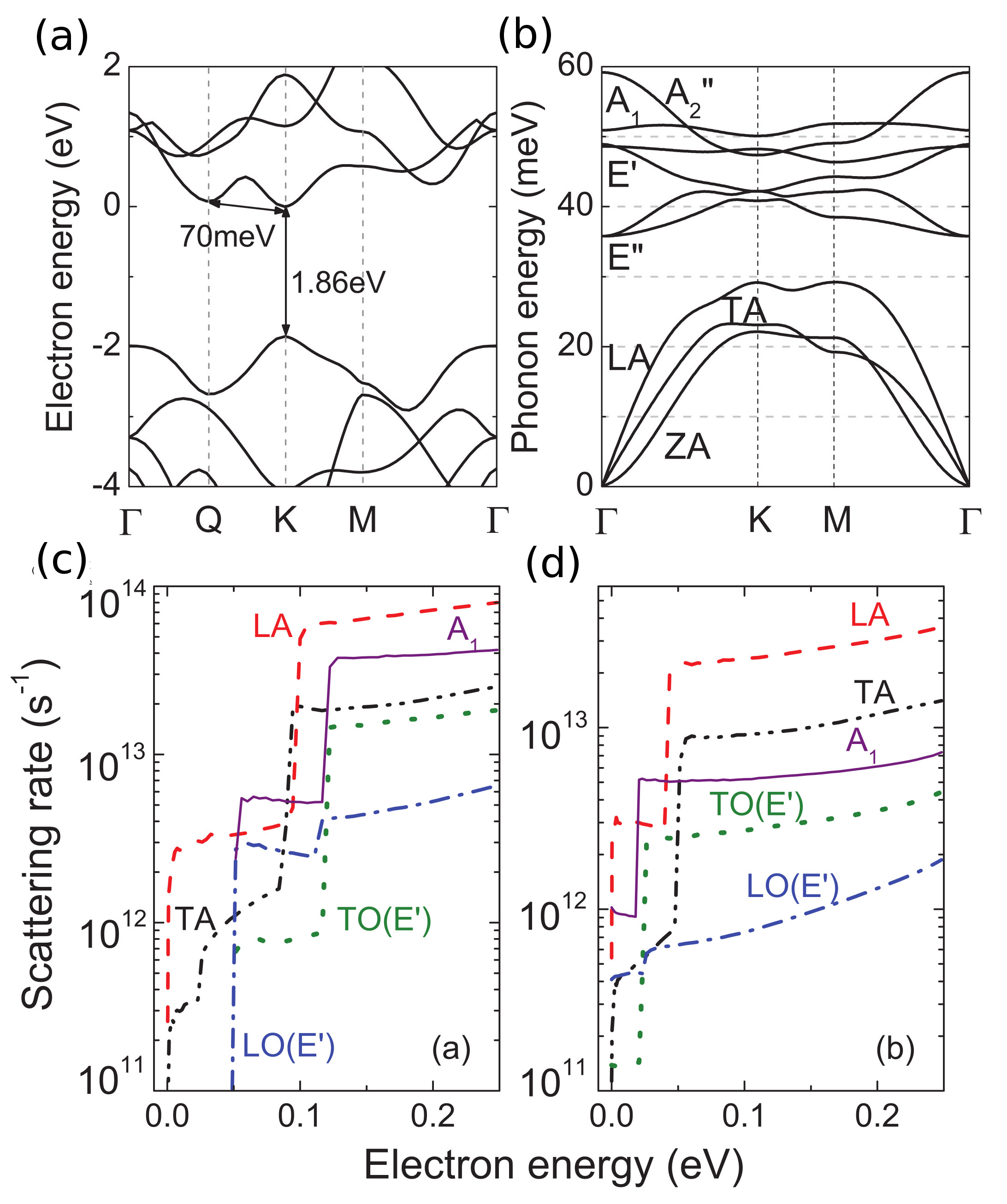}
\caption{The electron (a) and phonon (b) band structure of monolayer MoS$_2$ obtained
  from the local density approximation to DFT.
  (c,d) Scattering rates of $K$ valley electrons in  monolayer MoS$_2$ via emission (c) and absorption (d) of phonons
  at room temperature, based on \textit{ab initio} electron-phonon matrix elements and Fermi's golden rule.
  In the calculations of the scattering rate, the electron wavevector is assumed to be along the $K$-$\Gamma$ line.
  Adapted from~\cite{Li2013}; copyright (2013) by the American Physical Society.}
\label{fig:mobmos2}
\end{centering}
\end{figure}

Subsequently, Li~\textit{et al.}~\cite{Li2013} combined DFPT calculations of the EPI and full-band Monte Carlo simulations to obtain a
room-temperature mobility of 130~\cmVs{} for monolayer MoS$_2$.
In contrast to the work of Kaasbjerg~\textit{et al.}~\cite{Kaasbjerg2012a}, they found that the LA phonons provide the largest carrier scattering rates (see Fig.~\ref{fig:mobmos2}),
which the authors attributed to the strong intervalley scattering from the $K$/$K'$ valleys to the $Q$ valleys (another set of satellite valleys along the path $\Gamma$-$K$).
Subsequently, Restrepo~\textit{et al.}~\cite{Restrepo2014} solved the BTE using the SERTA where the momentum-resolved carrier relaxation time was calculated using first-principles electron-phonon matrix elements
and obtained a room-temperature mobility of 225~\cmVs{}.
Zhang~\textit{et al.}~\cite{Zhang2014} used deformation potential theory to calculate the LA-phonon-limited electron mobility of MoS$_2$ and obtained a mobility value of 340~\cmVs{}.
Li~\cite{Li2015} carried out first-principles calculations of the intrinsic charge transport properties of monolayer MoS$_2$
by iteratively solving the linearized BTE on dense electron and phonon momentum grids and by employing a linear interpolation of the electron-phonon matrix elements calculated with DFPT.
A room-temperature mobility of 150~\cmVs{} was obtained.
Gunst~\textit{et al.}~\cite{Gunst2016} also carried out first-principles calculations of the mobility of monolayer MoS$_2$ by solving the BTE using the MRTA.
The calculated room-temperature intravalley mobility was around 400~\cmVs{} at a carrier density of 3$\times$10$^{12}$~cm$^{-2}$.
We note that in all these studies, Fr\"ohlich interactions were neglected on the basis that their influence on carrier scattering in MoS$_2$ is small,
although Kaasbjerg~\textit{et al.}~\cite{Kaasbjerg2012a} had already demonstrated that Fr\"ohlich interactions also substantially contribute to the carrier scattering rate at room temperature.
More recently, Sohier~\textit{et al.}~\cite{Sohier2018} applied their first-principles approach for computing the transport properties of 2D materials to doped MoS$_2$,
where the effects of dimensionality and doping effects on the electron-phonon matrix elements were accounted for within DFPT.
They obtained a room-temperature carrier mobility of 144~\cmVs{} for $n$-doped MoS$_2$ at a carrier density of $5\times 10^{13}$~cm$^{-2}$.

It is worth noting that the intrinsic mobility of monolayer MoS$_2$ obtained by Kaasbjerg~\textit{et al.} (410~\cmVs{})~\cite{Kaasbjerg2012a}
is substantially higher than some of the later \textit{ab initio} studies~\cite{Li2013,Li2015}, which obtained values around 130-150~\cmVs{}.
Indeed, in the original study of Kaasbjerg~\textit{et al.}~\cite{Kaasbjerg2012a}, the carrier scattering rates were calculated using deformation potentials fitted to \textit{ab initio} scattering rates.
For the calculation of \textit{ab initio} scattering rates, the electron-phonon matrix elements were assumed to be independent of the wavevector $\mathbf{k}$ of the charge carriers
and to only depend on the phonon wavevector $\mathbf{q}$ and branch index $\nu$.
For the charge carriers in the two conduction band valleys at $K$ and $K'$, the electron-phonon matrix elements $g_{mn\nu}(\mathbf{k},\mathbf{q})$ at a generic wavevector $\mathbf{k}$ were considered to be
the same as those for charge carriers at the minimum of the corresponding valley.
It is not yet clear what is the effect of these approximations on the calculated mobility.
Piezoelectric scattering, which is present in monolayer MoS$_2$ due to the absence of inversion symmetry, was not taken into account in the study of Kaasbjerg~\textit{et al.}~\cite{Kaasbjerg2012a}
and also not in most later works.
However,  the authors subsequently considered the effect of piezoelectric interactions on the carrier mobility of monolayer MoS$_2$, which resulted in a revised room-temperature mobility
of 320~\cmVs{} at a carrier density of 10$^{11}$~cm$^{-2}$~\cite{Kaasbjerg2013}.

In addition to these differences, another important source of discrepancy may come from the different treatments of intervalley carrier scattering between the $K$ and $Q$ valleys~\cite{Li2013,Sohier2018}.
In the study of Kaasberjerg \textit{et al.}~\cite{Kaasbjerg2013}, since the $Q$ valleys are ${\sim}$0.2~eV above the $K$ valleys,
intervalley scattering from the conduction band edge at the $K$ valleys into the $Q$ valleys was not considered.
However, in the DFT calculations by Li~\textit{et al.}~\cite{Li2013}, the separation between the $K$ and $Q$ valleys was much smaller (70~meV) and consequently intervalley carrier scattering
between these valleys substantially contributed to the overall carrier scattering rate.
When the scattering between the $K$ and $Q$ valleys was not included, the carrier mobility obtained by Li~\textit{et al.}~\cite{Li2013} increased from 130~\cmVs{} to 320~\cmVs{},
which is closer to the value reported by Kaasbjerg~\textit{et al.} (410~\cmVs{})~\cite{Kaasbjerg2012a}.
It should be pointed out that it is difficult to accurately determine the energy separation between the $K$ and $Q$ valleys within the framework of DFT and that the energy separation
reported in other later studies~\cite{Li2015,Sohier2018} was more in line with the original work of Kaasberjerg \textit{et al.}~\cite{Kaasbjerg2012a}.
At any rate, the results by Li \textit{et al.}~\cite{Li2013} highlight the importance of a precise description of intervalley scattering for an accurate determination of the carrier mobility of transition metal dichalcogenides,
a point that was also emphasized in the study of Sohier \textit{et al.}~\cite{Sohier2018}.

\subsubsection{Indium selenide}

From an application point of view, the four 2D materials discussed so far each have their own advantages and disadvantages.
Graphene has extremely high carrier mobility even at room temperature~\cite{Bolotin2008}, yet its zero bandgap results in transistors with a very small on-off current ratio.
Strategies to open up a bandgap in graphene, such as by patterning graphene into nanoribbons~\cite{Li2008,Han2010} or by hydrogenation~\cite{Elias2009}, result in a reduced carrier mobility.
Silicene is considered to be more compatible with silicon-based electronics than other 2D materials.
However, silicene is typically synthesized on metal substrates under ultra-high vacuum conditions~\cite{Vogt2012,Feng2012,Fleurence12,Chen2012a}.
Since a post-synthesis transfer to a different substrate is needed for device fabrication and since silicene is unstable in air~\cite{Tao2015}, device fabrication and processing is challenging.
Phosphorene multilayers possess high carrier mobility, but their electronic transport properties are anisotropic~\cite{Liu2014} and their stability under ambient conditions is poor~\cite{CastellanosGomez2014,Island2015}.
Recently, 2D indium selenide (InSe) has emerged as a promising alternative for 2D nanoeletronics, as the experimentally measured electron mobility of multilayer InSe of up to 1,000~\cmVs{}
is comparable to that of phosphorene, while its thermodynamical stability is even higher~\cite{Bandurin2017}.
The atomistic structure of monolayer InSe is shown in Fig.~\ref{fig:structures}(k).
\begin{figure}[t]
\begin{centering}
\includegraphics[width=0.99\linewidth]{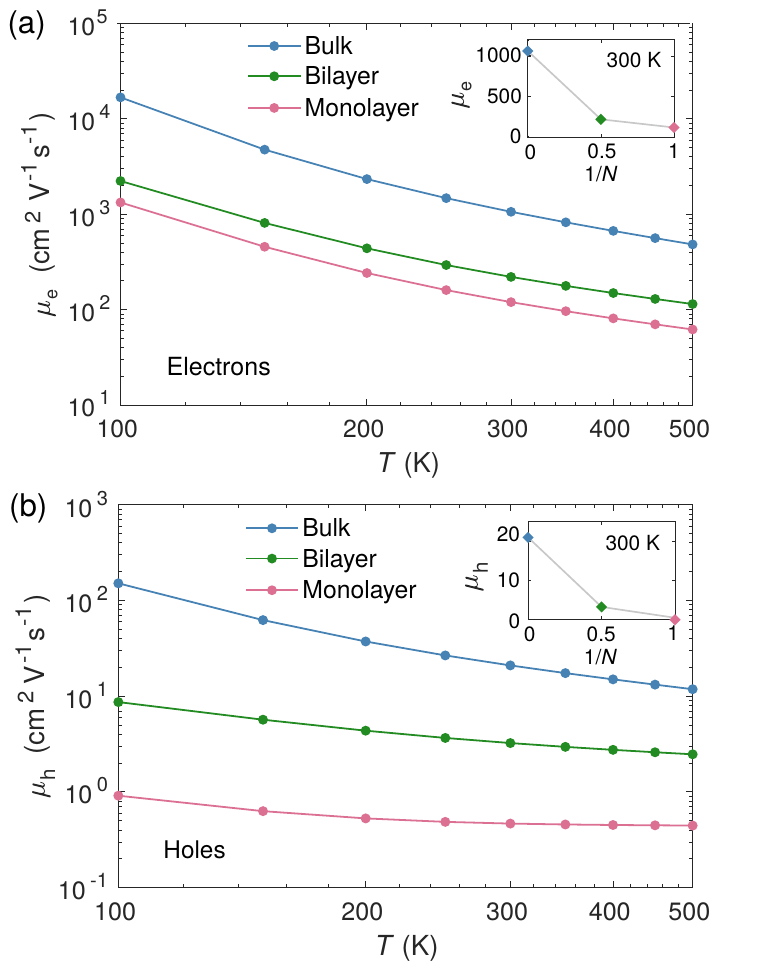}
  \caption{\label{fig:inse} Temperature dependence of the in-plane hole (a) and electron mobility (b)
  of monolayer, bilayer, and bulk InSe, using the \textit{ab initio} Boltzmann transport equation.
  The insets show the calculated room-temperature mobility vs. the reciprocal number of layers,
  $1/N$, for $N=1$, 2, and $\infty$.
  Adapted from~\cite{Li2019}; copyright (2019) by the American Chemical Society.}
\end{centering}
\end{figure}

Li~\textit{et al.}~\cite{Li2019} recently investigated the intrinsic carrier mobility of monolayer, bilayer, and bulk InSe using the \textit{ab initio} BTE.
The electron-phonon matrix elements were calculated via DFPT and interpolated onto very dense electron and phonon momentum grids using Wannier interpolation~\cite{Giustino2007,Mostofi2014,Ponce2016a}.
The BTE was solved in the SERTA~\cite{Ponce2018}.
As InSe is a polar semiconductor, the coupling of charge carriers to the long-wavelength polar LO phonons was found to be rather strong in InSe.
For 2D InSe, the long-range Fr\"ohlich interaction needs to be handled with care as the periodic boundary conditions typically employed in plane-wave \textit{ab initio} calculations can
induce spurious Fr\"ohlich interactions with neighboring image cells~\cite{Sohier2016,Sohier2017b}.

The authors calculated the intrinsic electron and hole mobility of monolayer, bilayer, and bulk InSe as a function temperature using the BTE (Fig.~\ref{fig:inse}).
The calculated electron mobilities were found to be 120, 220, and 1,060~\cmVs{}, respectively, in good agreement with available transport measurement data~\cite{Segura1984,Feng2014,Sucharitakul2015,Bandurin2017}.
In comparison, the hole mobility was found to be much smaller $\le$~20~\cmVs{}.
Li~\textit{et al.}~\cite{Li2019} analyzed the electron scattering mechanisms in detail and found that the Fr\"ohlich interaction dominated; on the other hand, hole scattering was dominated by Fr\"ohlich interaction only in the bulk case.
They also found that the significant layer dependence of the carrier mobility of InSe can neither be explained in terms of the layer dependence of the carrier effective mass
nor in terms of the layer dependence of the electron-phonon matrix elements.
Instead, they found that the thickness dependence originates from a decrease of the electronic density of states when InSe layers are stacked together, which reduces the phase space for carrier scattering.
The significant layer-dependent density of states in InSe results from the strong interlayer electronic coupling, which leads to a hybridization of band-edge states when wavefunctions of different layers come into contact.

Li~\textit{et al.}~\cite{Li2019} also generalized the dimensionality dependence of the carrier mobility of InSe to other 2D materials by developing a simple tight-binding model.
Using this model, they found that the carrier mobility of 2D materials is rather sensitive to the interlayer electronic coupling strength.
In particular, they proposed that for 2D semiconductors with non-negligible interlayer electronic coupling, there exists an intrinsic layer thickness below which the carrier mobility increases with the number of layers
and above which the mobility gradually reaches the bulk value.
As interlayer interaction is ubiquitous in layered materials, they suggested that van der Waals epitaxy could be employed to engineer the carrier mobility of 2D materials.

\subsection{High-throughput calculations of carrier mobilities}\label{Sec4.3}

In recent years, the development of automatic workflows for \textit{ab initio} calculations, combined with the increasing accessibility of supercomputing resources,
has brought about a new paradigm of materials discovery through high-throughput (HT) computation~\cite{Curtarolo2013}.
HT computations are designed for calculating a target property for a large set of materials with minimal user intervention.
There are multiple initiatives aiming to build large databases of materials properties through HT computations.
Well-known examples of such databases include the Materials Project~\cite{MP}, the AFLOW repository~\cite{Curtarolo2012}, the Materials Cloud~\cite{MaterialsCloud},
the Open Quantum database~\cite{Saal2013}, and the Harvard Clean Energy Project~\cite{Hachmann2011}.
At present, these databases have covered a wide range of materials properties including energetics, thermodynamics, electronic band structures~\cite{Setyawan2010}, elastic and piezoelectric tensors~\cite{deJong2015a,deJong2015b},
and phonon spectra~\cite{Petretto2018}.
It is natural to expect that HT databases will be expanded to include electronic transport properties in the near future.

From a practical standpoint, the implementation of \textit{ab initio} calculations of carrier mobilities in an HT framework is presently hampered by the heavy computational workload associated
with the evaluation of all possible electron-phonon scattering processes for charge carriers in solids.
Wannier interpolation of electron-phonon matrix elements~\cite{Giustino2007} has been rather successful in reducing the computational cost of mobility calculations,
yet the automatic generation of Wannier functions~\cite{Mustafa2015,Cances2017}, as needed in HT calculations, is still at an early stage.

Meanwhile, there have already been a few studies reporting HT calculations of electronic transport properties using simplified models of carrier scattering.
Ricci~\textit{et al.}~\cite{Ricci2017} generated a large dataset of transport properties for around 48,000 materials, based on DFT band structures and Boltzmann transport calculations within the c-SERTA.
In these calculations, the authors assumed that the carrier lifetime $\tau$ is a constant, empirical quantity, independent of carrier energy and transport direction.
This is clearly a strong approximation, as also noted by the authors.
Nevertheless, this approximation allowed the authors to focus only on the effects of the electronic band structure on the charge transport properties.
They calculated the ratio between the electrical conductivity $\sigma$ and the relaxation time $\tau$ for a wide range of materials, temperatures, and carrier densities.
To obtain the actual conductivity and carrier mobility, one would need to supply the empirical relaxation time $\tau$ for each individual material,
which could be obtained by fitting the calculation data to the experimentally measured conductivities.
In this sense, the conductivity database of Ricci~\textit{et al.}~\cite{Ricci2017} is semi-empirical.

\begin{figure*}[th]
  \centering
  \includegraphics[width=0.99\linewidth]{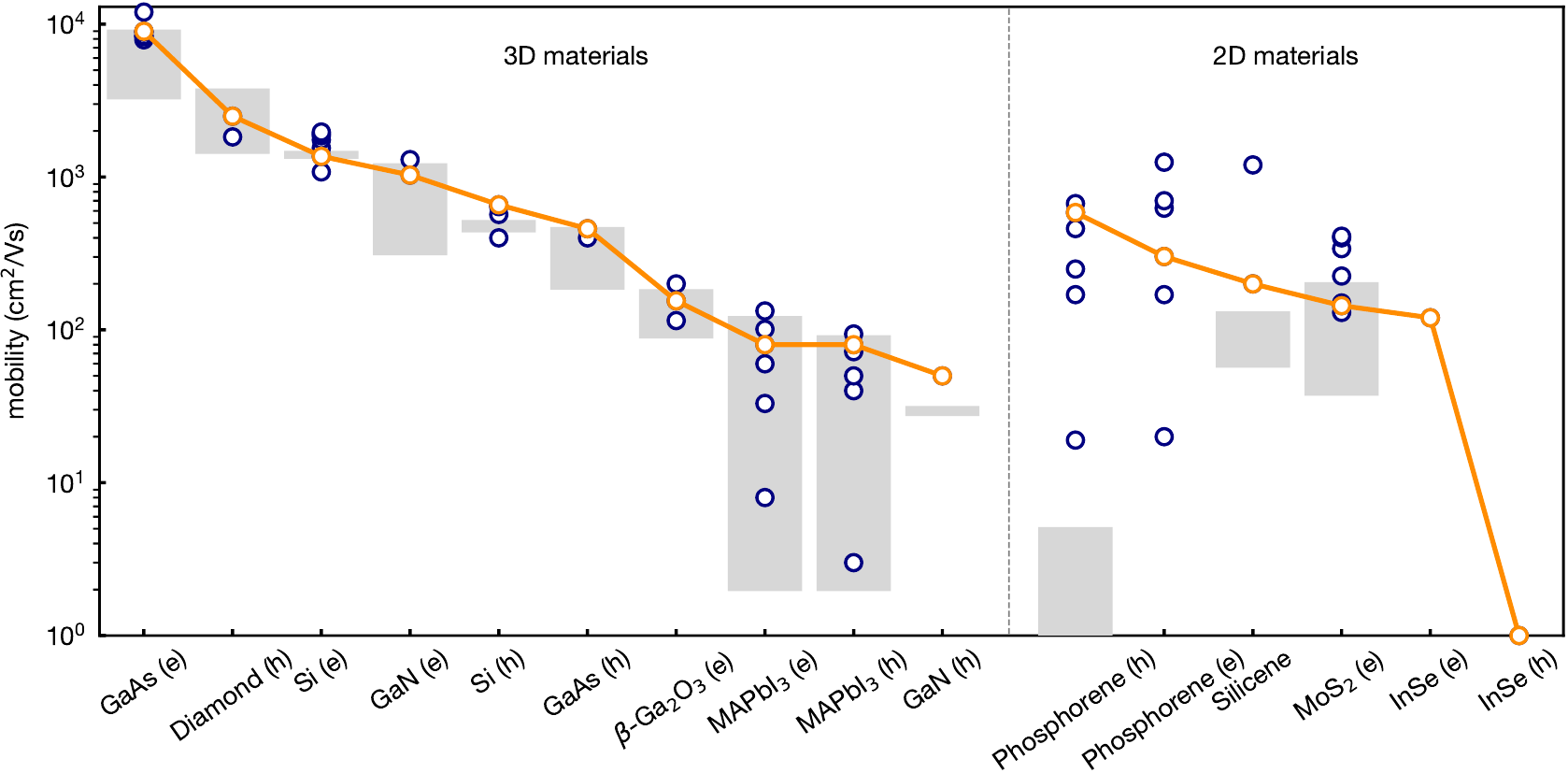}
  \caption{\label{fig:moball} Range of measured electron (e)
  and hole (h) mobilities at room temperature (shaded region), and calculated values (blue dots).
  The results highlighted in orange can be considered to be the most accurate calculations
  reported so far.
  Materials are sorted with the mobility decreasing towards the right hand side:
  electron mobility of GaAs, theory~\cite{Ma2018,Rode1970,Zhou2016,Rode1971,Liu2017,Arabshahi2008,Rode1971}
  and experiments~\cite{Madelung2003,Rode1971,Hicks1969};
  hole mobility of diamond, theory~\cite{Macheda2018,Pernot2010} and
  experiments~\cite{Isberg2005,Jansen2013,Gabrysch2011,Nesladek2008,Vavilov1976,Dean1965};
  electron mobility of silicon, theory~\cite{Restrepo2009,Ma2018,Li2015,Fiorentini2016,Luisier2009,Ponce2018,Luisier2009} and experiments~\cite{Ludwig1956,Cronemeyer1957,Li1977,Jacoboni1977};
  electron mobility of w-GaN, theory~\cite{Arabshahi2008,Ponce2019a} and
  experiments~\cite{Kyle2014,Goetz1998,Goetz1996,Ilegems1973};
  hole mobility of silicon, theory~\cite{Ponce2018,Luisier2009,Ma2018,Luisier2009}
  and experiments~\cite{Ludwig1956,Cronemeyer1957,Jacoboni1977,Dorkel1981};
  hole mobility of GaAs, theory~\cite{Ma2018,Scholz1995} and
  experiments~\cite{Mears1971,Madelung2003,Sotoodeh2000,Wenzel1998,Kim1991,Hill1970,Beaton2010};
  electron mobility of $\beta$-Ga$_2$O$_3$, theory~\cite{Ma2016,Kang2017,Ghosh2016}
  and experiments~\cite{Zhang2018,Wong2016};
  electron mobility of MAPbI$_3$, theory~\cite{Ponce2019} and
  experiments~\cite{Shrestha2018,Saidaminov2015,Milot2015,Dong2015,Karakus2015,Savenije2014,Shi2015a};
  hole mobility of MAPbI$_3$, theory~\cite{Ponce2019}
  and experiments~\cite{Shrestha2018,Saidaminov2015,Milot2015,Dong2015,Karakus2015,Savenije2014,Shi2015a};
  hole mobility of w-GaN, theory~\cite{Ponce2019a}
  and experiments~\cite{Arakawa2016,Horita2017};
  hole mobility of phosphorene, theory~\cite{Qiao2014,Sohier2018,Jin2016,Rudenko2016,Liao2015b,Gaddemane2018}
  and experiments~\cite{Cao2015};
  electron mobility of phosphorene, theory~\cite{Qiao2014,Rudenko2016,Trushkov2017,Sohier2018,Liao2015b,Gaddemane2018}; electron mobility of silicene, theory~\cite{Li2013,Gunst2016}
 and experiments~\cite{Tao2015};
  electron mobility of MoS$_2$, theory~\cite{Kaasbjerg2012a,Gunst2016,Zhang2014,Restrepo2014,Li2015,Sohier2018,Li2013} and experiments~\cite{Radisavljevic2011,Yu2016,Cui2015,Radisavljevic2013};
  electron mobility of InSe, theory~\cite{Li2019} and experiments~\cite{Li2019}.}
\end{figure*}

For 2D materials, there have been significant efforts in the community aimed at identifying all possible stable monolayer materials that can be isolated from bulk compounds
~\cite{Lebgue2013,Haastrup2018,Cheon2017,Mounet2018,Ashton2018}.
Notable large databases of 2D materials resulting from these efforts include the Computational 2D materials Database (C2DB) by Haastrup~\textit{et al.}~\cite{Haastrup2018,C2DB}
and the 2D structures and layered materials database hosted on the Materials Cloud~\cite{Mounet2018,MaterialsCloud}.
Sohier~\textit{et al.}~\cite{Sohier2018} recently developed a rigorous approach for HT computations of the charge transport properties of 2D materials based on an \textit{ab initio} description of the EPI.
As mentioned earlier, Wannier interpolation of electron-phonon matrix elements has been an essential technique for reducing the computational cost of \textit{ab initio} mobility calculations.
However, the development of Wannier interpolation in an HT framework has not yet been achieved.
Furthermore, for 2D materials, Wannier interpolation may not be critical since the sampling of the Brillouin zone is limited to two dimensions.
Therefore, by exploiting symmetry and identifying the pockets of electronic states relevant for carrier transport, a simple linear interpolation of the EPI and other relevant quantities
could be sufficient to achieve numerical convergence.
Sohier~\textit{et al.}~\cite{Sohier2018} calculated the carrier mobility based on DFT electronic band structures and the BTE approach.
They also took into account the effects of dimensionality and charge doping on the EPI using the recently developed DFPT framework for gated 2D materials~\cite{Sohier2017a}.

Sohier~\textit{et al.}~\cite{Sohier2018} applied their automatic framework to the calculation of carrier mobilities of six well-studied 2D materials
($n$-doped MoS$_2$ , WS$_2$ , WSe$_2$, arsenene, and pristine and $p$-doped phosphorene) at high doping levels ($\sim$10$^{13}$~cm$^{-2}$), obtaining mobility values similar to earlier \textit{ab initio} works.
This approach is promising for 2D materials, but it would be computationally more challenging for bulk systems.
Therefore, further developments in automatic interpolation techniques are still needed in order to realize efficient HT \textit{ab initio} mobility calculations for a wide range of materials.

\subsection{Predictive accuracy}\label{Sec4.4}

Figure~\ref{fig:moball} provides a summary of calculated as well as measured intrinsic mobilities at room temperature of all the compounds reviewed here.
%
In Section~\ref{sec:gaas}, we showed that GaAs has the largest electron mobility among bulk, three-dimensional semiconductors reviewed in this study, with experimental values ranging from 3,300 to 9,000~\cmVs{}~\cite{Madelung2003,Rode1971,Hicks1969} and calculations ranging from 7,900 to 12,000~\cmVs{}~\cite{Ma2018,Rode1970,Zhou2016,Rode1971,Liu2017,Arabshahi2008,Rode1971}.
The mobility of diamond has mostly been measured in boron-doped samples and room-temperature values range from 1,450 to 3,700~\cmVs{}~\cite{Isberg2005,Jansen2013,Gabrysch2011,Nesladek2008,Vavilov1976,Dean1965}.
In this case the calculations yielded 1,830~\cmVs{}~\cite{Pernot2010} and 2,500~\cmVs{}~\cite{Macheda2018}.
We could not find reports of electron mobility in $n$-type diamond.
The lowest mobility among the three-dimensional bulk semiconductors considered here is for holes in wurtzite GaN, with a measured value of approximately 30~\cmVs{}~\cite{Arakawa2016,Horita2017} and calculations yielding 50~\cmVs{}~\cite{Ponce2019a}.

In the case of 2D materials there are fewer experimental data to compare with, as discussed in Section~\ref{Sec4.2}.
Apart from graphene, whose large mobility is off the scale of Fig.~\ref{fig:moball}, MoS$_2$ is the 2D semiconductor which has most intensely been investigated in experiments.
The measured mobility of MoS$_2$ ranges from 40 to 200~\cmVs{}~\cite{Radisavljevic2011,Yu2016,Cui2015,Radisavljevic2013},
while the calculations span the 130 to 410~\cmVs{} range~\cite{Kaasbjerg2012a,Gunst2016,Zhang2014,Restrepo2014,Li2015,Sohier2018,Li2013}.
For this and other 2D materials, it appears that calculations tend to lie above the experimental ranges, as shown in Fig.~\ref{fig:moball}.
This trend could be a consequence of the fact that scattering mechanisms other than EPI are not included in the calculations or that the field is still relatively young and the gap between experiment and theory will gradually close as we move forward.

%
%
%
%
%
%
%


Overall we find the calculated mobilities from the \textit{ab initio} solution to the BTE to be in close agreement with experiment.
Typically calculated values are close to the upper end of the range of experimental data, although in some cases (such as the hole mobility of silicon, GaN, and MAPbI$_3$) the calculations overestimate the measurements.
One obvious reason for this overestimation is that only the scattering of electrons by phonons is included in the theory.
Additional scattering mechanisms, such as defect and ionized-impurity scattering, can further lead to a reduction of the mobility.
In addition, we note that the EPI matrix elements computed within DFPT are usually underestimated due to the band gap problem~\cite{Yin2013,Antonius2014}.
Therefore, everything else being kept the same, one would expect a reduced mobility from using EPI matrix elements evaluated with hybrid functionals or $GW$ calculations.
However, many-body corrections to the electronic band structure might counterbalance this effect,
hence it is difficult to predict the magnitude and sign of the changes to the mobility caused by many-body corrections to local or semilocal DFT functionals.

%
As expected, the hole mobility is smaller than the electron mobility.
Indeed, conduction-band electrons tend to be more mobile than valence band electrons due to the latter being typically more strongly bound.
%
This is usually connected with the fact that the valence band wavefunctions have a bonding character, while the conduction-band ones have an anti-bonding character, which also leads to
lower electron effective masses than hole effective masses~\cite{Gibbs2017,Ricci2017}.
%

%
%
%

%

From this review it emerges that an accurate description of EPI is critical to correctly interpret experimental data.
Simplified models based on constant scattering rates or models that include only portions of the EPI like acoustic-deformation potential, optical-deformation potential, piezoelectric
scattering, or Fr\"ohlich phonon scattering are certainly useful to rationalize trends, but they can lead to quantitatively (and seldom qualitatively) incorrect predictions, as emphasized recently~\cite{Fischetti2019}.
The entire theory detailed in Section~\ref{Sec2} relies on an unbiased, \textit{ab initio} description of the EPI and should be preferred over empirical models when addressing quantitatively predictive mobility calculations.



\section{New directions and opportunities}\label{Sec5}

\subsection{Spin transport}\label{Sec5.1}

Understanding spin equilibration and spin transport has a direct impact on the developement of many devices, ranging from spin field-effect transistors~\cite{Datta1990}, spin filters~\cite{Hanson2004},
and spin diodes~\cite{Solomon1976} to spin qubits~\cite{Sarma2001}.
The SOC is a destructive force for spin coherence but can also lead to interesting phenomena like the spin Hall effect~\cite{Balakrishnan2013} or weak antilocalization~\cite{McCann2012}.
In materials lacking inversion symmetry, the bands can undergo a \textit{Rashba splitting}, with the spin orientation of the electrons in the split bands rotating in opposite directions around the band extrema~\cite{Bihlmayer2015}.
%
In the case of materials with a strong Rashba splitting, the electrons can experience spin-momentum locking, which can prevent electron-phonon backscattering~\cite{Manchon2015}.
This suppression has been reported in scanning-tunneling-microscopy experiments~\cite{Roushan2009}.
This field of study has been coined \textit{spin-orbitronics} and is a specialized branch of spin transport.

\begin{figure}[t]
  \centering
  \includegraphics[width=0.99\linewidth]{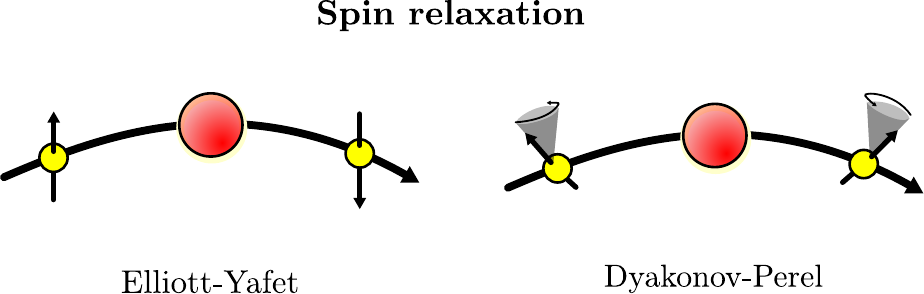}
  \caption{\label{fig:eydp}Schematic representation of the Elliott-Yafet and Dyakonov-Perel spin relaxation mechanisms.
  The yellow dots denote an electron before and after scattering from a scattering center (red dot).}
\end{figure}

Spin decoherence mainly results from the Elliott-Yafet (EY) and Dyakonov-Perel (DP) spin relaxation mechanisms~\cite{Zutic2004}.
The EY mechanism is based on the fact that a periodic lattice-induced SOC is modified by phonons and can directly couple to the spin-up and spin-down states of the electrons.
The associated spin relaxation rate $\tau_s$ due to phonon scattering can be estimated by integrating the spin-flip Eliashberg function defined as
\begin{multline}
  \! \alpha_{\mathrm{s}}^2 F(\omega) \! \equiv \! \frac{1}{\mathrm{DOS}^2(\varepsilon_{\mathrm{F}})}
 \! \sum_{mn\nu}\int \!\! \frac{\mathrm{d}^3 q}{\Omega_{\mathrm{BZ}}} \! \! \int \! \frac{\mathrm{d}^3 k}{\Omega_{\mathrm{BZ}}} \left| g_{m\uparrow n\downarrow \nu}(\mathbf{k},\mathbf{q}) \right|^2 \\
  \times  \delta(\omega -\omega_{\mathbf{q}\nu})
   \delta(\varepsilon_{\mathrm{F}}-\varepsilon_{m\uparrow\mathbf{k}+\mathbf{q}}) \delta(\varepsilon_{\mathrm{F}}-\varepsilon_{n\downarrow\mathbf{k}}),
\end{multline}
where $g_{m\uparrow n\downarrow \nu}(\mathbf{k},\mathbf{q})$ denotes the spin-resolved electron-phonon matrix element and DOS($\varepsilon_{\mathrm{F}}$) is the density of states at the Fermi level $\varepsilon_{\mathrm{F}}$~\cite{Fabian1999}.
In the case of the DP mechanism, one can use approximate models for the electron and phonon band structures, so that the spin relaxation rate of conduction band electrons with energy $\varepsilon$ can be simplified into~\cite{Pikus1984}
\begin{equation}
  \frac{1}{\tau^{\text{DP}}} = \gamma \tau_p(\varepsilon ) \alpha^2 \frac{\varepsilon^3}{\hbar^2 E_{\mathrm{g}}}.
\end{equation}
Here $\gamma$ is a constant that depends on the type of phonon involved, $\alpha$ is the SOC strength, $\tau_p$ the energy-dependent electron-phonon relaxation rate, and $E_{\mathrm{g}}$ the band gap energy.
In both cases, the spin coherence time will be long if the SOC is small and the band gap is large.
However, in the case of the EY mechanism, the spin coherence time will be long if the electron-phonon coupling is small, whereas for the DP mechanism, the faster the momentum relaxation, the slower the spin dephasing.
This is due to the fact that the EY effect happens \emph{during} the collision with a scattering center while the DP effect takes place \emph{in-between} collisions (see Fig.~\ref{fig:eydp}).

An additional benefit of implementing the spin-resolved electron-phonon matrix elements $g_{m\uparrow n\downarrow \nu}(\mathbf{k},\mathbf{q})$ in a public first-principles code will be to enable
an evaluation of Eq.~\eqref{eq:iterboltzmann} including the spin degree of freedom, which will allow the computation of transport properties with spin.
Indeed, some decay channels can become forbidden due to the orientation of the spins and this is expected to significantly increase the mobility.

In terms of existing first-principles works on spin decoherence, relatively few exist up to now.
Restrepo and Windl~\cite{Restrepo2012} studied the spin relaxation time of silicon, graphite, and diamond in 2012 using the Quantum ESPRESSO software suite~\cite{Giannozzi2017},
considering only the EY-dominated temperature regime; they found a relaxation time of about 5~ns for graphite and silicon and a much longer relaxation time of 180~ns for diamond.
In 2014, Kang and Choi~\cite{Kang2014,Kang2014a} computed the temperature dependence of the electron-spin relaxation rate due to piezoelectric phonon scattering in GaAs
and found a good agreement with experiment.

\subsection{Carrier mobility in topological materials}\label{Sec5.2}

Topological insulators (TIs) attracted considerable attention owing to the many new opportunities offered by the topological nature of their electronic wavefunctions and band structures~\cite{Hasan2010,Qi2011,Bansil2016}.
These materials are gapped in the bulk, hence bulk electrons cannot conduct electricity; however, the surfaces of 3D TIs or the edges of 2D TIs
support gapless conducting states that are protected by time-reversal symmetry (TRS)~\cite{Kane2005a,Kane2005b,Bernevig2006,Fu2007a,Fu2007b,Moore2007,Roy2009}.
The surface/edge states of TIs are spin-nondegenerate and exhibit spin-momentum locking, so that unless the TRS is broken, elastic backscattering of electrons is not allowed.
Due to this symmetry protection, the surface or edge states of TIs are highly conducting channels for charge and spin transport.
At low temperature, the absence of backscattering of electrons in the edge states of 2D TIs leads to a spin-polarized quantum conductance $G = 2e^2/h$~\cite{Konig2007,Brune2012}.
If transport in the ballistic regime can be maintained at device operation temperature, spin-polarized currents with very little dissipation can be generated on the surfaces of 3D TIs and at the edges of 2D TIs,
which has potential implications for spintronics applications.
Furthermore, the surface states of 3D TIs exhibit many exotic quantum mechanical properties originating from the helically spin-polarized Dirac fermions, which could be used to enable fault-tolerant quantum computing~\cite{Fu2008}.
A wide range of both 3D and 2D TIs have so far been theoretically proposed and experimentally realized~\cite{Bansil2016}.

In terms of charge transport in TIs, it should be emphasized that, while single-particle elastic backscattering by perturbations that preserve TRS is forbidden,
inelastic scattering via electron-electron interaction~\cite{Kane2005a,Schmidt2012} and EPI~\cite{Budich2012} is still possible.
These mechanisms give rise to a finite conductivity and a temperature-dependent deviation from the quantized conductance~\cite{Kane2005a,Schmidt2012}.
For the 2D surface states of 3D TIs, in addition to the aforementioned inelastic scattering processes, elastic scattering at an angle less than $\pi$ is also possible.
Furthermore, backscattering from TRS-breaking defects such as magnetic impurities is not symmetry-protected.
If the size of a TI sample is small enough such that the wavefunctions on opposite surfaces or edges overlap, inter-edge (in 2D) or inter-surface (in 3D) scattering can occur.
All these intrinsic and extrinsic scattering sources will limit the conductivity and mobility of the surface and edge states at finite temperature. 
Indeed, experiments on Bi$_2$Se$_3$ demonstrated that topological protection does not always result in high mobilities~\cite{Butch2010,Kim2012,Wu2016}.

Due to the spin-momentum locking of the surface/edge states in TIs, a finite electron mean free path corresponds to an equal and finite spin diffusion length~\cite{OjedaAristizabal2012}.
Hence, the computation of the carrier lifetime and mobility is not only relevant for potential electronics applications that exploit the high conductivity of surface/edge states,
but also for spintronics applications that exploit the spin degrees of freedom of charge carriers, as discussed in Sec.~\ref{Sec5.1}.
Future developments of first-principles methods for computing the carrier mobility and transport properties of TI surface/edge states will need to treat the effects of EPI,
electron-electron interactions, and impurity/defect scattering by considering the charge and spin degree of freedom on an equal footing.
Formalisms that go beyond the BTE may also prove necessary in order to capture the quantum coherence effects of the Dirac fermions in the ballistic transport regime~\cite{Culcer2010,Li2012,Burkov2010}.
The availability of such first-principles transport approaches would enable the quantitative study of coupled charge and spin transport in realistic TI systems,
which will be useful for the development of practical spintronics and other applications.

In addition to the study of surface and edge states, it should be emphasized that the bulk transport properties of TIs are interesting and important in their own right.
This is because the contribution from bulk states to the conductance is almost always present at finite temperature, as most TIs are small bandgap semiconductors
with thermally populated charge carriers, and the unintentional doping from defects and impurities is difficult to avoid.
A better understanding of bulk transport would help disentangle the contributions from bulk and surface states to the overall charge transport of TIs.
What is perhaps even more interesting, is that most TI materials contain heavy elements and exhibit strong SOC, which results in band inversion and highly non-parabolic bulk band dispersions.
The manifestation of such band structures and strong SOC effects on the bulk carrier dynamics is an intriguing topic that has only started to be explored from first principles~\cite{Liu2018}.
In this respect, it is worth mentioning that many TIs, such as bismuth antimony (Bi$_{1-x}$Sb$_x$), bismuth telluride (Bi$_2$Te$_3$), antimony telluride (Sb$_2$Te$_3$), and bismuth selenide (Bi$_2$Se$_3$)
~\cite{Hsieh2008,Zhang2009,Hsieh2009,Chen2009} are also excellent thermoelectric materials~\cite{Xu2017}.
It is anticipated that a deeper understanding of bulk carrier transport in TIs could also lead to new insights that will enable the design of better thermoelectrics,
in addition to electronics and spintronics applications.

\subsection{Berry phases in carrier transport}\label{Sec5.3}

For a quantum system described by a Hamiltonian $\hat{H}(\mathbf{R})$, where $\bf{R}$ is a set of parameters that parameterize the Hamiltonian,
the adiabatic evolution of an eigenstates $|n(\mathbf{R})\rangle$ along a loop $\mathcal{C}$ in parameter space will acquire a gauge-invariant, geometric phase factor $\exp[{i\gamma_n(\mathcal{C})}]$,
in addition to a dynamical phase factor \cite{Berry1984,Xiao2010}.
This geometric phase $\gamma_n(\mathcal{C})$ is known as the \emph{Berry phase}.
The Berry phase can be written as a loop integral in parameter space~\cite{Xiao2010}:
\begin{equation}
  \gamma_n(\mathcal{C}) = \oint_{\mathcal{C}} \mathrm{d} \mathbf{R} \cdot \mathcal{A}_n(\mathbf{R}),
\end{equation}
where $\mathcal{A}_n(\mathbf{R})$ is a vector-valued function called the \emph{Berry connection} or the \emph{Berry vector potential}
\begin{equation}
  \mathcal{A}_n(\mathbf{R}) = i\langle n(\mathbf{R}) | \frac{\partial}{\partial \mathbf{R}} | n(\mathbf{R}) \rangle.
\end{equation}
From the Berry vector potential, the \emph{Berry curvature} can be defined.
In a 3D parameter space, the curvature is given by
\begin{equation}\label{eq:BerryCurvature}
  \boldsymbol{\Omega}_n(\mathbf{R}) = \frac{\partial}{\partial \mathbf{R}} \times \mathcal{A}_n(\mathbf{R}).
\end{equation}
Based on Stokes' theorem, the Berry phase can be written in terms of the surface integral of the Berry curvature:
\begin{equation}
  \gamma_n(\mathcal{C}) = \int_{\mathcal{S}} \mathrm{d} \mathbf{S} \cdot \boldsymbol{\Omega}_n(\mathbf{R}).
\end{equation}
For electrons in crystalline solids, the cell-periodic part of the Bloch states $|n\mathbf{k}\rangle$, $|u_{n\mathbf{k}}\rangle \equiv \mathrm{e}^{-i\mathbf{k}\cdot{\mathbf{r}}}|n\mathbf{k}\rangle$, are the eigenstates of the Bloch Hamiltonian
$\hat{H}_{\mathbf{k}} \equiv e^{-i\mathbf{k}\cdot{\mathbf{r}}}\hat{H}e^{i\mathbf{k}\cdot{\mathbf{r}}}$, where $\hat{H}$ is a single-particle Hamiltonian that is crystal periodic.
The Berry connection of band $n$ is then given by
\begin{equation}
  \mathcal{A}_n(\bm{k}) = i\langle u_{n\mathbf{k}} | \frac{\partial}{\partial \mathbf{k}} | u_{n\mathbf{k}} \rangle.
\end{equation}
Based on Eq.~\eqref{eq:BerryCurvature}, the Berry curvature $\boldsymbol{\Omega}_{n\mathbf{k}}$ is then given by
\begin{equation}
  \boldsymbol{\Omega}_{n\mathbf{k}} =  i \left\langle \frac{\partial u_{n\mathbf{k}}}{\partial \mathbf{k}} \right|
  \times \left| \frac{\partial u_{n\mathbf{k}}}{\partial \mathbf{k}} \right\rangle,
\end{equation}
where we used the fact that the curl of a gradient vanishes identically.
The Berry curvature can also be written as a summation over the eigenstates~\cite{Xiao2010,Xu2014}, which is more convenient for numerical calculations:
\begin{equation}
  \boldsymbol{\Omega}_{n\mathbf{k}} = i \hbar^2 \sum_{m \neq n}
  \frac{\mathbf{v}_{nm\mathbf{k}} \times \mathbf{v}_{mn\mathbf{k}}}{(\varepsilon_{n\mathbf{k}}-\varepsilon_{m\mathbf{k}})^2},
\end{equation}
where $\varepsilon_{n\mathbf{k}}$ denotes the band energy and $\mathbf{v}_{mn\mathbf{k}}\equiv \langle m\mathbf{k} |\hat{\mathbf{p}}/m| n\mathbf{k} \rangle$ are the matrix elements of the canonical velocity operator.

The Berry phase has a wide range of effects on the electronic properties of materials~\cite{Xiao2010}.
In particular, the Berry curvature behaves like an effective magnetic field in momentum space and can directly influence the electron dynamics.
A salient example in carrier transport is that the Berry curvature introduces an anomalous velocity term~\cite{Xiao2010} into the equations of motion of the Bloch electrons~\cite{Sundaram1999}.
This term is responsible for the anomalous Hall effect \cite{Nagaosa2010}, which refers to the appearance of a large spontaneous Hall current in a ferromagnet when an external electric field is applied.
First-principles calculations of the DC anomalous Hall conductivity in ferromagnets based on the Berry curvature formalism have been carried out by various authors~\cite{Fang2003,Yao2004,Wang2006,Wang2007}.

Aside from the anomalous Hall conductivity, it should be noted that most existing transport calculations that neglect the anomalous velocity term were in reasonable agreement with experiment.
This should not be surprising, because for crystals that possess both time-reversal symmetry and spatial inversion symmetry, the Berry curvature vanishes throughout the Brillouin zone~\cite{Xiao2010}.
Moreover for crystals that possess only time-reversal symmetry, the integral of the Berry curvature over the Brillouin zone vanishes.
However, if the time-reversal symmetry is broken by magnetic ordering or if the spatial inversion symmetry is absent in the crystal structure or broken by external perturbations,
the Berry phase effect becomes important.
The recent rise of the field of valleytronics~\cite{Schaibley2016,Vitale2018} originates from the observation that certain transition-metal dichalcogenides lack inversion symmetry
in their monolayer form and possess inequivalent valleys in their electronic structure, with the associated Berry curvatures of equal magnitude but opposite sign.
Such a valley-dependent Berry curvature can be combined with valley-selective carrier pumping~\cite{Mak2012,Zeng2012,Cao2012} to realize devices that use the valley degree of freedom to encode and process information.
Self-consistent, first-principles calculations of charge and spin transport in materials for valleytronics that take into account both the Berry phase effect and various intra- and intervalley carrier scattering mechanisms
present both an opportunity and a challenge for transport simulations.

In addition to modifying the carrier velocity, the Berry phase has profound effects on the scattering and lifetime of charge carriers.
Indeed, it has been shown that for charge carriers, a Berry phase change of $\pi$ under the rotation of the electron wavefunction in $\mathbf{k}$-space leads to the absence of backscattering~\cite{Ando1998}.
This effect contributes to the large carrier mobility of graphene and carbon nanotubes.
As \textit{ab initio} calculations can provide abundant and detailed information on the Berry curvature associated with the electronic structure and microscopic carrier scattering events,
more insights can be gained from the study of the correlation between the Berry phase and the carrier mobility.
This also suggests that in the future, the Berry phase of materials might be engineered to optimize charge and spin transport.

\subsection{Transport in correlated electron systems}\label{Sec5.4}

Correlated electron systems are typically transition metal-based compounds with partially filled $d$ and $f$ shells, where electrons occupy orbitals with narrow spatial extension.
Electrons in these spatially confined orbitals experience strong Coulomb repulsion and their behavior cannot be described accurately using a mean-field theory of independent particles.
The complex competition between the charge, spin, and orbital degrees of freedom in correlated electron systems gives rise to a multitude of exotic phenomena and transport properties,
such as high-temperature superconductivity~\cite{Lee2006}, metal-insulator transitions (MITs)~\cite{Mott1968,Imada1998}, and colossal magnetoresistance~\cite{Jin1994,Salamon2001}.
These properties make correlated-electron materials promising candidates for applications in new generations of electronics and spintronics.
A growing effort is focused on exploring novel device applications of correlated oxide materials that exhibit MITs under external fields~\cite{Yang2011,Zhou2013}.
Field-effect transistors based on an electrostatic doping-induced MIT in correlated materials such as SrTiO$_3$~\cite{Shibuya2007}, KTaO$_3$~\cite{Yoshikawa2009}, La$_2$CuO$_4$~\cite{Misewich2000},
and VO$_2$~\cite{Yang2012,Ruzmetov2009} have been explored experimentally.
These kinds of Mott field-effect transistors exploit the sudden change of the free carrier density across the MIT, which changes the channel conductance by orders of magnitude.
This could lead to devices with lower power consumption and higher switching frequency than silicon-based devices~\cite{Zhou2013}.
The measured carrier mobilities of correlated-electron materials, however, are typically small (less than 10~\cmVs{} at room temperature),
consistent with the localized nature of the $d$ and $f$ orbitals and the associated narrow bands and heavy effective masses.

Understanding and predicting the transport properties of correlated systems from first principles constitutes an outstanding challenge for materials modeling.
While much progress has been made in the development of first-principles methods for computing the mobility of charge carriers in materials that do not exhibit strong electronic correlation,
these methods are mostly based on DFT and DFPT with a semi-local description of exchange and correlation.
However, DFT fails for strongly correlated systems such as Mott insulators~\cite{Mott1968,Imada1998}, due to its tendency to delocalize electrons.
To better account for the onsite Coulomb interaction in correlated insulators, the DFT+$U$ method \cite{Anisimov1997a} is usually employed,
which combines semi-local DFT with a Hubbard $U$ correction for the onsite Coulomb interaction.
The DFT+$U$ method is computationally efficient and has been successful in describing the insulating ground states and the long-range magnetic order of many correlated insulators.
It is expected that combining DFT+$U$ with \textit{ab initio} transport methods will lead to insights on the transport properties of certain classes of correlated-electron systems.
Indeed, a growing amount of studies indicate that the inclusion of electronic correlation effects within the DFT+$U$ framework corrects the underestimation of the EPI
in correlated-electron systems in semi-local DFT~\cite{Giustino2017,Zhang2007,Floris2011,Hong2012}, resulting in better agreement with experiments.

In addition to the DFT+$U$ method, two other approaches have been successful in improving the description of the electronic structure of correlated systems: hybrid functionals~\cite{Perdew1996,Heyd2003}
and the $GW$ method~\cite{Aryasetiawan1998}.
Hybrid functionals incorporate a fraction of the exact exchange energy from Hartree-Fock theory into the semi-local DFT exchange-correlation energy,
while the $GW$ method treats nonlocal and dynamical electron-electron correlations within many-body perturbation theory.
Both approaches lead to a more accurate determination of the electronic band gap and the improved description of the screening properties enables more accurate calculations
of the EPI~\cite{Giustino2017}.
Existing applications of hybrid functionals and the $GW$ method to calculate electron-phonon-related properties have so far mainly focused on a more accurate description of
the superconducting transition temperature in correlated systems~\cite{Yin2013,Komelj2015,Li2019a}.
It is however expected that the same methods could also improve the accuracy of carrier mobility calculations in correlated-electron systems.

The computational efficiency of the DFT+$U$ method has made it a popular choice for first-principles calculations of correlated materials.
However DFT+$U$ represents a static approximation to electronic correlation in solids and as a result it is not capable of describing correlated metallic states
as well as the dynamic transfer of spectral weight across the MIT.
These challenges have been addressed with the development of dynamical mean field theory (DMFT)~\cite{Georges1996,Kotliar2004}.
In essence, DMFT maps a lattice problem of interacting electrons onto a self-consistent single-site model in which a single impurity atom interacts with a reservoir of noninteracting electrons
that represent the rest of the crystal.
Electrons may hop in and out of the impurity site via the hybridization between electrons of the atoms and the bath.
This mapping becomes exact in the limit of a high number of spatial dimensions or coordination number of the lattice.
DMFT has been combined with band structure methods~\cite{Anisimov1997b} such as DFT within the local density approximation (LDA), leading to the so-called LDA+DMFT scheme~\cite{Kotliar2006}.
The LDA+DMFT method has been rather successful for understanding the electronic structure and phase evolution of many correlated-electron materials, such as doped Mott insulators \cite{Marianetti2004},
heavy-fermion systems~\cite{Shim2007}, and iron-based superconductors~\cite{Yin2011a,Yin2011b}.
Developing \textit{ab initio} transport approaches based on DMFT could lead to significant advances in the accuracy of first-principles calculations of carrier mobilities in strongly correlated electron systems.

\section{Summary and outlook}\label{Sec6}

In this manuscript we presented a summary of the current state-of-the-art theoretical description of electronic transport in solids
as well as an overview of the computational studies of several semiconductors and future challenges for first-principles calculations.

On the theory side, we first derived the Boltzmann transport equation from a non-equilibrium Green's function formalism.
Then we discussed some common approximations to the linearized Boltzmann transport equation, namely
the momentum relaxation time approximation, the self-energy relaxation time approximation, and the lowest-order variational approximation.
In each case we made an effort to identify the key approximations involved.
For completeness we provided a brief discussion of the Kubo formalism as an alternative approach.
In Figure~\ref{fig:overview} we made an attempt to establish the relations between the various approximations in use in the literature and where they stand in terms of accuracy and computational cost.

On the front of computational methods, we provided a brief overview of the existing implementations in Table~\ref{table1}.
We then summarized existing computational studies on the transport properties of several three-dimensional and two-dimensional semiconductors.
In particular, we reviewed works on silicon, diamond, GaAs, GaN, $\beta$-Ga$_2$O$_3$, MAPbI$_3$, graphene, silicene, phosphorene, MoS$_2$, and InSe.
In all cases, we identified the integration over all possible phonon momenta in the Boltzmann transport equation as the key challenge for obtaining accurate results
and we highlighted the importance of using very fine wavevector grids in the Brillouin zone, as provided for example by Wannier interpolation.
A comparison of the most recent calculations of carrier mobility with experiments revealed that current \textit{ab initio} methods are fairly predictive
and that the computed mobilities typically lie near the upper end of the experimental range in many cases.
In the case of bulk three-dimensional materials the agreement between theory and experiments is impressive, while for 2D materials some discrepancies remain.
It is expected that by upgrading the computational methodology from standard density function theory to higher-level approaches such as hybrid functionals, GW, and dynamical mean field theory
and by improving at the same time band structures, phonon dispersions, and electron-phonon matrix elements, the gap between theory and experiment will close in the near future.

Finally, we discussed a few possible avenues for future research in the field of first-principles calculation of electronic transport.
One of the most promising areas is the field of quantum materials, in particular topological insulators, and the related field of spintronics.
These calculations will require an accurate description of band topology, Berry curvature, and electron-phonon interactions in the presence of spin-orbit coupling.
While these aspects have not been explored in great detail until now, they do not pose particular challenges from a computational standpoint.
Another interesting class of systems which will require further theoretical and computational developments is provided by strongly correlated systems.
Here, the accurate description of the electronic structure and of electron-phonon interactions remain challenging, but much progress is being made in this area.
Therefore we expect exciting new developments soon.
As many-body techniques become more widely used in electronic-structure calculations, we can expect that in the near future we will be able to explore transport properties from a many-body perspective, for example starting from the Kadanoff-Baym approach summarized in Sec.~\ref{Sec2.1}.

Overall, these seem to be exciting times for computational research on the transport properties of advanced materials and we hope that this review will provide a useful reference frame for future work in this area.

\section*{Acknowledgments}

The authors would like to thank
Nicola Bonini,
Wu Li,
Francesco Macheda,
Roxana Margine,
and Jonathan Yates for stimulating discussions.
This work was supported by the Leverhulme Trust (Grant~RL-2012-001),
the UK Engineering and Physical Sciences Research Council (grant No.~EP/M020517/1),
the Graphene Flagship (Horizon 2020 Grant No.~785219 - GrapheneCore2),
the Marie Sk\l{}odowska-Curie Grant Agreement No.~743580,
and the PRACE-15 and PRACE-17 resources MareNostrum at BSC-CNS.


\appendix

\addtocontents{toc}{\protect\addtolength{\protect\cftsecnumwidth}{47pt}}

\section{Equation of motion for the lesser Green's function}\label{appendix}

Here we provide more details about how to obtain Eq.~\eqref{eq:kbe} from Dyson's equation on the contour, Eq.~\eqref{eq:dysoncontour}.
From the definitions of the Green's function on the contour Eq.~\eqref{eq:greensfuncontour}, the contour ordering symbol,
and the ordering of the three parts of the contour, it follows that we can identify seven unique types of Green's functions,
depending on which part of the contour the two arguments $z_1$ and $z_2$ are located on:
\begin{align}\label{eq:Gdefinitions}
  G^>(\mathbf{r}_1,\mathbf{r}_2;t_1,t_2)                   & \!\equiv\! G(\mathbf{r}_1,\mathbf{r}_2;z_1\!=\!t_{1+},z_2\!=\!t_{2-}), \\
  G^<(\mathbf{r}_1,\mathbf{r}_2;t_1,t_2)                   & \!\equiv\! G(\mathbf{r}_1,\mathbf{r}_2;z_1\!=\!t_{1-},z_2\!=\!t_{2+}), \\
  G^{\mathrm{T}}(\mathbf{r}_1,\mathbf{r}_2;t_1,t_2)            & \!\equiv\! G(\mathbf{r}_1,\mathbf{r}_2;z_1\!=\!t_{1-},z_2\!=\!t_{2-}), \\
  G^{\overline{\mathrm{T}}}(\mathbf{r}_1,\mathbf{r}_2;t_1,t_2) & \!\equiv\! G(\mathbf{r}_1,\mathbf{r}_2;z_1\!=\!t_{1+},z_2\!=\!t_{2+}), \\
  G^{\lceil}(\mathbf{r}_1,\mathbf{r}_2;\tau,t)             & \!\equiv\! G(\mathbf{r}_1,\mathbf{r}_2;z_1\!=\!t_0-i\tau,z_2\!=\!t), \! \\
  G^{\rceil}(\mathbf{r}_1,\mathbf{r}_2;t,\tau)             & \!\equiv\! G(\mathbf{r}_1,\mathbf{r}_2;z_1\!=\!t,z_2\!=\!t_0-i\tau), \! \\
  G^{\mathrm{M}}(\mathbf{r}_1,\mathbf{r}_2;\tau_1,\tau_2)      & \!\equiv\! \nonumber \\
                                            G(\mathbf{r}_1,&\mathbf{r}_2;z_1\!=\!t_0-i\tau_1,z_2\!=\!t_0-i\tau_2),
\end{align}
where $t_{1+(-)}$ denotes the time $t_1$ on the contour branch $\gamma_{+(-)}$ from Fig.~\ref{fig:contour}.
We take the limit $t_0\to-\infty$, which corresponds to the approximation that the system has thermalized
with the surrounding heat bath in the distant past and that there is no correlation between
processes during the thermalization and at times we are interested in.
Mathematically, it can be shown using the Riemann-Lebesgue lemma that in this limit $G^{\lceil,\rceil}\to 0$.
This leads to the function $G^{\mathrm{M}}$ becoming decoupled from the other Green's functions.
The remaining four functions can be written explicitly in terms of the electron field operators:
\begin{align}
  G^>(\mathbf{r}_1,\mathbf{r}_2;t_1,t_2)                   =& \frac{-i}{\hbar}\left\langle\hat{\psi}_{\mathrm{H}}(\mathbf{r}_1,t_1)\hat{\psi}^{\dagger}_{\mathrm{H}}(\mathbf{r}_2,t_2)\right\rangle, \\
  G^<(\mathbf{r}_1,\mathbf{r}_2;t_1,t_2)                   =& \frac{i}{\hbar}\left\langle\hat{\psi}^{\dagger}_{\mathrm{H}}(\mathbf{r}_2,t_2)\hat{\psi}_{\mathrm{H}}(\mathbf{r}_1,t_1)\right\rangle, \\
  G^{\mathrm{T}}(\mathbf{r}_1,\mathbf{r}_2;t_1,t_2)            =& \theta(t_1-t_2)G^>(\mathbf{r}_1,\mathbf{r}_2;t_1,t_2) \nonumber \\
                                           +& \theta(t_2-t_1)G^<(\mathbf{r}_1,\mathbf{r}_2;t_1,t_2), \\
  G^{\overline{\mathrm{T}}}(\mathbf{r}_1,\mathbf{r}_2;t_1,t_2) =& \theta(t_1-t_2)G^<(\mathbf{r}_1,\mathbf{r}_2;t_1,t_2) \nonumber \\
                                           +& \theta(t_2-t_1)G^>(\mathbf{r}_1,\mathbf{r}_2;t_1,t_2),
\end{align}
where $\theta(t)$ denotes the Heaviside step function, and $G^<$
differs by a sign from $G^>$ due to the anti-commuting nature of the electron field operators.
The four functions above are commonly referred to as the \emph{greater}, \emph{lesser}, \emph{time-ordered}, and \emph{anti-time-ordered} Green's functions.
It is convenient to replace the latter two functions by the two linear combinations
\begin{align}
  G^{\mathrm{R}}(\mathbf{r}_1,\mathbf{r}_2;t_1,t_2) = & G^{\mathrm{T}}(\mathbf{r}_1,\mathbf{r}_2;t_1,t_2) \nonumber \\
                                  & - G^<(\mathbf{r}_1,\mathbf{r}_2;t_1,t_2), \\
  G^{\mathrm{A}}(\mathbf{r}_1,\mathbf{r}_2;t_1,t_2) = & G^<(\mathbf{r}_1,\mathbf{r}_2;t_1,t_2) \nonumber \\
                                  & - G^{\overline{\mathrm{T}}}(\mathbf{r}_1,\mathbf{r}_2;t_1,t_2) ,
\end{align}
called the \emph{retarded} and \emph{advanced} Green's function, respectively.
They can also equivalently be written as
\begin{align}
  G^{\mathrm{R}}(\mathbf{r}_1,\mathbf{r}_2;t_1,t_2) =& \theta(t_1-t_2)\big[ G^>(\mathbf{r}_1,\mathbf{r}_2;t_1,t_2) \nonumber \\
                                 & - G^<(\mathbf{r}_1,\mathbf{r}_2;t_1,t_2) \big], \label{eq:Gretared}\\
  G^{\mathrm{A}}(\mathbf{r}_1,\mathbf{r}_2;t_1,t_2) =& \theta(t_2-t_1)\big[ G^<(\mathbf{r}_1,\mathbf{r}_2;t_1,t_2) \nonumber \\
                                 & - G^>(\mathbf{r}_1,\mathbf{r}_2;t_1,t_2) \big]. \label{eq:Gadvanced}
\end{align}
With these definitions, we can write down Dyson's equation for $z_1=t_{1,-}$ and $z_2=t_{2,+}$.
Using the Langreth rules~\cite{Stefanucci2013,Langreth1970}, we find:
\begin{align}\label{eq:dysonlesser}
  G^<(1,2)                  &= G_0^<(1,2)   \nonumber \\
 + \int \mathrm{d} 3 \int \mathrm{d}^3 & r_4   \Big[   G_0^<(1,3) \Sigma^{\delta}(\mathbf{r}_3,\mathbf{r}_4;t_3) G^{\mathrm{A}}(\mathbf{r}_4,\mathbf{r}_2;t_3,t_2) \nonumber \\
                            &+ G_0^{\mathrm{R}}(1,3) \Sigma^{\delta}(\mathbf{r}_3,\mathbf{r}_4;t_3) G^<(\mathbf{r}_4,\mathbf{r}_2;t_3,t_2) \Big] \nonumber  \\
      + \int \mathrm{d} 3 \int \mathrm{d} 4 &   \Big[ G_0^<(1,3) \Sigma^{\mathrm{A}}(3,4) G^{\mathrm{A}}(4,2) \nonumber \\
                            &+ G_0^{\mathrm{R}}(1,3) \Sigma^<(3,4) G^{\mathrm{A}}(4,2) \nonumber \\
                            &+ G_0^{\mathrm{R}}(1,3) \Sigma^{\mathrm{R}}(3,4) G^<(4,2) \Big],
\end{align}
where the definitions of the different component functions of $\Sigma$ follow those for $G$,
and we allowed for the possibility of a time-diagonal self-energy $\Sigma^{\delta}(\mathbf{r}_1,\mathbf{r}_2;t)$.
The latter arises, for example, from the leading-order coupling to a time-dependent external field or from the Hartree electron self-energy.
%
%
We also use the notation $1\equiv(\mathbf{r}_1,t_1)$ and $\int \mathrm{d} 1 \equiv \int_{-\infty}^{+\infty} \mathrm{d} t_1 \int \mathrm{d}^3 r_1$.
Note that in the case of the unperturbed Green's function $G_0$, as defined in Eq.~\eqref{eq:g0},
 the thermal and quantum average $\langle \ldots \rangle$ is evaluated with the exactly diagonalizable weight operator $\exp(-\hat{H}_0/k_B T)/Z_0$; furthermore, the field operators in $G^{<,>}_0$ are in the interaction picture,
$\hat{\psi}_{\mathrm{I}}(\mathbf{r},t) = \mathrm{e}^{\frac{i}{\hbar}\hat{H}_0t} \hat{\psi}(\mathbf{r}) \mathrm{e}^{\frac{-i}{\hbar}\hat{H}_0t}$, instead of the Heisenberg picture.

In order to simplify Eq.~\eqref{eq:dysonlesser} and arrive at an equation of motion
for $G^{<}$, we introduce the inverse of the non-interacting Green's function
\begin{equation}\label{eq:g0minus1}
  G^{-1}_0(1,2) = \delta(1,2) \Big[ i \hbar \frac{\partial }{\partial t_2} - h_0(\mathbf{r}_2,-i\hbar\nabla_2) \Big],
\end{equation}
where $h_0$ was defined in Eq.~\eqref{eq:h0}.
We can then make use of the fact that
\begin{align}
  \int \mathrm{d} 2 \, G^{-1}_0(1,2) G^{>,<}_0(2,3) &= 0, \label{eq:identity1}\\
  \int \mathrm{d} 2 \, G^{-1}_0(1,2) G^{\mathrm{R},\mathrm{A}}_0(2,3) &= \delta(1,3), \label{eq:identity2}
\end{align}
which follows directly from the definitions of $\hat{H}_0$ and $h_0$,
from $\partial \theta(t-t')/\partial t = \delta(t-t')$, and from the
anti-commutation relations for the electronic field operators:
\begin{align}
  \big\{\hat{\psi}(\mathbf{r}_1),\hat{\psi}(\mathbf{r}_2)\big\}     &= \big\{\hat{\psi}^{\dagger}(\mathbf{r}_1),\hat{\psi}^{\dagger}(\mathbf{r}_2)\big\} = 0, \\
  \big\{\hat{\psi}(\mathbf{r}_1),\hat{\psi}^{\dagger}(\mathbf{r}_2)\big\} &= \delta^{(3)}(\mathbf{r}_1-\mathbf{r}_2).
\end{align}
The latter also hold at equal times in the interaction and Heisenberg pictures.

We now multiply both sides of Eq.~\eqref{eq:dysonlesser} from the left with $G^{-1}_0$ from Eq.~\eqref{eq:g0minus1},
make use of Eqs.~\eqref{eq:identity1} and \eqref{eq:identity2}, and integrate over $(\mathbf{r}_1,t_1)$.
After relabeling some space-time coordinates we find that Eq.~\eqref{eq:dysonlesser} becomes:
\begin{multline}\label{eq:eomlesser1}
   \int \! \mathrm{d} 3 \, G^{-1}_0(1,3) G^<(3,2) \\
    = \int \! \mathrm{d}^3 r_3  \Sigma^{\delta}(\mathbf{r}_1,\mathbf{r}_3;t_1) G^<(\mathbf{r}_3,\mathbf{r}_2;t_1,t_2) \\
      \! + \! \int \! \mathrm{d} 3  \left[ \Sigma^<(1,3) G^{\mathrm{A}}(3,2) \!+\! \Sigma^{\mathrm{R}}(1,3) G^<(3,2) \right].
\end{multline}
We note that Eq.~\eqref{eq:dysonlesser} also holds if $G_0$ and $G$ are interchanged in the terms involving the
self-energy on the right-hand side.
We can obtain a similar equation to Eq.~\eqref{eq:eomlesser1} by applying $G^{-1}_0$ from the right to this interchanged version of Eq.~\eqref{eq:dysonlesser}
and obtain:
\begin{multline}\label{eq:eomlesser2}
      \int \! \mathrm{d} 3 G^<(1,3) G^{-1}_0(3,2) \\
    =\int \! \mathrm{d}^3 r_3  G^<(\mathbf{r}_1,\mathbf{r}_3;t_1,t_2) \Sigma^{\delta}(\mathbf{r}_3,\mathbf{r}_2;t_2) \\
      \! +\! \int \! \mathrm{d} 3 \left[ G^<(1,3) \Sigma^{\mathrm{A}}(3,2) \!+\! G^{\mathrm{R}}(1,3) \Sigma^<(3,2) \right].
\end{multline}
Finally, we subtract Eq.~\eqref{eq:eomlesser2} from \eqref{eq:eomlesser1} and evaluate the resulting equation at equal times $t_1=t_2=t$,
since only the time-diagonal $G^<$ is needed to calculate the current density, Eq.~\eqref{eq:currentdensity}.
Using the definitions of $G^{-1}_0$, $G^{\mathrm{R,A}}$, and $\Sigma^{\mathrm{R,A}}$, and the multi-dimensional chain rule for the derivative $\partial/\partial t$
then yields the Kadanoff-Baym equation of motion for $G^<$, Eq.~\eqref{eq:kbe}.


\section{Current-current correlation function on the Keldysh-Schwinger contour}\label{appendix2}

Within the Keldysh-Schwinger contour formalism, we calculate the linear response
of the current density to an external electric field by expanding the exponential factor
$\exp [-i/\hbar \int_{\gamma} \mathrm{d} z \,\hat{H}(z)]$ in powers of the external Hamiltonian.
To linear order in $\hat{H}_{\mathrm{ext}}(z)$, the expectation value of the current density reads
\begin{align}\label{eq:currentexpansion}
     \mathbf{J}(\mathbf{r},z) &= \frac{1}{Z} \mathrm{tr} \Big\{ \mathcal{T}_{\mathrm{C}}
    \mathrm{e}^{-\frac{i}{\hbar} \int_{\gamma} \mathrm{d} z' \, \hat{H}(z')} \hat{\mathbf{J}}(\mathbf{r},z)  \Big\} \\
    & \simeq \big\langle \hat{\mathbf{J}}(\mathbf{r},z) \big\rangle_{\mathrm{eq}}
            -\frac{i}{\hbar} \int_{\gamma} \mathrm{d} z' \Big\{ \big\langle \hat{\mathbf{J}}(\mathbf{r},z) \hat{H}_{\mathrm{ext}}(z') \big\rangle_{\mathrm{eq}} \nonumber \\
          & \quad \quad  - \big\langle \hat{\mathbf{J}}(\mathbf{r},z) \big\rangle_{\mathrm{eq}}
            \big\langle \hat{H}_{\mathrm{ext}}(z') \big\rangle_{\mathrm{eq}} \Big\}, \label{eq:currentexpansion2}
\end{align}
where we introduced the short-hand notation
\begin{multline}
  \big\langle \hat{O}_1(z_1)\hat{O}_2(z_2) \big\rangle_{\mathrm{eq}} \\
  \equiv \frac{1}{Z_{\mathrm{eq}}} \mathrm{tr} \Big\{ \mathcal{T}_{\mathrm{C}}
   \mathrm{e}^{-\frac{i}{\hbar} \int_{\gamma} \mathrm{d} z \, [\hat{H}_{\mathrm{eq}}]_{z}} \hat{O}_1(z_1) \hat{O}_2(z_2)  \Big\},
\end{multline}
and identified the partition function without external fields as $Z_{\mathrm{eq}} \equiv \mathrm{tr}\{\exp[-\hat{H}_{\mathrm{eq}}/k_B T]\}$.
Note that the third term in Eq.~\eqref{eq:currentexpansion2} arises from an expansion of the partition function $Z$.
This can be accomplished by writing it in the form
\begin{align}
  Z = \mathrm{tr}\Big\{  \mathrm{e}^{-\beta \hat{H}(t_0)} \Big\} &= \mathrm{tr}\Big\{ \mathcal{T}_{\mathrm{C}}   \mathrm{e}^{ \frac{-i}{\hbar} \int_{\gamma_{\mathrm{M}}} \mathrm{d} z \hat{H}(z)} \Big\}, \\
                                                    &= \mathrm{tr}\Big\{ \mathcal{T}_{\mathrm{C}}   \mathrm{e}^{ \frac{-i}{\hbar} \int_{\gamma} \mathrm{d} z \hat{H}(z)} \Big\},
\end{align}
where we made use of the fact that, in the absence of any other operators, the exponentials involving the integrals along $\gamma_-$ and $\gamma_+$ cancel each other.
Retaining only terms linear in the vector potential $\mathbf{A}(z)$, the expectation value of the total current density
in the cartesian direction $\alpha$ on the contour reads
\begin{multline}\label{eq:currentcontour}
  J_{\alpha}(\mathbf{r},z) = \big\langle [\hat{J}^{(\mathrm{p})}(\mathbf{r})]_z \big\rangle_{\mathrm{eq}}
  + \big\langle \hat{J}^{(\mathrm{d})}(\mathbf{r},z) \big\rangle_{\mathrm{eq}} \\
  + \frac{i}{\hbar} \sum_{\beta} \int_{\gamma} \mathrm{d} z' \, A_{\beta}(z') \int \mathrm{d}^3 r' \, \mathcal{J}_{\alpha,\beta}^{(\textrm{p})}(\mathbf{r},\mathbf{r}';z,z'),
\end{multline}
where
\begin{multline}\label{eq:correlationcorrelationfct}
  \mathcal{J}_{\alpha,\beta}^{(\textrm{p})}(\mathbf{r},\mathbf{r}';z,z') \equiv \big\langle [\hat{J}_{\alpha}^{(\mathrm{p})}(\mathbf{r})]_z  [\hat{J}_{\beta}^{(\mathrm{p})}(\mathbf{r}')]_{z'} \big\rangle_{\mathrm{eq}} \\
  - \big\langle [\hat{J}_{\alpha}^{(\mathrm{p})}(\mathbf{r})]_z \big\rangle_{\mathrm{eq}}
    \big\langle [\hat{J}_{\beta}^{(\mathrm{p})}(\mathbf{r}')]_{z'} \big\rangle_{\mathrm{eq}},
\end{multline}
is the connected part of the current-current correlation function.
To get an expression for the expectation value of the current density at time $t$,
we evaluate Eq.~\eqref{eq:currentcontour} at $z=t_-$ or $z=t_+$, and obtain in both cases
\begin{multline}
  J_{\alpha}(\mathbf{r},t) = J^{(\mathrm{p})}_{0,\alpha}(\mathbf{r},t) + \frac{e}{m} A_{\alpha}(t) \varrho_0(\mathbf{r},t) \\
  + \frac{i}{\hbar} \sum_{\beta} \int_{-\infty}^{+\infty} \mathrm{d} t' \, A_{\beta}(t') \mathcal{J}_{\alpha,\beta}^{(\textrm{p}),\textrm{R}} (\mathbf{r},\mathbf{r}';t,t').
\end{multline}
Here we identified the expectation values of the current density and the charge density in the absence of external fields
and furthermore identified the \emph{retarded} component of the current-current correlation function as
\begin{multline}
  \mathcal{J}_{\alpha,\beta}^{(\textrm{p}),\textrm{R}}(\mathbf{r},\mathbf{r}';t,t')  = \mathcal{J}_{\alpha,\beta}^{(\textrm{p})}(\mathbf{r},\mathbf{r}';z=t_-,z'={t'_-}) \\
  - \mathcal{J}_{\alpha,\beta}^{(\textrm{p})}(\mathbf{r},\mathbf{r}';z=t_-,z'={t'_+}),
\end{multline}
in analogy to the definition for the electronic Green's function provided in \ref{appendix}.

\section{Kubo formula in the independent-particle approximation}\label{indepparaapprox}

The correlation function on the contour, Eq.~\eqref{eq:correlationcorrelationfct}, can be written in terms of the electron field operators as
\begin{multline}\label{eq:correlationfunctionfield}
  \mathcal{J}_{\alpha,\beta}^{(\textrm{p})}(\mathbf{r},\mathbf{r}';z,z') =
  -\frac{e^2 \hbar^2}{m^2} \lim_{\substack{\tilde{\mathbf{r}} \to \mathbf{r} \\ \tilde{\mathbf{r}}' \to \mathbf{r}'}} \frac{\partial}{\partial r_{\alpha}} \frac{\partial}{\partial r'_{\beta}} \\
  \Big\{ \big\langle [\hat{\psi}^{\dagger}(\tilde{\mathbf{r}})]_z [\hat{\psi}(\mathbf{r})]_z  [\hat{\psi}^{\dagger}(\tilde{\mathbf{r}}')]_{z'} [\hat{\psi}(\mathbf{r}')]_{z'} \big\rangle_{\mathrm{eq}} \\
  \! -\big\langle [\hat{\psi}^{\dagger}(\tilde{\mathbf{r}})]_z [\hat{\psi}(\mathbf{r})]_z \big\rangle_{\mathrm{eq}}  \big\langle [\hat{\psi}^{\dagger}(\tilde{\mathbf{r}}')]_{z'} [\hat{\psi}(\mathbf{r}')]_{z'} \big\rangle_{\mathrm{eq}} \Big\}.
\end{multline}
In the \emph{independent-particle approximation} (IPA), the two-particle correlation function in the second line of Eq.~\eqref{eq:correlationfunctionfield} is approximated
by the sum of two products of two one-particle correlation functions each,
corresponding to the two possible pairwise pairings of the field operators.
One of the two products is canceled by the term in the third line, so that the independent-particle version
of the current-current correlation function reads
\begin{multline}
  \mathcal{J}_{\alpha,\beta}^{(\textrm{p})}(\mathbf{r},\mathbf{r}';z,z') \stackrel{\text{IPA}}{\approx}
  - \frac{e^2 \hbar^4}{m^2} \Big[ \frac{\partial}{\partial \mathbf{r}_{\alpha}} G(\mathbf{r},\mathbf{r}';z,z') \Big] \\
  \times \Big[ \frac{\partial}{\partial \mathbf{r}'_{\beta}} G(\mathbf{r}',\mathbf{r};z',z) \Big],
\end{multline}
where we accounted for an extra minus sign arising from the anti-commuting nature of the field operators.
Here, the one-particle Green's function is understood to be defined with respect to the Hamiltonian $\hat{H}_{\mathrm{eq}}$.
We expand the one-particle Green's function in the basis of known eigenstates of $\hat{H}_0$, and retain only the diagonal terms:
\begin{equation}
  \!\! \! \! G(\mathbf{r},\mathbf{r}';z,z') \! \approx \! \sum_n \!\! \int \! \frac{\mathrm{d}^3 k}{\Omega_{\mathrm{BZ}}} \, \varphi_{n\mathbf{k}}(\mathbf{r}) \varphi^*_{n\mathbf{k}}(\mathbf{r}') G_{n\mathbf{k}}(z,z').
\end{equation}
This yields the following compact expression for the spatially integrated current-current correlation function in the IPA:
\begin{multline}
  \frac{1}{V} \! \int \! \mathrm{d}^3 r \! \int \! \mathrm{d}^3 r'   \mathcal{J}_{\alpha,\beta}^{(\textrm{p})}(\mathbf{r},\mathbf{r}';z,z') \approx \frac{e^2 \hbar^2}{V_{\mathrm{uc}}}  \\
  \times \sum_{mn} \int \frac{\mathrm{d}^3 k}{\Omega_{\mathrm{BZ}}}
  v^{\alpha}_{mn\mathbf{k}} v^{\beta}_{nm\mathbf{k}} G_{n\mathbf{k}}(z,z')G_{m\mathbf{k}}(z',z),
\end{multline}
where we identified the previously defined velocity matrix elements $v^{\alpha}_{mn\mathbf{k}}$.
We can then obtain the retarded part of the current-current correlation function
in terms of the components of the one-particle Green's functions defined in \ref{appendix}:
\begin{multline}
  \frac{1}{V} \!\! \int \!\! \mathrm{d}^3 r \!\! \int \!\! \mathrm{d}^3 r' \mathcal{J}_{\alpha,\beta}^{(\textrm{p})}(\mathbf{r},\mathbf{r}';t,t') \approx \frac{e^2 \hbar^2}{V_{\mathrm{uc}}} \sum_{mn} \! \int \!\!  \frac{\mathrm{d}^3 k}{\Omega_{\mathrm{BZ}}} v^{\alpha}_{mn\mathbf{k}}  \\
 \times \! v^{\beta}_{nm\mathbf{k}} [G^{\mathrm{R}}_{n\mathbf{k}}\!(t,t')G^<_{m\mathbf{k}}\!(t',t) \!+\! G^<_{n\mathbf{k}}\!(t,t')G^{\mathrm{A}}_{m\mathbf{k}}\!(t',t)].
\end{multline}
Lastly, we use the spectral representation of the retarded, advanced, and lesser one-particle Green's functions~\cite{Kita2010},
\begin{align}
  G^<_{n\mathbf{k}}(t,t') &= \frac{i}{\hbar} \int \mathrm{d} \omega \mathcal{A}_{n\mathbf{k}}(\omega ) f(\hbar\omega ) \mathrm{e}^{-i \omega (t-t')} \\
  G^{\mathrm{R},\mathrm{A}}_{n\mathbf{k}}(t,t') &= \frac{1}{\hbar} \! \int \! \frac{\mathrm{d} \omega }{2 \pi} \! \int \! \mathrm{d} \omega '
                                   \frac{\mathcal{A}_{n\mathbf{k}}(\omega ')}{\omega - \omega ' \pm i \eta} \mathrm{e}^{-i \omega (t-t')},
\end{align}
where $\eta = 0^+$ denotes a positive infinitesimal, and $\mathcal{A}_{n\mathbf{k}}(\omega ) \geq 0$ is the electronic spectral function,
to arrive at Eq.~\eqref{eq:sigmaIPA}.

\section*{References}


\providecommand{\newblock}{}

\end{document}